\documentclass[aps,prd,amsmath,floats,floatfix, twocolumn, superscriptaddress,nofootinbib,showpacs,longbibliography]{revtex4-1}
\usepackage[T1]{fontenc}
\usepackage[utf8]{inputenc}
\usepackage{lmodern}
\usepackage{mathtools}
\usepackage{enumitem}
\usepackage{verbatim}
\usepackage{multirow}
\usepackage[dvipsnames, usenames]{xcolor}
\definecolor{linkcolor}{rgb}{0.0,0.3,0.5}
\usepackage[hypertexnames=false, unicode, colorlinks=true, linkcolor=linkcolor,
citecolor=linkcolor, filecolor=linkcolor,urlcolor=linkcolor,
pdfusetitle]{hyperref}
\usepackage{orcidlink}
\usepackage[all]{hypcap}
\usepackage{graphicx}
\usepackage{amssymb}
\usepackage[normalem]{ulem} 
\usepackage{bm} 
\usepackage[english]{babel}
\usepackage{blindtext}
\usepackage{ragged2e}
\usepackage[labelfont={small,bf}, font=footnotesize, justification=justified,format=plain,singlelinecheck=off]{caption}

\usepackage[font=footnotesize]{subcaption}
\usepackage{amsmath}
\usepackage{physics}
\usepackage{xspace}
\DeclareMathOperator*{\argmin}{arg~min}

\def\Sref#1{Sect.~\ref{#1}\xspace}
\def\Fref#1{Fig.~\ref{#1}\xspace}

\def\Eref#1{Eq.~\ref{#1}\xspace}

\begin{document}

\title{Unveiling Microlensing Biases in Testing General Relativity with Gravitational Waves}

\author{Anuj Mishra\orcidlink{0000-0002-2580-2339}}
\email{anuj@iucaa.in}
\affiliation{Inter-University Centre for Astronomy and Astrophysics, Post Bag 4, Ganeshkhind, Pune 411007, India}

\author{N. V. Krishnendu}
\email{krishnendu.nv@icts.res.in}
\affiliation{International Centre for Theoretical Sciences, Tata Institute of Fundamental Research, Bengaluru - 560089, India}

\author{Apratim Ganguly\orcidlink{0000-0001-7394-0755}}
\email{apratim@iucaa.in}
\affiliation{Inter-University Centre for Astronomy and Astrophysics, Post Bag 4, Ganeshkhind, Pune 411007, India}


\begin{abstract}
{\footnotesize Gravitational waves (GW) from chirping binary black holes (BBHs) provide unique opportunities to test general relativity (GR) in the strong-field regime.
However, testing GR can be challenging when incomplete physical modeling of the expected signal gives rise to systematic biases.
In this study, we investigate the potential influence of wave effects in gravitational lensing (which we refer to as microlensing)
on tests of GR using GWs for the first time.
We utilize an isolated point-lens model for microlensing with the lens mass ranging from $10-10^5~$M$_\odot$ and base our conclusions on an astrophysically motivated population of BBHs in the LIGO-Virgo detector network.
Our analysis centers on two theory-agnostic tests of gravity: the inspiral-merger-ringdown consistency test (IMRCT) and the parameterized tests, providing insights into deviations from GR across different evolutionary phases of GW signals: inspiral, intermediate, and merger-ringdown. 
Our findings reveal two key insights: First, microlensing can significantly bias GR tests, with a confidence level exceeding $5\sigma$. 
Notably, substantial deviations from GR $(\sigma > 3)$ tend to align with a strong preference for microlensing over an unlensed signal, underscoring the need for microlensing analysis before claiming any erroneous GR deviations.
Nonetheless, we do encounter scenarios where deviations from GR remain significant ($1 < \sigma < 3$), yet the Bayes factor lacks the strength to confidently assert microlensing. Second, deviations from GR correlate with pronounced interference effects, which appear when the GW frequency ($f_\mathrm{GW}$) aligns with the inverse time delay between microlens-induced images ($t_\mathrm{d}$).
These false deviations peak in the wave-dominated region 
and fade where $f_\mathrm{GW}\cdot t_\mathrm{d}$ significantly deviates from unity.
Particularly, in the geometrical optics regime $(f_\mathrm{GW}\cdot t_\mathrm{d}\gg 1)$, biases remain minimal despite instances of strong microlensing effects. 
Our findings apply broadly to any microlensing scenario, extending beyond specific models and parameter spaces, as we relate the observed biases to the fundamental characteristics of lensing.
}
\end{abstract}

\maketitle

\section{\label{sec:intro}Introduction}
General Relativity (GR), proposed by Einstein over a century ago, has withstood numerous strong field gravity tests and remains the most successful theory of gravity to date. These tests encompass a wide range of phenomena, from precise solar system measurements~\cite{Will:2014kxa} to the observation of binary pulsars~\cite{Wex:2014nva}, and most notably, the direct detection of gravitational waves (GWs) by the LIGO~\cite{LIGOScientific:2014pky} and Virgo~\cite{VIRGO:2014yos} detectors. The groundbreaking observations from the first three observing runs of Advanced LIGO and Advanced Virgo~\cite{LIGOScientific:2018mvr,LIGOScientific:2020ibl,LIGOScientific:2021usb,LIGOScientific:2021djp} have provided a unique opportunity to explore the characteristics of gravity within the highly nonlinear and dynamic regime~\cite{GW150914-TGR-LVC, GW170817_TGR_LVC, GWTC-1-TGR-LVC,GWTC-2-TGR-LVC,GWTC-3-TGR-LVK}, thus subjecting Einstein's theory of gravity to novel scrutiny. The remarkable agreement between these observations and the predictions of GR has consistently reaffirmed the theory's robustness and validity.

The LIGO-Virgo-KAGRA (LVK) collaboration has been instrumental in conducting rigorous tests of GR using GWs. These tests primarily rely on GR to identify potential deviations from its predictions, as no reliable modified gravity waveform models are currently available which can describe the complete inspiral-merger-ringdown dynamics of a compact binary system. It involves comparing the data collected by GW detectors with theoretical waveform models. While such tests have effectively probed GR in various aspects, they also exhibit sensitivity to unaccounted-for physical phenomena, including eccentricity and the presence of exotic compact objects, in addition to potential beyond-GR physics. Hence, it is imperative that these models incorporate all relevant physics to avoid any potential false biases in the tests.

One particularly intriguing source of bias in tests of GR arises from the propagation effect of lensing of GWs when they traverse the vicinity of massive objects. The observed alterations encompass a spectrum of manifestations, including repeated events, phase shifts, variations in amplitude, the emergence of beating patterns, and distortions. From probing the relative speeds of GWs with respect to light~\cite{Collett:2016dey,Fan:2016swi}, modified velocity dispersion~\cite{Yang:2018bdf,Chung:2021rcu} and propagation~\cite{Mukherjee:2019wcg,Mukherjee:2019wfw,Finke:2021znb,Iacovelli:2022tlw}, to understanding the nature of GW polarization in alternate theories of gravity with additional degrees of freedom~\cite{Goyal:2020bkm,MaganaHernandez:2022ayv} leading to birefringence~\cite{Goyal:2023uvm}, lensing can be a useful tool for GR tests.

In this paper, we will focus on the lensing scenario where a GW signal traverses the spacetime near an isolated compact object (microlens) such that the (micro-)images produced by the microlens are not resolved temporally, leading to either interference effects or superposition of images. This phenomenon, which we refer to as \textit{microlensing}\footnote{Some authors specifically differentiate the phenomenon where multiple images overlap from interference effects, referring to it as \textit{millilensing}~\citep[e.g.,][]{Liu:2023ikc}. Millilensing can also result in a modulated signal ~\citep[see Fig. 1 in][]{Janquart:2023mvf}. In our study, while we primarily focus on the lensing parameter space relevant for inducing wave effects, we use the term "geometric optics regime" within our microlensing parameter space for cases where signals can be assumed to overlap as a result of lensing.}, may lead to significant frequency-dependent effects if the frequency of the signal is comparable to the inverse time delay between microimages~\citep{1986ApJ...307...30D, Nakamura:1997sw, Baraldo:1999ny, Nakamura:1999uwi, Bulashenko:2021fes, Leung:2023lmq, mishra2023exploring}. 
These microlenses can be in the form of stars, stellar remnants, or even compact dark matter, etc., in the mass-range $\sim 1-10^5~$M$_\odot$. This mass range is particularly pertinent for ground-based detectors searching for microlensing signatures and aligns with the parameter space used by the LVK collaboration~\citep{LIGOScientific:2021izm,LIGOScientific:2023bwz}.
We explore the underlying physics of these lensing phenomena and investigate how they can affect the observed GW signals and their potential to introduce biases in the tests of GR.
We note that the search for such microlensing signatures can shed light on the fraction of compact dark matter objects in this mass range~\citep{Jung:2017flg, Basak:2021ten,Diego:2019rzc}.
In this study, we restrict ourselves to isolated point lenses and do not consider complex scenarios of microlensing, such as microlenses embedded in a strong lens potential~\citep{Diego:2019lcd, Diego:2019rzc, Seo:2021ucd, Mishra:2021xzz, Meena:2022unp, Meena:2023qdq}.
We also restrict ourselves to ground-based detectors. However, wave-optics effects in the context of space-based detectors can also affect the inferred parameters~\citep{Caliskan:2022hbu, Caliskan:2023zqm} and might lead to biases in the tests of GR.
We also note that we do not consider lensing scenarios where the images are completely separated out, i.e., the time delay between the images is larger than the duration of the signal (strong-lensing cases).

We focus on two fundamental types of tests: consistency tests and parameterized tests, each designed to scrutinize GR in distinct ways. 
Consistency tests, as the name suggests, focus on evaluating the adherence of observed GW signals to the expected behavior prescribed by GR, without invoking specific parameterizations of deviations from the theory. These tests serve as a measure of the self-consistency of the signal or its overall consistency with the available data. In this study, we use the inspiral-merger-ringdown consistency test (IMRCT)~\cite{Ghosh2016,Ghosh2017}. The IMRCT seeks to ascertain the consistency between the low-frequency and high-frequency components of the GW signal, shedding light on potential departures from GR~\cite{GW150914-TGR-LVC,GWTC-1-TGR-LVC,GWTC-2-TGR-LVC,GWTC-3-TGR-LVK}. In contrast to consistency tests, parameterized tests are tailored to invoke specific parametrizations that are particularly suited to uncover deviations from GR rooted in distinct physical effects. For instance, here we deploy the parameterized tests of post-Newtonian (PN) coefficients which are designed to be sensitive to physical effects manifesting at different PN orders~\cite{ParametrizedTests_Arun2006a,ParametrizedTests_Arun2006b,PPE_Yunes2009,TIGER_2011,Yunes:2016jcc}, and analytical phenomenological coefficients to model the intermediate and merger-ringdown regimes~\cite{GWTC-2-TGR-LVC}.

In this study, we explore the possibility of microlensing-induced biases in tests of GR.
We first analyze simulated GW150914-like microlensed signals, which serve as our initial motivation for the exploration of these biases and enable us to search for any possible patterns in the distortions. We find that the biases are pronounced where the wave effects are dominated but do not correlate with the detection statistic that the LVK collaboration uses for microlensing searches. We further validate our findings for a population of simulated GW signals.
Through a rigorous examination of this lensing scenario, we contribute to the ongoing effort to ensure the accuracy and robustness of GR tests conducted in the strong gravitational field regime. To our knowledge, this is the first such study that looks into the biases in tests of GR due to frequency-dependent modulations coming from wave-optics effects.
However, we do note that some previous studies have explicitly looked into the effect of Morse phase shift in type-II strong lensing scenarios, which is also a consequence of the wave-optics effect capable of biasing inferred parameters~\citep{Ezquiaga:2022nak, Janquart:2021nus, Vijaykumar:2022dlp}. These phase shifts can introduce distortions that may be degenerate with GW propagation involving a modified dispersion relation, potentially leading to biases in tests of GR~\cite{Ezquiaga:2022nak}.

This paper is organized as follows: In Sect. \ref{sec:basics}, we discuss the basics of microlensing and the two tests of GR that we adopt for our study. We also give details of our computational setup. In Sect. \ref{sec:results_and_discussions}, we discuss the results from our analysis and conclude in Sect. \ref{sec:conclusion}. We consistently report all mass related quantities, such as $M_\mathrm{Lz}$, in units of the solar mass (M$_\odot$).

\section{\label{sec:basics}Basics: Microlensing, Tests of General Relativity and Computational Setup}
\subsection{\label{subsec:basics_microlensing}Microlensing}
Gravitational lensing (GL) is a well-established phenomenon in gravitational physics that occurs when signals, such as light or GWs, 
are influenced by the gravitational field of massive objects. 
In case of strong-lensing, where geometrical (or ray) optics hold, multiple images of the same source are formed, which, for electromagnetic (EM) waves, has been extensively studied and observed in various astrophysical contexts \citep{mollerach2002gravitational, 1992grle.book.....S}. In the context of GWs, these images can be separated by a time delay of a few hours to several months \citep{Oguri:2018muv, More:2021kpb}. However, the lensing of GWs is yet to be observed \citep{Hannuksela:2019kle, Dai:2020tpj, LIGOScientific:2021izm, LIGOScientific:2023bwz}.

While the underlying theory of lensing is similar for both EM waves and GWs, their implications can have striking differences. 
The primary source of difference arises from the fact that in the case of GW propagation and detection, the phase information is preserved. 
This preservation of phase information contrasts with EM waves, where we typically measure flux due to the incoherent radiations emitted by astrophysical sources.
As a result, the lensing of EM waves differs from that of GWs, even within the regime where geometrical optics is valid.
For instance, when a  gravitational lens leads to the formation of multiple images, apart from affecting the amplification of the signal, it also induces a characteristic constant phase shift known as `Morse phase' \citep{1992grle.book.....S, Dai:2017huk, Ezquiaga:2022nak, Janquart:2021nus, Bulashenko:2021fes}. Whereas in the case of GWs, this phase shift can be explicitly measured for saddle-point images.
Moreover, wave effects can arise in the lensing of GWs. For instance, the intervening compact object(s) in the mass range~$\sim (1,~10^5)~{\rm M_\odot}$ can lead to interference effects in GWs detected in the ground-based detectors. 
This is because the presence of an intervening lens can lead to the formation of (micro-)images such that the wavelength of the signal $\lambda_{\rm GW}$ (or frequency $f_{\rm GW}$) is comparable to the time-delay between these micro-images $t_\mathrm{d}$ (in natural units), i.e., when $f_{\rm GW} t_\mathrm{d}\sim 1$~\citep[e.g.,][]{Nakamura:1999uwi, Takahashi:2003ix}. 
This condition can also be loosely understood as occurring when the wavelength of the signal is comparable to the Schwarzschild radius $R_\mathrm{Sch}$ of the lens (because $t_\mathrm{d}\sim R_\mathrm{Sch}/c$, where $c$ is the speed of light).

In such a scenario, the signal is affected
by a frequency-dependent amplification factor, defined as the ratio of the lensed signal to the unlensed signal in the frequency domain. 
It is obtained from the diffraction integral~\citep[e.g.,][]{1992grle.book.....S, goodman2005introduction},
\begin{equation}
F(f, \pmb{y}) = \frac{f}{i}\int d^2\pmb{x} ~ e^{i2\pi f t_{\rm d}(\pmb{x}, \pmb{y})} ,
\label{eq:amp_fac}
\end{equation}
where $f$ is the frequency of the signal; $\pmb{x}\equiv(x_1, x_2)$ and $\pmb{y}\equiv(y_1, y_2)$ denote the dimensionless lens and source plane coordinates, respectively. 
The (double) integral is performed over the lens plane; $t_{\rm d}(\pmb{x}, \pmb{y})$ denotes the time delay function (measured with respect to its unlensed counterpart) given as
\begin{equation}
    t_{\rm d}(\pmb{x}, \pmb{y}) = \frac{(1+z_{\rm l})\xi_0^2}{c}\frac{D_{\rm s}}{D_{\rm l} D_{\rm ds}} 
    \left[\frac{1}{2}(\pmb{x} -\pmb{y})^2 - \psi\left(\pmb{x}\right)
    + \phi_{\rm m}\left(y\right)\right]\,,
\label{eq:time_delay}
\end{equation}
where $D_{\rm l}$, $D_{\rm s}$, and $D_{\rm ds}$ respectively denote the angular diameter distances between the observer and the lens/deflector, the observer and the source, and the lens and the source. The lens redshift is denoted as $z_{\rm l}$, $\xi_0$ serves as an arbitrary scale factor to make coordinates dimensionless, $\psi(\pmb{x})$ is the Fermat lens potential, $\phi_m\left(y\right)$ is a constant independent of lens properties, and $y$ represents the impact parameter defined as the magnitude of the vector $\pmb{y}$.
Using \Eref{eq:amp_fac}, the microlensed waveform $\tilde{h}_\mathrm{ML}(f)$ is obtained from the unlensed waveform $\tilde{h}_\mathrm{UL}(f)$ as
\begin{equation}
h_{\text{ML}}(f; \{\pmb{\lambda},\pmb{\lambda}_\mathrm{lens}\}) = F(f; \pmb{\lambda}_\mathrm{lens})\cdot h_{\text{UL}}(f; \pmb{\lambda}), 
\label{eq:microlensed_waveform}
\end{equation}
where $f$ represents the frequency of the GW signal, $\pmb{\lambda}$ denotes the set of parameters defining the unlensed GW signal, and $\pmb{\lambda}_\mathrm{lens}$ denotes the lens-related parameters governing lensing effects.

The diffraction integral, \Eref{eq:amp_fac}, can be solved analytically only for some trivial lens models, such as an isolated point-mass lens model \citep{Takahashi:2003ix}, where $\psi\left(\pmb{x}\right)=\log(|\pmb{x}|)$ and the amplification factor takes the form~\citep{Takahashi:2003ix}
\begin{equation}
    \begin{split}
        F\left(\omega,y\right) = \exp\bigg\{\frac{\pi \omega}{4} + 
        \frac{i\omega}{2}\left[\ln\left(\frac{\omega}{2}\right) - 
        2\phi_{\rm m}\left(y\right)\right]\bigg\} \\
        \times\ \Gamma\left(1-\frac{i\omega}{2}\right)
        {}_{1}{F}_{1}\left(\frac{i\omega}{2},1;\frac{i\omega y^2}{2}\right),
    \end{split}
    \label{eq:amp fac point}
\end{equation}
where $\omega$ represents the dimensionless frequency that depends solely on the redshifted lens mass, $M_\mathrm{Lz}$, for a given dimensionful frequency $f$, expressed as $\omega = 8\pi GM_{\rm Lz} f/c^3$, and $\phi_{\rm m}(y)$ is a frequency-independent quantity depending only on $y$, given by $\phi_{\rm m}(y) = (x_{\rm m} - y)^2/2 - \ln(x_{\rm m})$, where $x_{\rm m} = \left(y+\sqrt{y^2+4}\right)/2$.
The term $_{1}{F}_{1}$ denotes the confluent hyper-geometric function, and $\Gamma$ denotes the gamma function.
The scale factor, $\xi_0$, has been chosen as equivalent to the Einstein radius of the point mass lens. 
Therefore, the amplification factor due to an intervening isolated point mass lens can be expressed in terms of only two parameters: the redshifted lens mass $(M_\mathrm{Lz})$ and the impact parameter $(y)$.
 

\begin{figure}
    \centering
    \includegraphics[width = 0.49\textwidth]{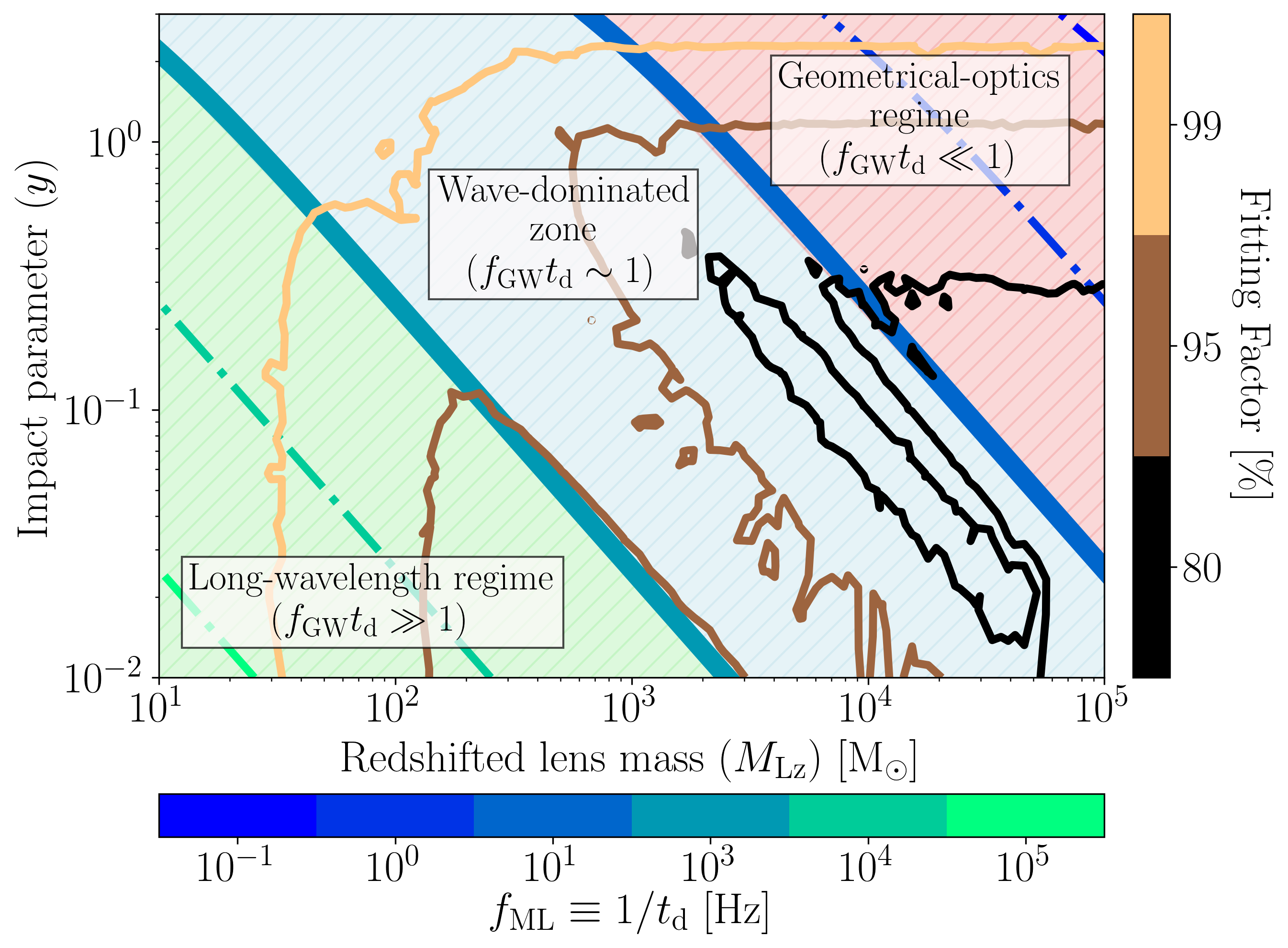}
    \caption{\justifying
    Figure illustrating the contrast in trends between the variation of characteristic micro-lensing frequency, $f_{\rm ML}$, and the fitting factor, $FF$, value when microlensed signals are recovered using unlensed templates. $f_{\rm ML}$  indicates the onset of significant microlensing effects (see \Eref{eq:fML_definition}), whereas $FF$ is related to the  Bayes Factor in favour of microlensing against the unlensed hypothesis (see \Eref{eq:bayes_factor_estimate}). The analysis is conducted within the parameter space relevant to isolated point lenses, focusing on the current-generation ground-based detectors.
    The $f_{\rm ML}$ values are shown in \textit{Blue - Green colorbar}.  Contours at $10$ and $1000$ Hz denote the rough transition regions, dividing the parameter space into three zones. (i) Long-Wavelength Regime where the GW frequency ($f_\mathrm{GW}$) is significantly lower than $f_\mathrm{ML}$, meaning $f_\mathrm{GW} \ll f_\mathrm{ML}$, resulting in minimal interaction between GWs and the microlens, as seen from the light green colored region of the figure.
    (ii) Wave Dominated Zone: This is the region where $f_\mathrm{GW} \sim f_\mathrm{ML}$, leading to substantial interference effects on GWs, depicted with light blue color. (iii) Geometrical-Optics Regime: The right-most region in the figure is shown with light coral color where $f_\mathrm{GW} \gg f_\mathrm{ML}$. This region is inclusive of milli-lensing and strong-lensing scenarios. \textit{Blue - Green colorbar} represent $FF$ contours where the parameters of the source binary are kept fixed to a non-spinning equal mass binary having $60~$M$_\odot$. For each microlensed WF, the $FF$ values are computed by maximizing its match with the unlensed WFs within the parameter space of component masses and aligned spins.
    }
    \label{fig:fML_contours}
\end{figure}

\begin{figure*}
    \centering
    \includegraphics[width=\textwidth]{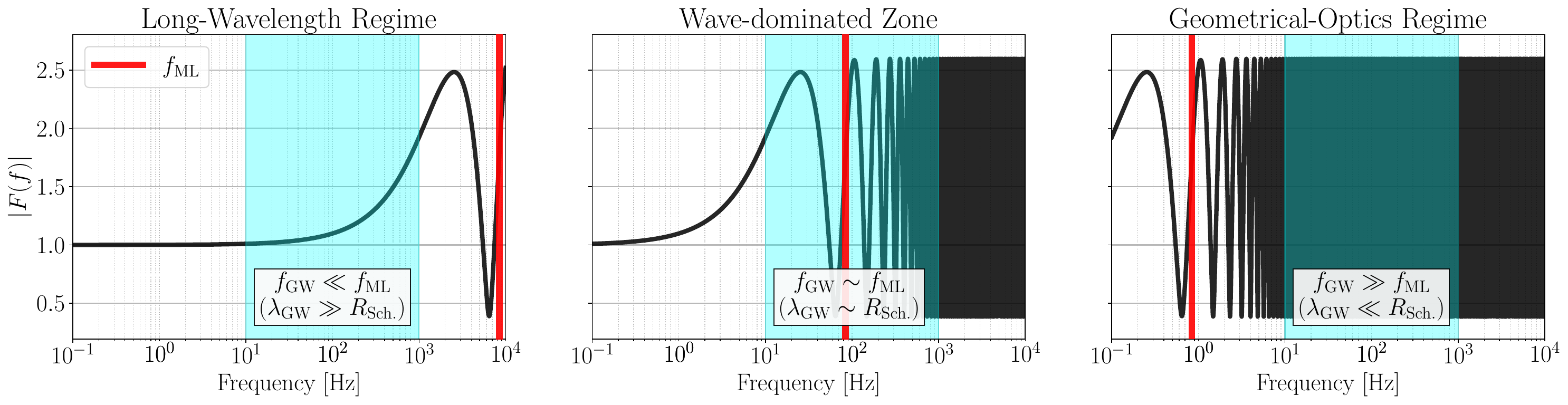}
    \caption{\justifying Illustration of the nature of microlensing modulations in the three regimes. The absolute part of the amplification factor, $|F(f)|$, is plotted against frequency (black lines). Microlens parameters are fixed at $y=0.3$ and $M_\mathrm{Lz}/{\rm M}_\odot \in \{10,~10^3,~10^5\}$, arranged from left to right. The blue-shaded region represents the sensitivity band of current ground-based detectors, approximately $10-100~$Hz, for reference. The red vertical line indicates the characteristic frequency for the onset of significant microlensing effects, $f_\mathrm{ML}$. The panels are divided based on whether $f_\mathrm{ML}>1000~$Hz (\textit{left panel}, long-wavelength regime), $f_\mathrm{ML}\in(10-1000)~$Hz (\textit{middle panel}, wave-dominated zone), or $f_\mathrm{ML}< 10~$Hz (\textit{right panel}, geometrical optics regime).}
    \label{fig:fML_visualise}
\end{figure*}

In the case of the isolated point-mass lens model,
the time delay between the two (micro) images is determined only by  $M_\mathrm{Lz}$ and $y$, given by \citep{schneider1986two,1992grle.book.....S}:
\begin{equation}
    t_\mathrm{d}(M_\mathrm{Lz},~y) = 4 M_\mathrm{Lz} \left[\frac{y+\sqrt{y^2+4}}{2} + \ln\left( \frac{\sqrt{y^2+4}+y}{\sqrt{y^2+4}-y}  \right)\right].
    \label{eq:time_delta_PML}
\end{equation}
By examining how the GW frequency, $f_{\rm GW}$, compares to the time delay caused by the gravitational lens, $t_\mathrm{d}$, we can categorize  $M_\mathrm{Lz}-y$ space into three distinct regimes \citep{mishra2023exploring, Bondarescu:2022srx}:
\begin{enumerate}[label=\roman*.]
\item \textit{Long-wavelength regime}: In this regime, $f_{\rm GW}\cdot t_\mathrm{d}\ll 1$ ($\lambda_\mathrm{GW}\gg R_\mathrm{Sch}$), 
leading to only a minimal interaction
between the signal and the lens. The amplification factor in this regime can be approximated as \cite{Tambalo:2022plm}: 
\begin{equation}
    F(\omega,~y) \simeq 1 + \frac{\omega}{2}[\pi - i(2\log(y) - \mathrm{Ei}(i\omega y^2/2)] + \mathcal{O}(\omega^2),
    \label{eq:F(f)_low_freq_limit}
\end{equation}
where $\mathrm{Ei}(i\omega y^2/2)$ is the exponential integral.
Note that, in the leading order, $F(\omega,~y) \propto 1+\omega$. Thus, for $\omega\ll 1$, there is no effect of lensing on signals as $F(\omega,~y)=1$. However, as $f_{\rm GW}\cdot t_\mathrm{d}$ approaches unity while still being much less than it, the presence of non-zero $\omega$ can lead to a frequency-dependent amplification of the signal. 
That is why some authors also term this region as the `amplification' regime \cite{Bondarescu:2022srx}.

    
\item \textit{Wave-dominated zone}: In this zone,  $f_{\rm GW}\cdot t_\mathrm{d}\sim 1$ (where interference effects will be dominating). This leads to a frequency-dependent modulation of the signal which in case of a point-lens is governed by \Eref{eq:amp fac point}.

\item \textit{Geometrical Optics regime}: If~$f_{\rm GW}\cdot t_d \gg 1$ in Eq.~\eqref{eq:amp_fac}, the integral becomes highly oscillatory, and only the stationary point of the integrand have a non-zero contribution to the integral. In such cases, Eq.~\eqref{eq:amp_fac} can be written as
\begin{equation}
    F\left(f\right)\big|_{\rm geo} = \sum_j \sqrt{|\mu_j|} \exp\left(2\pi i f t_{\rm d, j} - i \pi n_j \right),
    \label{eq:amp fac point geo}
\end{equation}
where $\mu_j$ and $t_{d,j}$ are, respectively, the magnification factor and the time delay for the $j$-th image. Also, $n_j$ is the Morse index, with values 0, 1/2, and 1 for stationary points corresponding to the minima, saddle and maxima of the time-delay surface, respectively. In the case of point mass lens, it reduces to \cite{Takahashi:2003ix}
\begin{equation}
    F(f) = |\mu_+|^{1/2} - i |\mu_-|^{1/2} e^{i 2\pi f t_\mathrm{d}},
    \label{eq:F(f)_high_freq_limit}
\end{equation}
where the magnification of each image is $|\mu_\pm| = 1/2 \pm (y^2 +2)/(2y\sqrt{y^2 + 4}) $ and the time delay between the two microimages is given by \Eref{eq:time_delta_PML}.
Physically, \Eref{eq:F(f)_high_freq_limit} implies that in the high-frequency limit, the modifications due to a point lens can be considered as a trivial superposition of two signals which differ only by a constant amplitude, a phase shift of $\pi/2$, and a time-delay value smaller than the chirp time of the signal\footnote{We exclude cases of strong lensing, where images are entirely separated, by requiring the time-delay value to be less than the chirp time of the signal. Consequently, while we do take into account cases of ``millilensing," we do not include instances of strong lensing in our analysis.}.
\end{enumerate}

One can explicitly find the characteristic frequency $f_\mathrm{ML}$ where the wave effects dominate for a given $M_\mathrm{Lz}$ and $y$, given by
\begin{equation}
    f_\mathrm{ML}(M_\mathrm{Lz},~y) \equiv \frac{1}{t_\mathrm{d}(M_\mathrm{Lz},~y)}.
    \label{eq:fML_definition}
\end{equation}

Considering that current ground-based detectors (LIGO and Virgo) are primarily sensitive in the band $\sim(10-1000)~$Hz, we can illustrate the variation of $f_\mathrm{ML}$ in the $M_\mathrm{Lz}-y$ parameter space as shown by contours corresponding to the Blue-Green colorbar in \Fref{fig:fML_contours}. 
Here, we roughly demarcate $M_\mathrm{Lz}-y$ space based on whether $f_\mathrm{ML}>1000~$Hz (long-wavelength regime; depicted by light green colored region), $f_\mathrm{ML}\in(10,~1000)~$Hz (wave dominated zone; depicted by light blue colored region), or  $f_\mathrm{ML}<10~$Hz (geometrical optics regime; depicted by light coral colored region). 
Additionally, we also illustrate the variation of the fitting factor $FF$ in the figure, which we discuss later in \ref{subsubsec:match_and_model}.
Throughout this work, we will primarily work within the parameter space range depicted in the figure: $M_\mathrm{Lz}/{\rm M}_\odot\in(10,~10^5)$ and $y\in(0.01, 3)$, unless otherwise stated. This parameter space is relevant for searching microlensing signatures due to an isolated point lens in the context of ground-based detectors. 

In \Fref{fig:fML_visualise}, we explicitly illustrate the nature of microlensing modulations in the three regimes. The absolute part of the amplification factor, $|F(f)|$, is plotted against frequency (black lines), where the microlens parameters are fixed at $y=0.3$ and $M_\mathrm{Lz}/{\rm M}_\odot \in \{10,~10^3,~10^5\}$, arranged from left to right. The blue-shaded region represents the sensitivity band of current ground-based detectors, $\sim10-100~$Hz, for reference. 
In the long-wavelength regime (left panel), we observe solely frequency-dependent amplification, as discussed in relation to \Eref{eq:F(f)_low_freq_limit}.
Moving to the middle panel, we notice that $f_\mathrm{ML}$ indeed marks the point where interference effects begin to dominate. This point typically occurs shortly after the first peak and trough of the amplification factor. It is worth pointing out that the characteristic location of $f_\mathrm{ML}$ (lying between the first two peaks) remains consistent across different lensing parameters.
In contrast, the right panel illustrates highly oscillatory modulations that eventually lead to an averaging effect (geometrical optics limit). This behavior eventually results in frequency-independent effects.

When significant, microlensing modulations can lead to systematic biases in the estimation of source parameters, especially when the intervening microlens belong to the wave-dominated zone \cite{mishra2023exploring}.
When microlensed GW signals are recovered with the usual unlensed GW signals, the phase modulations can bias the inference of intrinsic parameters, especially the in-plane spin components (or precession). Among extrinsic parameters, the recoveries of luminosity distance are affected the most. In contrast, the trigger time and the sky-position parameters, right ascension (RA) and declination (Dec) ($\alpha$ and $\delta$), are the mostly well-recovered source parameters. 

The micro-lensing-induced systematic biases could be mistakenly interpreted as evidence of a deviation from GR. This potential misinterpretation arises from the necessity of incorporating micro-lensing effects into the model during the GW signal analysis. Additionally, it is worth noting that the standard tests of GR routinely applied to observed GW signals~\cite{GWTC-3-TGR-LVK} are liable to a range of systematic effects~\cite{Narayan:2023vhm, Saini:2022igm, Bhat:2022amc}. 
While there have been studies addressing systematic biases that can lead to misinterpretations of GR, the impact of micro-lensing remains unexplored. To address this gap, we conduct an in-depth simulation study, focusing on two theory-agnostic GR tests, to assess how micro-lensing effects may influence the identification of non-GR signatures in observed GW data.

\subsection{\label{subsec:basics_tgr}Tests of General Relativity (TGR)}

\subsubsection{\label{subsubsec:imrct}Inspiral-Merger-Ringdown Consistency Test (IMRCT)}

The IMRCT is one of the standard consistency tests of GR that is performed on real GW events. The test relies on checking the consistency between the measurements of mass and spin of the remnant BH inferred independently from the inspiral and post-inspiral parts of the GW signal. The demarcation between the inspiral and the post-inspiral regime is typically done by employing a cutoff frequency $f^\mathrm{IMR}_\mathrm{c}$ in the frequency domain, which is the dominant mode GW frequency $f_\mathrm{ISCO}$ of the innermost stable circular orbit (ISCO) of the remnant Kerr BH with mass $M_f$ and dimensionless spin $\chi_f$. In this work, the ISCO frequency of the final Kerr black hole is measured using the median values of the 1D-marginalized posteriors of the final mass and spin inferred from the full IMR waveform\footnote{We note that the LVK analyses use a more complicated method to compute $f_\mathrm{ISCO}$ from the posteriors obtained from IMR, see, e.g.,~\citep{GWTC-1-TGR-LVC,GWTC-3-TGR-LVK}.}~\citep{Ghosh2016,Ghosh2017}. To calculate the final mass and spin, we use the NR-calibrated fits as detailed in Ref.~\cite{LIGOScientific:2017bnn,GWTC-2-TGR-LVC, Final_mass_and_spin_public} where the method extends the final spin fit of aligned-spin binaries to a precessing case (see Eqn. 1 of Ref.~\cite{Final_mass_and_spin_public} for details).

To elaborate, we independently infer the posterior distributions of $M_f$ and $\chi_f$ from both the inspiral and the post-inspiral parts of the signal by using the augmented NR calibrated final state fits~\cite{Hofmann:2016yih,Jimenez-Forteza:2016oae,Healy:2016lce}. To constrain possible deviations from GR, a set of fractional deviation parameters $\Delta M_{f}/\overline{M}_{f}$ and $\Delta \chi_{f}/\overline{\chi}_{f}$ are defined, where

\begin{equation}
    \frac{\Delta M_{f}}{\overline{M}_{f}}\equiv 2\frac{M_{f}^\mathrm{insp}-M_{f}^\mathrm{post-insp}}{M_{f}^\mathrm{insp}+M_{f}^\mathrm{post-insp}}
\end{equation}
and 

\begin{equation}
    \frac{\Delta \chi_{f}}{\overline{\chi}_{f}}\equiv 2\frac{\chi_{f}^\mathrm{insp}-\chi_{f}^\mathrm{post-insp}}{\chi_{f}^\mathrm{insp}+\chi_{f}^\mathrm{post-insp}},
\end{equation}
where $\overline{M_f}$ and $\overline{\chi_f}$ denote the mean values and the superscripts denote the inspiral (insp) and the post-inspiral (postinsp) portions of the signal.
If a given GW signal is consistent with GR, the fractional deviations $\Delta M_{f}/\overline{M}_{f}$ and $\Delta \chi_{f}/\overline{\chi}_{f}$ should vanish in the ideal situation. We estimate the posteriors on these fractional deviation parameters and check the consistency with GR. 

To check the consistency of inspiral and post-inspiral measurements statistically, we use the GR quantile $Q_{\rm GR}$. In this case, the GR quantile for the 2-D distribution of final mass and spin parameters measures the fraction of the posterior samples that contain the GR value. Specifically, we utilize the \texttt{summarytgr} executable within \texttt{PESummary}~\cite{Hoy:2020vys} to calculate the GR deviations.
Notice that a smaller value denotes better agreement with GR, meaning that a large fraction of samples are contained in a small iso-probability contour. We use \texttt{IMRPhenomXPHM} \cite{PhenomXPHMPratten2020} as our waveform approximant for both injection and recovery of the IMRCT analysis throughout.

\subsubsection{\label{subsubsec:tiger}Parameterized test}
The dynamics of a BBH system can be described in three regimes: the inspiral, the merger and the ringdown. In the inspiral, the post-Newtonian formalism accurately describes the evolution, which is an expansion in the velocity parameter $v/c\ll1$, where $c$ is the velocity of light~\cite{Blanchet2013_LivingReview}. In contrast, one needs numerical relativity techniques to model the dynamics towards the merger regime~\cite{Pretorius:2007nq, Duez:2018jaf}, where the strong-field dynamics play a critical role. In the post-merger phase, BH perturbation theory provides a good description~\cite{Berti:2005ys}. 

Parameterized tests of post-Newtonian (PPN) theory are among the multiple ways of performing theory-agnostic tests of GR. Even though the PPN tests were restricted to the inspiral regime~\cite{ParametrizedTests_Arun2006a,ParametrizedTests_Arun2006b, Agathos:2013upa}, later on, they were also extended to the post-inspiral regime by employing the phenomenological waveform models~\citep{GWTC-1-TGR-LVC, GWTC-2-TGR-LVC, Meidam:2017dgf, GW150914-TGR-LVC, GW170817_TGR_LVC,GWTC-1-TGR-LVC,GWTC-2-TGR-LVC}. 
The PPN tests introduce a novel way of looking for deviations in the inspiral-merger-ringdown coefficients from the GR model by appropriately incorporating parameterized deviation coefficients in the model.
We denote the inspiral coefficients as $\{\chi_i\}$\footnote{Not to be confused with the spin parameter.}, and the post-inspiral coefficients as $\{\alpha_i, \beta_i\}$, where $\{\beta_i\}$ accounts for deviations mainly in the intermediate phase while $\{\alpha_i\}$ in the merger-ringdown phase.
The parameterized deviations to these coefficients are of the form,  $\chi_i \rightarrow (1+ \delta\hat{\chi}_i)\chi_i$, $\alpha_i\rightarrow (1+ \delta\hat{\alpha}_i)\alpha_i$ and $\beta_i\rightarrow (1+ \delta\hat{\beta}_i)\beta_i$, where $\delta\hat{\chi}_i$, $\delta\hat{\alpha}_i$ and $\delta\hat{\beta}_i$ are the fractional deviations from the GR coefficient.
In the case of phenomenological waveform models, the post-Newtonian inspiral coefficients are extended further by fitting to numerical relativity coefficients~\cite{Ajith:2007kx,Ajith:2009bn,Santamaria:2010yb,Husa:2015iqa,Khan:2015jqa, Garcia-Quiros:2020qpx,Pratten:2020fqn}. 
Besides the inspiral, the parameters $\delta\hat{\beta}_i$ explicitly capture deformations in the NR calibrated coefficients $\beta_i$ in the intermediate regime, whereas the parameters $\delta\hat{\alpha}_i$ describe deformations of the merger-ringdown coefficients $\alpha_i$ obtained 
by calibration to NR~\cite{Meidam:2017dgf,GWTC-1-TGR-LVC}.
The data provide constraints on these parameterized deviations from GR and can be used to test the consistency of GR. That means we obtain posterior probability distributions on each of the parameterized deviation coefficients. 
If the data is consistent with GR, the distribution will peak at zero, the GR value. 
Any deviation from  GR can hint at the presence of an alternative theory model or missing physics in our models, such as various systematic effects. 
To quantify such a possible deviation from GR, we use $\sigma_{\rm GR}$, defined as, $\sigma_{\rm GR}=(\mu-\mu_{\rm GR})/\sigma$. Here $\mu$ is the mean of the distribution, $\mu_{\rm GR}$ is the GR value, and $\sigma$ is the standard deviation of the posterior distribution. Since $\mu_{\rm GR}=0$, we simply get $\sigma_{\rm GR}=\mu/\sigma$.

In this study, we restrict to two inspiral coefficients, $\delta\hat{\chi}_0$ (0PN) and $\delta\hat{\chi}_4$ (2PN), respectively, which appear in the Newtonian and the second post-Newtonian order in the inspiral phase of the waveform. 
Moreover, we consider two coefficients in the post-inspiral regime: $\delta\hat{\beta}_2$ from the intermediate phase and $\delta\hat{\alpha}_2$ from the merger-ringdown phase.
We employ \texttt{IMRPhenomPv2}\footnote{Though there are recent phenomenological models with more accurate spin-precesion description~\cite{PhenomXPHMPratten2020,Pratten:2020fqn} and sub-dominant modes~\cite{Garcia-Quiros:2020qpx}, we stick to \texttt{IMRPhenomPv2} waveform model for our analysis.} waveform model for this study, which is a dominant-mode model for precessing BBHs where the inspiral coefficients are obtained from PN theory and calibration to NR waveforms, and the post-inspiral coefficients by fitting to NR-simulations~\cite{Hannam:2013oca,Husa:2015iqa,Khan:2015jqa,PhenomPv2_Technical_document}.
Consequently, the inspiral phase is defined up to a cutoff frequency $f_\mathrm{cut} = f^\mathrm{PAR}_\mathrm{c} = 0.018/M$, where $M$ is the total mass of the binary~\citep{Husa:2015iqa, Khan:2015jqa}.

\subsection{\label{subsec:basics_gw_analysis}GW data analysis and Parameter Estimation Setup}
Here we discuss some general tools and setup that are common throughout this work. 

\subsubsection{\label{subsubsec:injections}Injections}

To study the effect of microlensing on the tests of GR using GWs, we consider simulated binary black hole (BBH) events, called \textit{injections}. These injections are done in the detector network of LIGO-Virgo with the projected O4 sensitivity\footnote{For LIGO detectors, we used the PSD given in \url{https://dcc.ligo.org/public/0165/T2000012/002/aligo_O4high.txt}. While for Virgo, we used the PSD available at \url{https://dcc.ligo.org/public/0165/T2000012/002/avirgo_O4high_NEW.txt}.} \citep{Abbott:2020search}. To avoid noise systematics, i.e., biases due to specific noise realizations, we do not include noise in our injections (assuming the realization of Gaussian noise that yields zero noise in the data containing the signal).  

Firstly, in \Sref{subsec:GW150914}, we consider a set of $30$ zero-noise microlensed BBH injections, consisting of GW150914-like non-spinning signals with added microlensing effects. 
The component masses and the extrinsic parameters of the signals are kept similar to those of GW150914\footnote{The posterior samples can be found at \url{https://zenodo.org/records/6513631}. Specifically, we used the median values of 1D-marginalized posteriors from the `C01:Mixed' channel in `IGWN-GWTC2p1-v2-GW150914\_095045\_PEDataRelease\_mixed\_cosmo.h5'.}, except for the luminosity distance, which is tuned to fix the network optimal signal-to-noise ratio (SNR) to $50$ for each injection in the detector network of advanced LIGO at Hanford, Livingston and, Virgo. In addition, the spin components are kept zero to avoid possible systematic errors in the inference of the deviation parameters from GR and make sure any biases arise predominantly from microlensing effects. This is because one needs to have even higher SNR values of $\gtrsim 10^2$ to reliably estimate both the spin components \citep{Purrer:2017uch}. The microlens parameters used for generating the injections correspond to $\log_{10}M_\mathrm{Lz} \in \{1,~2,~3,~4,~5\}$ and $y\in\{0.01,~0.05,~0.1,~0.5,~1.0,~3.0\}$, making $30$ injections in total.

In \Sref{subsec:population_study}, to robustly analyze the effect of microlensing on tests of GR, we consider a population of simulated microlensed signals. We generate a mock GW data set of $\sim 2.5\times 10^4$ microlensed BBH signals, where the mass and spin priors are derived from the inferred population model of GWTC-3 data \citep{LIGOScientific:2021djp}: the source-frame component masses are sampled from the Power Law + Peak distribution and spin parameters are drawn from a Gaussian isotropic distribution.  
We assume the Madau-Dickison profile for the merger rate density in the universe \citep{Madau:2014bja,Fishbach:2018edt}, with the source redshift range set to $z_\mathrm{S} \in (0.001,~10)$. The upper limit of $z_{\rm S}=10$ is set because isolated point lenses tend to amplify the signals, increasing their optimal SNR and, consequently, their detection horizon~\cite {mishra2023exploring}. 
To sample microlens parameters, we assume a log-uniform prior in $M_\mathrm{Lz}$ and a linear power-law prior for $y$, where: 
\begin{equation}
\begin{aligned}
    p(\log_{10}M_\mathrm{Lz}) &\propto \text{Uniform}(1,~5), \\
    p(y) &\propto y,~~y\in(0.01,~3.00).\
    \label{eq:mlz_y_priors_v1}
\end{aligned}
\end{equation}
The motivation to use $p(y) \propto y$ comes from geometry and isotropy~\citep{Lai:2018rto}.
The other parameters were sampled from the usual prior distributions: isotropic sky location and orientation, and uniform polarization angle. We put an observed network SNR, $\rho_\mathrm{opt}^{\rm N}$, threshold of $8$ when using the unlensed templates for recovery in the LIGO-Virgo detector network. We further impose an additional requirement that $\mathrm{SNR}>6$ in both the inspiral and post-inspiral parts of the signal. This requirement ensures that we have an adequate amount of information for our analyses in both signal regimes. The demarcation between these two regimes is done via a cutoff frequency $f_\mathrm{cut}$ in the frequency domain. We define $f_\mathrm{cut} = f_\mathrm{ISCO}$ for IMRCT and $f_\mathrm{cut} = f_\mathrm{C}^\mathrm{PAR}$ for the parameterized test, as explained in \Sref{subsec:basics_tgr}.
For performing IMRCT, these SNR thresholds are consistent with the previous analyses~\cite{GWTC-3-TGR-LVK,GWTC-2-TGR-LVC}.  
However, we note that for parameterized tests, our SNR thresholds are stricter. For example, to study the deviations in the coefficients corresponding to the inspiral regime, such as $\delta\hat{\chi}_0$ and $\delta\hat{\chi}_4$ considered here, one only needs to have the inspiral $\mathrm{SNR}>6$. 
A stricter condition of $\mathrm{SNR}>6$ in both the inspiral and post-inspiral regimes is employed so that each injection can be analyzed for all four testing GR parameters.

\subsubsection{\label{subsubsec:pe_setup}Parameter Estimation}
Throughout this work, we use the publicly available Bayesian inference library \texttt{Bilby} \cite{Ashton:2018jfp} for performing parameter estimation runs. Specifically, we use the \texttt{dynesty}~\cite{speagle2020dynesty} nested sampler with the `acceptance-walk' method for the Markov-Chain Monte-Carlo (MCMC) evolution as implemented in \texttt{Bilby}, along with the sampler settings of nlive$=1024$, n-accept$=60$ and n-parallel$=2$ per injection. The lower frequency limit of the likelihood evaluation is set to $f_\mathrm{low}=20~$Hz, which is also the reference frequency. 

Furthermore, in order to compute microlensing effects, whether for generating microlensed injections or inferring microlens parameters, we employ a custom frequency domain source model, which incorporates the two microlensing parameters $M_\mathrm{Lz}$ and $y$ in addition to the standard 15 BBH parameters and is made publicly available via \texttt{Python/Cython} package \texttt{GWMAT}~(Mishra, A., 2023, in prep.). 

While inferring microlens parameters in our simulated microlensed signals with injected $M_\mathrm{Lz}\equiv M_\mathrm{Lz}^\mathrm{true}$ and $y\equiv y^\mathrm{true}$, we assume a log-uniform prior in $M_\mathrm{Lz}$ and a linear power-law prior for $y$, where: 
\begin{equation}
p(\log_{10}M_\mathrm{Lz}) \propto
\begin{cases}
    \text{Uniform}(-1,~5),& \text{if } \log_{10}M_\mathrm{Lz}^\mathrm{true}<3\\
    \text{Uniform}(-1,~7),& \text{otherwise}
\end{cases}
\end{equation}
\begin{equation}
p(y) \propto y, \text{where}
\begin{cases}
    y\in(0.001,~3.00),& \text{when }y^\mathrm{true}<1 \\
    y\in(0.001,~5.00),& \text{otherwise. }
    \label{eq:mlz_y_priors_v2}
\end{cases}    
\end{equation}
We also use these priors while generating a population of microlensed signals.

For the parameterized test of GR, we independently vary each of the four deviation parameters instead of simultaneously varying them, aligning with the methodology employed in the LVK catalog analyses~\citep{GWTC-2-TGR-LVC, GWTC-3-TGR-LVK} and similar studies~\citep{Narayan:2023vhm}. 
However, we do note that in an alternative theory of gravity, one would anticipate modifications to all post-Newtonian (PN) coefficients beyond a certain order. 
This choice of varying only one parameter at a time is made to avoid uninformative results, as observed in cases where multiple parameters are allowed to vary simultaneously, as illustrated for $\mathrm{GW150914}$ in~\cite{GW150914-TGR-LVC}. 
Nevertheless, earlier studies~\citep{Meidam:2017dgf,Johnson-McDaniel:2021yge} have demonstrated that even when varying a single testing parameter, it is still possible to detect deviations from GR that modify multiple PN coefficients (even when this testing parameter is itself not modified).

\subsubsection{\label{subsubsec:match_and_model}Model Comparison}
To study whether a given GW event is microlensed or not, one can analyze the data under the assumption of two hypotheses: (i) Null-hypothesis $\mathcal{H}_\mathrm{UL}$: the signal is unlensed, 
, and (ii) $\mathcal{H}_\mathrm{ML}$: the signal is microlensed. 
Under the assumption that the two models are equally likely, we set their prior probabilities to be equal (i.e., $p(\mathcal{H}_\mathrm{UL})=p(\mathcal{H}_\mathrm{ML})$). One then compares the two models using the Bayes factor formalism, where the Bayesian evidence  (or marginal likelihood) $\mathcal{Z}$ of $\mathcal{H}_\mathrm{ML}$ is compared against that of $\mathcal{H}_\mathrm{UL}$, given by 
\begin{equation}
    \mathcal{B}^\mathrm{ML}_\mathrm{UL} \equiv \frac{p(\mathcal{H}_\mathrm{ML}|\mathcal{D})}{p(\mathcal{H}_\mathrm{UL}|\mathcal{D})} = \frac{\mathcal{Z}_\mathrm{ML}}{\mathcal{Z}_\mathrm{UL}},
\end{equation}
where $\mathcal{D}$ represents data. 
As mentioned in the previous subsection~\ref{subsubsec:pe_setup}, we utilize the \texttt{dynesty}~\cite{speagle2020dynesty} nested sampler as implemented in \texttt{Bilby}~\citep{Ashton:2018jfp} to compute both the Bayesian evidence and inferring parameters, unless otherwise noted.  
However, since computing Bayesian evidence using a nested sampling algorithm is usually computationally expensive, one can also employ an estimation method \citep{Cornish:2011ys, Vallisneri:2012qq} for sufficiently high SNR values, where the leading order term in (log) Bayes factor can be expressed as
\begin{equation}
    \log_{10} \mathcal{B}^\mathrm{ML}_\mathrm{UL} \approx \frac{1}{2\ln{10}}\rho_\mathrm{res}^2 .
    \label{eq:bayes_factor_estimate}
\end{equation}
Here, $\rho_\mathrm{res}^2 = (1 - FF^2)\rho_\mathrm{ML}^2$ represents the SNR of the \textit{residual} waveform after the best fit unlensed waveform has been subtracted out from the true microlensed WF. $\rho_\mathrm{ML}$ is the SNR when the true microlensed waveform is used to extract the signal, and $FF$ represents the fitting factor value \citep{Sathyaprakash:1991mt, Ajith:2012mn}, indicating the \emph{match} between the best-fit unlensed waveform and the true microlensed waveform (or equivalently, $FF = \rho_\mathrm{UL}/ \rho_\mathrm{ML}$).
The match $\mathfrak{M}$ between two waveforms, $h_1(t)$ and $h_2(t)$, is defined as the noise-weighted inner product, maximized over phase $(\phi)$ and time $(t)$:
\begin{equation}
\begin{aligned}
    \mathfrak{M}(h_1, h_2) &\equiv \underset{t,\phi}{\mathrm{max}} \bra{h_1}\ket{h_2}\\
    & \equiv \underset{t,\phi}{\mathrm{max}} ~ 4 \mathcal{R}\int_{f_\mathrm{low}}^{f_\mathrm{high}} df \frac{\tilde{h}_1^*(f) \tilde{h}_2(f)}{S_\mathrm{n}(f)} \cdot e^{i ( 2\pi f t + \phi)},
    \label{eq:match_func_def}
\end{aligned}    
\end{equation}
where $f_\mathrm{low}$ and $f_\mathrm{high}$ are the lower and upper-frequency cutoffs of the integral. Here, the tilde represents the Fourier transform, the asterisk denotes the complex conjugate, and $S_\mathrm{n}(f)$ is the one-sided noise power spectral density (PSD). 
We only employ this estimation method when we study a population of $\mathcal{O}(10^4)$ signals in \Sref{subsubsec:pop_imrct_study}.

In \Fref{fig:fML_contours}, in addition to showing the trend of $f_\mathrm{ML}$, we also illustrate the trend of variation of $FF$ in the $M_\mathrm{Lz}-y$ parameter space. For each microlensed signal, the binary parameters are kept fixed to a non-spinning equal mass binary having $60~$M$_\odot$\footnote{We note that the general trend of the variation of the fitting factor values in the $M_\mathrm{Lz}-y$ parameter space should not depend on the source properties because the $FF$ depends only weakly on the binary parameters relative to the lensing parameters~\citep[see Fig.~2 in][]{mishra2023exploring}.}, and the $FF$ value is obtained by maximizing the match between the microlensed WF and the 4-dimensional aligned-spin template WFs, modelled by the two component masses and their aligned spins. The match is maximized using the Nelder-Mead algorithm as implemented in the `optimization' module of the \texttt{Scipy} library \citep{2020SciPy-NMeth}. As we can see, the trend of variation of $FF$ is different from that of the variation of $f_\mathrm{ML}$. Notably, the $FF$ values roughly increase as we move towards the bottom-right corner of the parameter space, with increasing $M_\mathrm{Lz}$ and $1/y$ values. Conversely, the $f_\mathrm{ML}$ values increase as we move from the top-right corner of the $M_\mathrm{Lz}-y$ parameter space towards the bottom-left corner. 
It is important to note that the region where wave effects dominate differs from the area associated with low match values (high Bayes factor values in favour of microlensing).
This observation plays a crucial role in the interpretation of our results.

It is important to note that the region where wave effects dominate differs from the area associated with low match values (high Bayes factor values in favour of microlensing).
This observation plays a crucial role in the interpretation of our results. 

\section{\label{sec:results_and_discussions}Results and Discussions}
\begin{figure}
     \centering
     \begin{subfigure}[b]{0.49\textwidth}
         \centering
             \includegraphics[width=\textwidth]{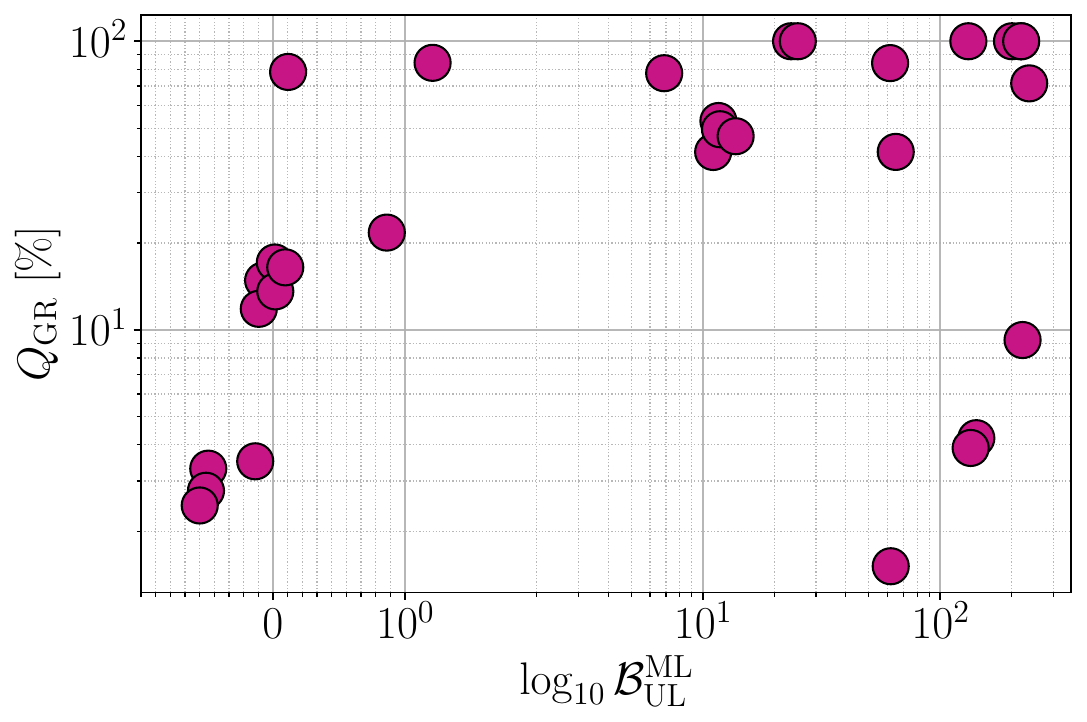}
         \caption{\justifying $\mathcal{B}^\mathrm{ML}_\mathrm{UL}$ vs. $Q_\mathrm{GR}$}
         \label{fig:BLU_vs_QGR_gw150914_IMRCT}
     \end{subfigure}
     \hfill
     \begin{subfigure}[b]{0.49\textwidth}
         \centering
         \includegraphics[width=\textwidth]{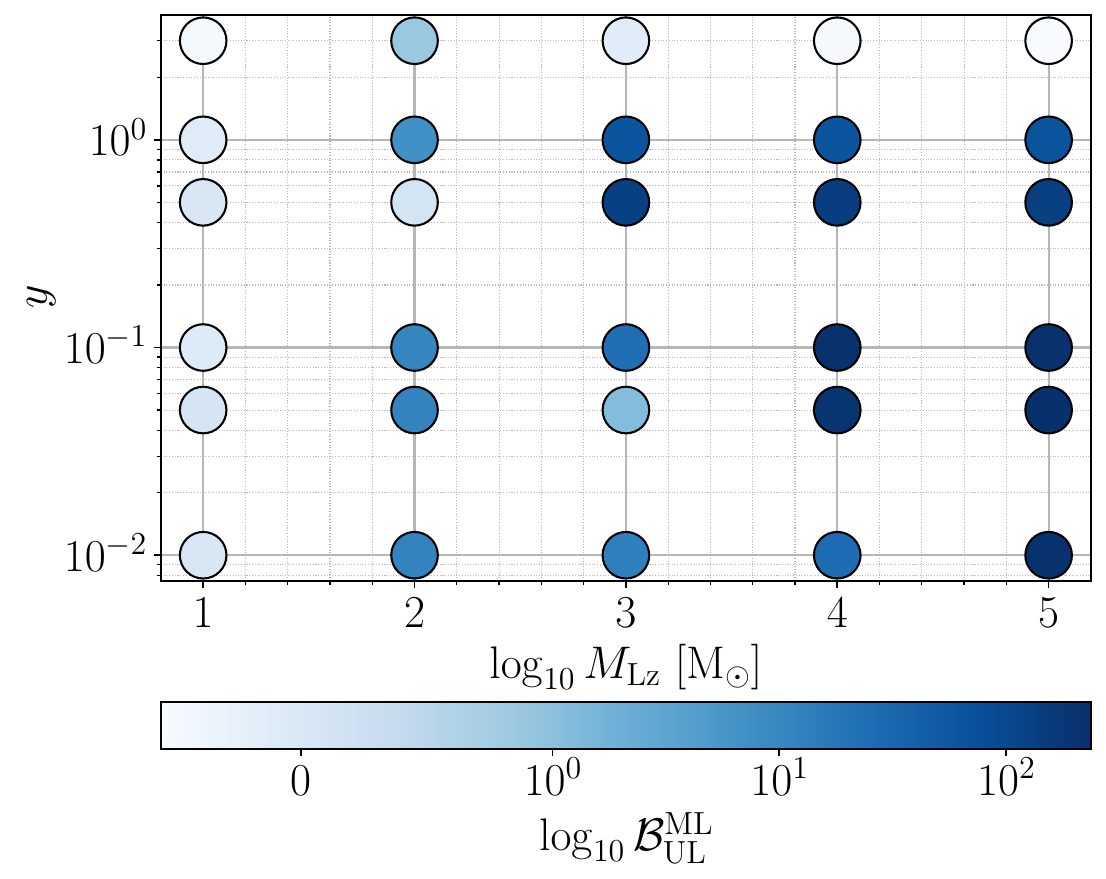}
         \caption{\justifying $\log_{10}\mathcal{B}^\mathrm{ML}_\mathrm{UL}$ dependence on $M_\mathrm{Lz}$-$y$ parameter space.}
         \label{fig:MLz-y_vs_BLU_gw150914_IMRCT}
     \end{subfigure}     
     \hfill
     \begin{subfigure}[b]{0.49\textwidth}
         \centering
         \includegraphics[width=\textwidth]{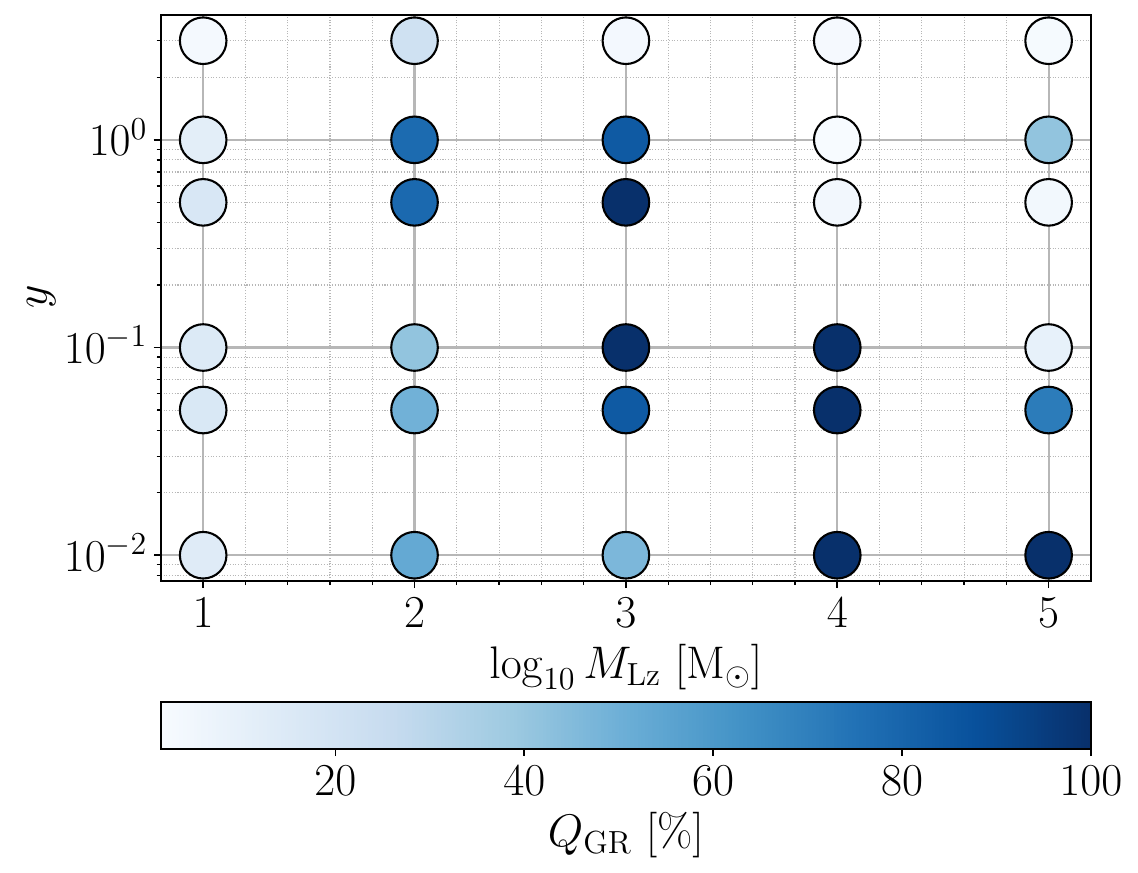}
         \caption{\justifying $Q_\mathrm{GR}$ dependence on $M_\mathrm{Lz}$-$y$ parameter space.}
         \label{fig:MLz-y_vs_QGR_gw150914_IMRCT}
     \end{subfigure}
        \caption{\justifying IMRCT results for GW150914-like microlensed injections, highlighting deviations from GR ($Q_\mathrm{GR}$), alongside the microlensing Bayes factor ($\log_{10}\mathcal{B}^{\rm ML}_{\rm UL}$) for reference.}
        \label{fig:gw150914_IMRCT}
\end{figure}

\begin{figure*}
     \centering
     \begin{subfigure}[b]{0.47\textwidth}
         \centering
             \includegraphics[width=\textwidth]{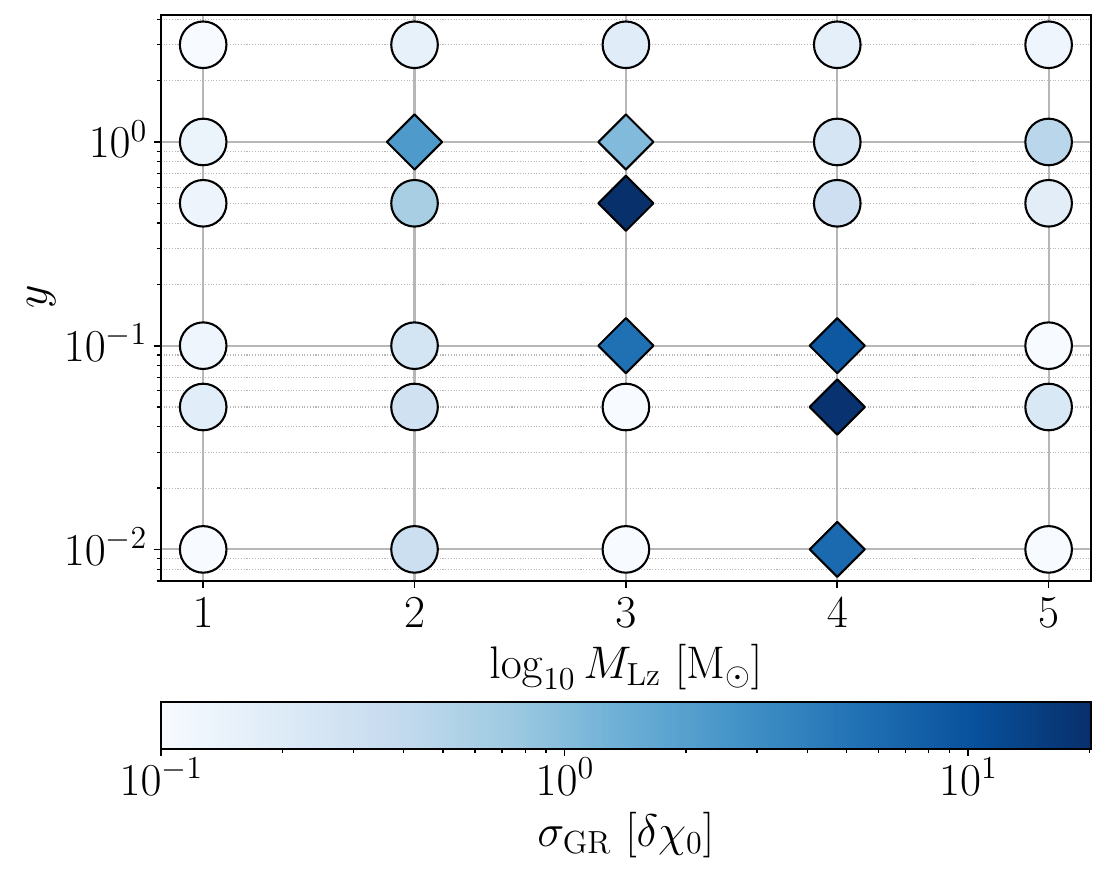}
         \caption*{}
         \label{fig:dchi_0_gw150914_TIGER}
     \end{subfigure}
     \begin{subfigure}[b]{0.47\textwidth}
         \centering
         \includegraphics[width=\textwidth]{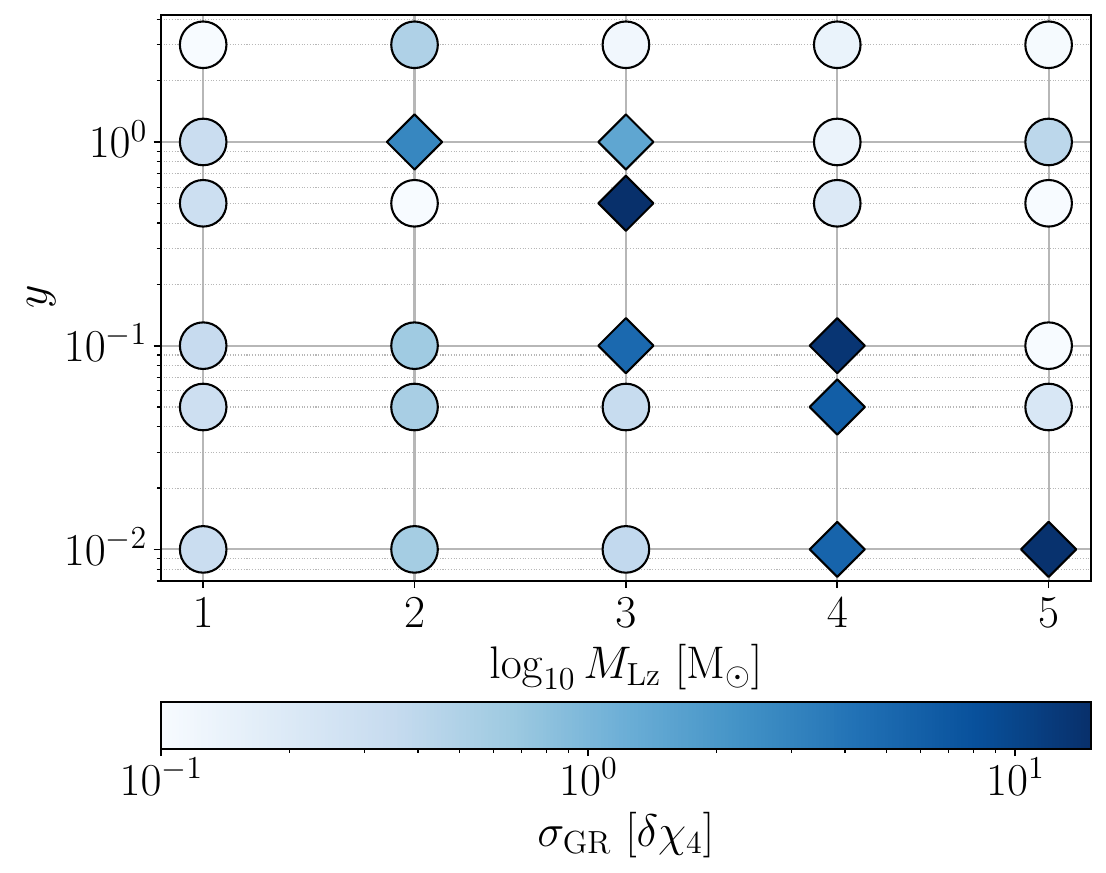}
         \caption*{}
         \label{fig:dchi_4_gw150914_TIGER}
     \end{subfigure}     
     \begin{subfigure}[b]{0.47\textwidth}
         \centering
         \includegraphics[width=\textwidth]{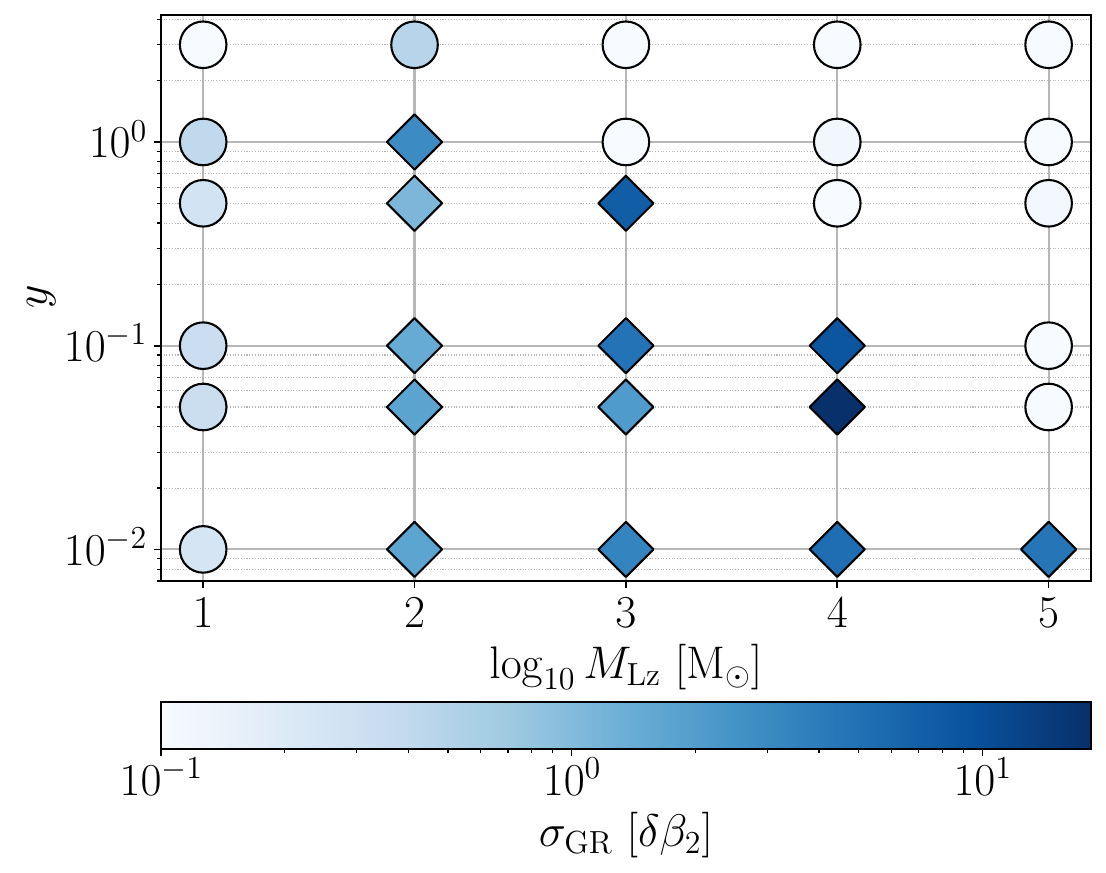}
         \caption*{}
         \label{fig:dalpha_2_gw150914_TIGER}
     \end{subfigure}
     \begin{subfigure}[b]{0.47\textwidth}
         \centering
         \includegraphics[width=\textwidth]{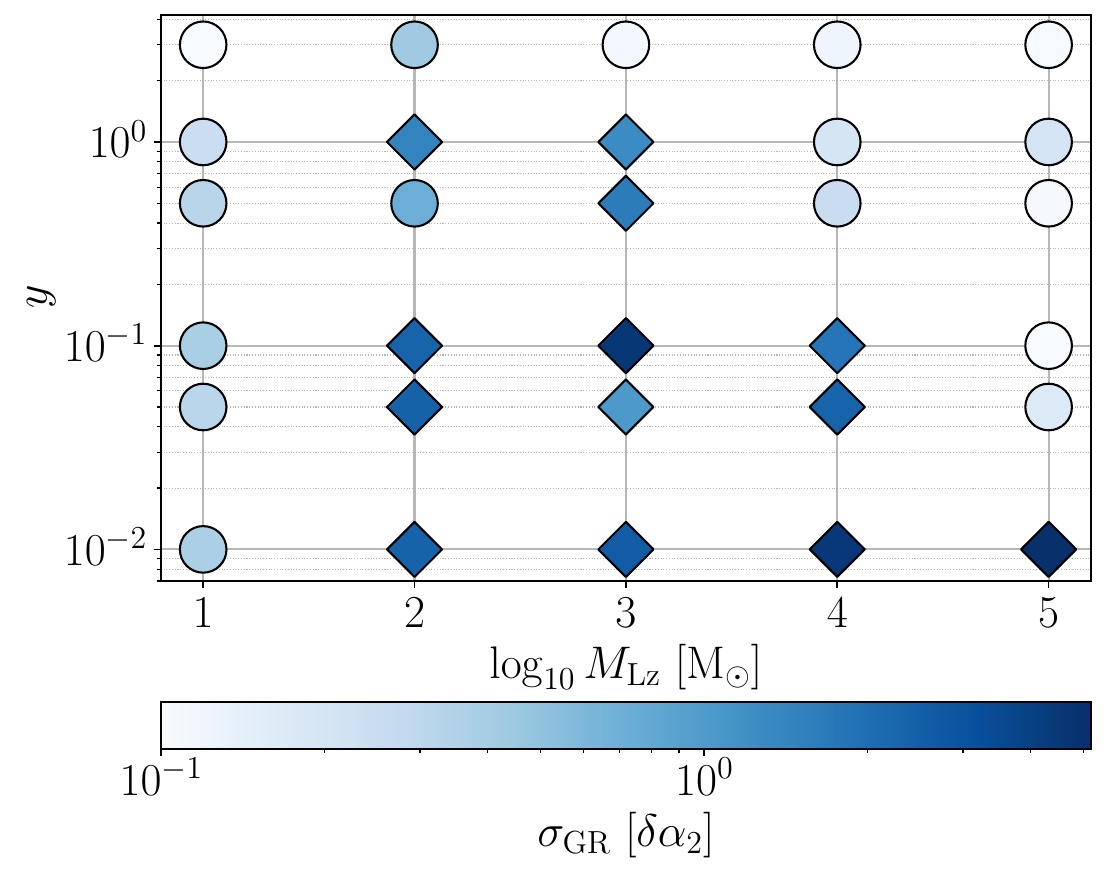}
         \caption*{}
         \label{fig:dbeta_2_gw150914_TIGER}
     \end{subfigure}     
        \caption{\justifying Bias in the parameterized test of GR in GW150914-like microlensed injections for four deviation parameters $\{\delta\hat{\chi}_0,~\delta\hat{\chi}_4,~\delta\hat{\alpha}_2,~\delta\hat{\beta}_2\}$. Each of the four panels displays the Gaussian sigma values at which GR is excluded ($\sigma_\mathrm{GR}$) in one of the four parameterized parameters, as indicated in the colorbar, within the $\log_{10}M_\mathrm{Lz}-y$ parameter space. The larger the magnitude of $\sigma_\mathrm{GR}$, the greater the significance of deviations from GR. For better visualization, we mark the cases with significant deviations ($\sigma_\mathrm{GR}>1$) using a diamond marker instead of the circular markers.}
        \label{fig:gw150914_TIGER}
\end{figure*}

\begin{figure*}
    \centering
    \includegraphics[width=1.05\textwidth]{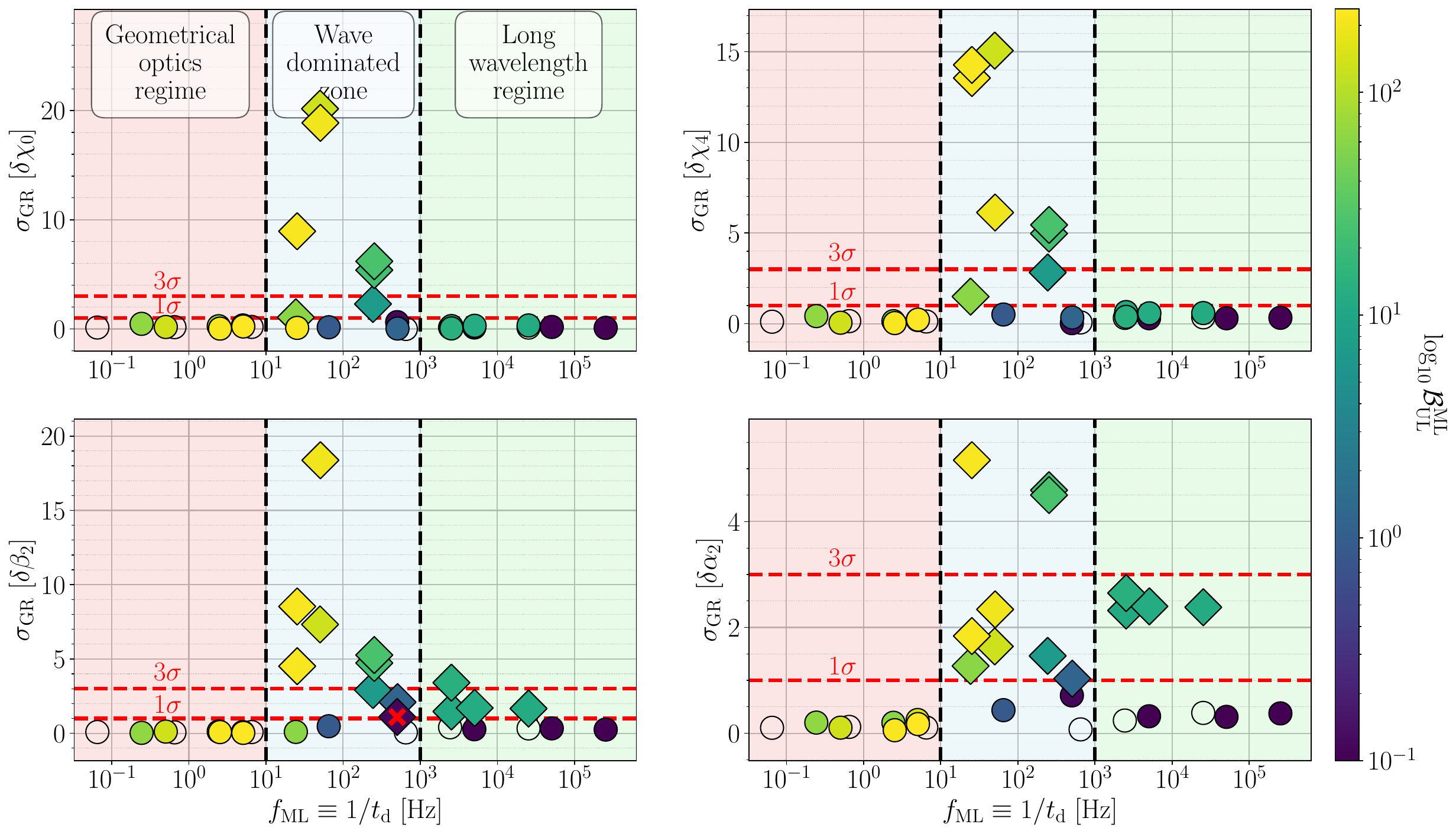}
    \caption{\justifying Similar to \Fref{fig:gw150914_TIGER}, but with $M_\mathrm{Lz}-y$ parameter space suppressed into a one-dimensional representation using $f_\mathrm{ML}$ (see \Eref{eq:fML_definition}). The figure highlights three scenarios: the long-wavelength regime, the wave-dominated zone, and the geometrical optics regime. The colormap represents $\log_{10}\mathcal{B}^{\rm ML}_{\rm UL}$ values for each injection. Transparent markers with black edges indicate cases where $\log_{10}\mathcal{B}^{\rm ML}_{\rm UL}<0$. Dotted-red lines indicate the $1\sigma$ and $3\sigma$ significance of deviations. Diamond markers indicate significant deviations $(\sigma_\mathrm{GR}>1)$, and cases meeting $\log_{10}\mathcal{B}^{\rm ML}_{\rm UL}<1$ (indicating stealth bias) have an added red cross.}
    \label{fig:gw150914_TIGER_with_fML}
\end{figure*}

\subsection{\label{subsec:GW150914}GW150914-like Microlensed Injections}
As described in \Sref{subsubsec:injections}, here we consider zer-noise GW150914-like non-spinning injections with added microlensing effects. The microlens parameters used for generating the injections correspond to $\log_{10}M_\mathrm{Lz} \in \{1,~2,~3,~4,~5\}$ and $y\in\{0.01,~0.05,~0.1,~0.5,~1.0,~3.0\}$, making $30$ injections in total. These ideal microlensed BBH injections serve as our initial motivation for the exploration of microlensing-induced biases in tests of GR and enable us to search for patterns in the distortions, helping us identify the intriguing microlensing parameter space where deviations become more prominent.

\subsubsection{\label{subsubsec:GW150914_imrct_study}IMRCT}
We conduct six sets of parameter estimation runs for all the $30$ microlensed injections. In the case of IMRCT, each set includes runs for the inspiral part, post-inspiral part, and the full IMR signal, for both unlensed and microlensed hypotheses. The inspiral and post-inspiral regions of our injections are demarcated by the cutoff frequency of  $f^\mathrm{IMR}_\mathrm{c}=f_\mathrm{ISCO}=128~$Hz.\footnote{We note that in the ideal case, one may determine this cutoff frequency from the mass and spin estimates of the binary considering the full-IMR analysis. However, even when computed using un-lensed IMR recoveries, the value of $f_\mathrm{ISCO}$ for our injections typically remains close to $\sim 130~$Hz, except for a few exceptional cases where it varies between, roughly, $115~$Hz and $160~$Hz. It is important to note that the choice of the cutoff frequency itself serves as an arbitrary threshold for distinguishing the inspiral phase from the post-inspiral phase. Previous studies have demonstrated that small variations in the cutoff frequency do not have a significant impact on the IMRCT results~\cite{Ghosh2016}. Therefore, we believe that these variations in $f_\mathrm{ISCO}$ do not substantially affect our findings. Moreover, our choice of a high SNR of $50$ ensures that there is sufficient information content in both the inspiral and post-inspiral phases for accurately inferring the final mass and spin.} 

In Figure \ref{fig:gw150914_IMRCT}, we present the results of the IMR consistency test for GW150914-like microlensed injections. 
We focus on identifying deviations from GR using the GR quantile value, denoted as $Q_\mathrm{GR}$. 
We also illustrate the logarithmic Bayes factor $\log_{10}\mathcal{B}^{\rm ML}_{\rm UL}$ in favor of the microlensed hypothesis over the unlensed hypothesis for the full IMR signal, for reference.
Firstly, in Figure \ref{fig:BLU_vs_QGR_gw150914_IMRCT}, we depict the relationship between $Q_\mathrm{GR}$
and $\log_{10}\mathcal{B}^{\rm ML}_{\rm UL}$. 
We immediately observe that microlensing effects can introduce bias in the IMR consistency test, resulting in high $Q_\mathrm{GR}$ values that surpass $90\%$ in some cases. However, despite noting several occurrences of high $Q_\mathrm{GR}$ values $(>70\%)$ when $\log_{10}\mathcal{B}^{\rm ML}_{\rm UL}$ is also high $(>1)$, we do not discern any significant correlation between the two quantities.
In other words, there are instances where $\log_{10}\mathcal{B}^{\rm ML}_{\rm UL}$ is low~($<1$), yet $Q_\mathrm{GR}$ is high~($>70\%$), and vice versa, where $Q_\mathrm{GR}$ is low~($<10\%$), even when $\log_{10}\mathcal{B}^{\rm ML}_{\rm UL}$ is high~($>1$).


Upon closer inspection, we find that the dependence of $Q_\mathrm{GR}$ and $\log_{10}\mathcal{B}^{\rm ML}_{\rm UL}$ values on the $M_\mathrm{Lz}-y$ parameter space shows a characteristic trend that is notably different between the two (see Figs. \ref{fig:MLz-y_vs_BLU_gw150914_IMRCT} and \ref{fig:MLz-y_vs_QGR_gw150914_IMRCT}). 
Specifically, $\log_{10}\mathcal{B}^{\rm ML}_{\rm UL}$ adheres to the expected behavior, monotonically increasing as we increase the lens mass while reducing the impact parameter. For instance, in \Fref{fig:MLz-y_vs_QGR_gw150914_IMRCT}, the $\log_{10}\mathcal{B}^{\rm ML}_{\rm UL}$ is negative where the microlensing effects are weak (for injections having either $\log_{10}M_\mathrm{Lz}=1$ or $y=3$) and increases as we move towards the bottom-right region in the $\log_{10}M_\mathrm{Lz}-y$ plane, where it reach values $\mathcal{O}(10^2)$, indicating strong preference for microlensing.
In contrast, the variation in the $Q_\mathrm{GR}$ values does not mirror the trend as exhibited by $\log_{10}\mathcal{B}^{\rm ML}_{\rm UL}$, as significant $Q_\mathrm{GR}$ values (exceeding $50\%$) are primarily confined within a diagonal region spanning from the top-left to the bottom-right corner. Within the diagonal region, the  $Q_\mathrm{GR}$ values are high for high lens mass and low impact parameters (bottom-right region of the diagonal) and decrease as we move up in the diagonal towards the top-left direction. 

As shown in Figure \ref{fig:fML_contours} and elaborated upon in \Sref{subsec:basics_microlensing}, the diagonal region in the $\log_{10}M_\mathrm{Lz}-y$ parameter space corresponds to the range where wave effects dominate within the sensitivity band of ground-based detectors. In practical terms, the microlens parameters lying in this diagonal region would lead to interference patterns for signals in the frequency range of roughly $10$-$1000~$Hz. We observe that $Q_\mathrm{GR}$ is significant primarily within the `wave-dominated zone' and decreases as it moves toward the `long-wavelength regime'. Conversely, in the `geometric-optics regime' (the parameter space where the geometric-optics approximation tends to hold), we often encounter cases where deviations from GR are minimal ($Q_\mathrm{GR}<10\%$) albeit the lensing effects being very high $\log_{10}\mathcal{B}^{\rm ML}_{\rm UL}\sim\mathcal{O}(10)$. 

Therefore, from \Fref{fig:MLz-y_vs_QGR_gw150914_IMRCT}, \textit{it is evident that deviations from GR appear to occur primarily when interference effects are pronounced}. 
We will further validate this claim using a parameterized test as well as by studying a simulated population similar to GWTC-3 in the following sections.



\subsubsection{\label{subsubsec:GW150914_ppn_study}Parameterized test of
GR}
To further investigate the apparent deviation from GR due to microlensing effects in the inspiral, intermediate, and merger-ringdown phases of the signal, we perform the parameterized test of GR on our microlensed injection set. As discussed in \Sref{subsubsec:tiger}, we utilize four parameterization variables: $\delta\hat{\chi}_0$, $\delta\hat{\chi}_4$, $\delta\hat{\beta}_2$, and $\delta\hat{\alpha}_2$. 



In \Fref{fig:gw150914_TIGER}, we show the results of the parameterized test conducted on our microlensed injection set. Each of the four panels displays the Gaussian sigma values at which GR is excluded ($\sigma_\mathrm{GR}$; see \Sref{subsubsec:tiger}) in one of the four parameterized parameters, as indicated in the colorbar, within the $\log_{10}M_\mathrm{Lz}-y$ parameter space. The larger the magnitude of $\sigma_\mathrm{GR}$, the greater the significance of deviations from GR. For better visualization, we mark the cases with significant deviations ($\sigma_\mathrm{GR}>1$) using a diamond marker instead of the circular markers.
For all four parameters, we observe that the deviations are primarily significant within the wave-dominated zone and gradually decay as we move further into the long-wavelength regime. Meanwhile, the geometrical optics regime doesn't give any significant deviations. This pattern resembles our observation of the $Q_\mathrm{GR}$ value in \Fref{fig:MLz-y_vs_QGR_gw150914_IMRCT}.

To robustly illustrate the relationship between the deviations from GR and the characteristic frequency $f_\mathrm{ML}$ at which wave effects caused by a microlens are expected to become more pronounced, we suppress the two-dimensional parameter space of $M_\mathrm{Lz}-y$, shown in \Fref{fig:gw150914_TIGER}, into a one-dimensional representation using $f_\mathrm{ML}$.
We then represent the sigma deviation values in \Fref{fig:gw150914_TIGER_with_fML}, explicitly emphasizing three distinct regions: the long-wavelength regime, the wave-dominated zone, and the geometrical-optics regime. 
Moreover, we also depict the $\log_{10}\mathcal{B}^{\rm ML}_{\rm UL}$ values for each injection using a colormap for comparison between the deviations from GR with the overall strength of microlensing. 
The transparent markers that are not colored and only contain a black edge (unfilled circles)
are cases where $\log_{10}\mathcal{B}^{\rm ML}_{\rm UL}<0$ are cases where $\log_{10}\mathcal{B}^{\rm ML}_{\rm UL}<0$.
We also mark the $1\sigma$ and the $3\sigma$ deviations with dotted-red lines for better visualization.

For all four parameters, we clearly observe that the deviations from GR primarily increase in the wave-dominated zone (blue shaded region in the middle of each panel), including several cases with $\sigma_\mathrm{GR}\gtrsim 5$\footnote{We note that based on $\mathcal{O}(10^4)$ sample points in our posterior distributions, a value greater than $3$ Gaussian sigma cannot be stated with certainty~\citep{Narayan:2023vhm}. Nonetheless, we still quote the actual Gaussian sigma values derived from our distributions, along with this cautionary note.}
for all the parameters. 
Similarly, within the long-wavelength regime, we only observe significant deviations ($\sigma_\mathrm{GR}>1$) for parameters that measure deviations in the post-inspiral regime, i.e., $\delta\hat{\alpha}_2$ and $\delta\hat{\beta}_2$, where we also observed slight deviations around the intersection of the wave-dominated zone and the long-wavelength regime (also see Appendix \ref{appendix:fML_closer_look}).
Within the geometrical-optics regime, we do not notice any significant deviations for any parameter despite encountering several cases with high values of $\log_{10}\mathcal{B}^{\rm ML}_{\rm UL}$, including values $\mathcal{O}(10^2)$.

It's also worth noting that in all cases where high biases are observed ($\sigma_\mathrm{GR}>3$), we consistently find a high Bayes factor ($\log_{10}\mathcal{B}^{\rm ML}_{\rm UL}>1$) in favor of microlensing.
Conversely, there is only a single instance (marked with a red cross in the bottom-left panel) where significant biases are present ($1<\sigma_\mathrm{GR}<3$), but the Bayes factor remains notably low ($\log_{10}\mathcal{B}^{\rm ML}_{\rm UL}<0.5$). 
\textit{This suggests that events showing high deviations ($\sigma_\mathrm{GR}>3$) from GR must also be analyzed for the presence of microlensing features before claiming any (erroneous) GR deviations.}
However, to establish this conclusion more firmly, we conduct a more robust study using a population of microlensed signals, as detailed in Section \ref{subsubsec:pop_tiger_study}.

\begin{figure}
     \centering
     \begin{subfigure}[b]{0.48\textwidth}
        \centering
        \includegraphics[width=\textwidth]{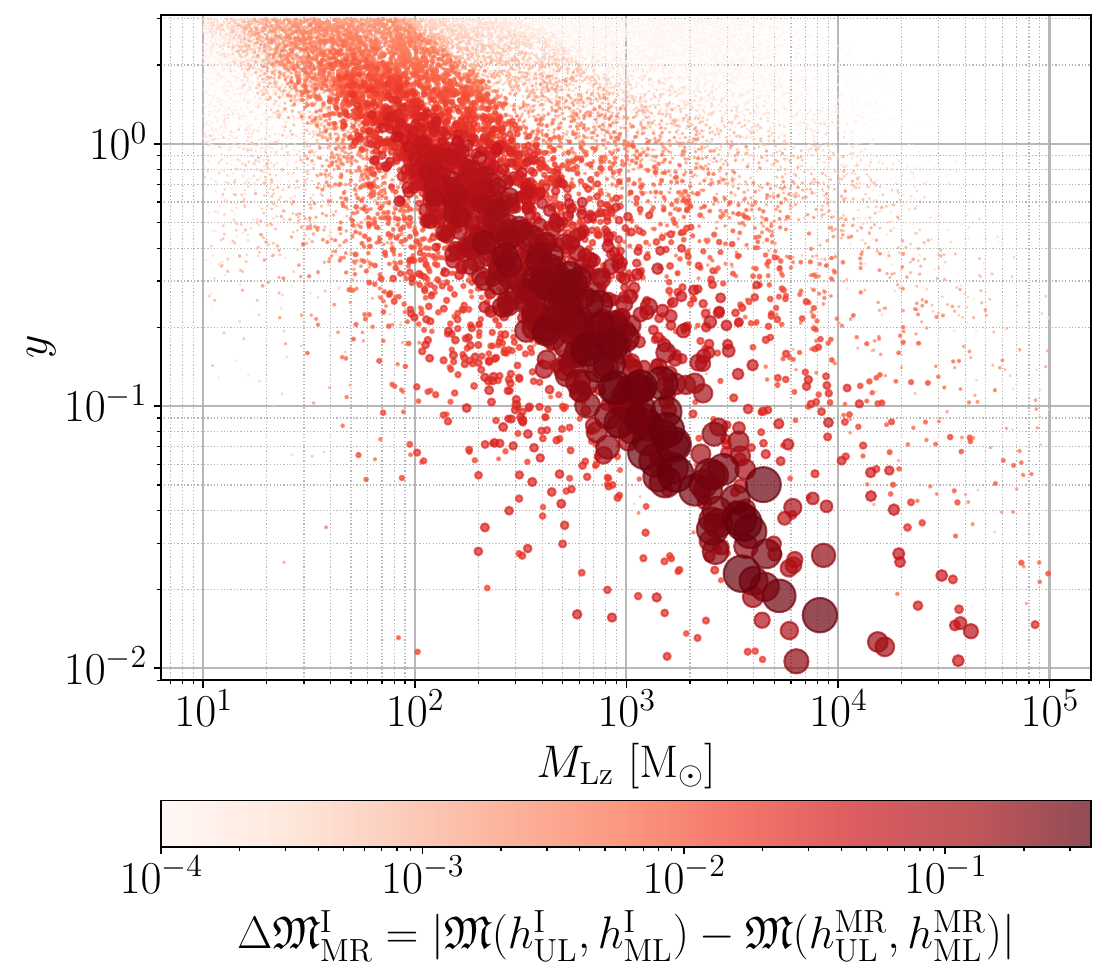}
        \caption*{}
        \label{fig:pop_IMRCT_v1}
     \end{subfigure}
     \begin{subfigure}[b]{0.48\textwidth}
        \centering
        \includegraphics[width=\textwidth]{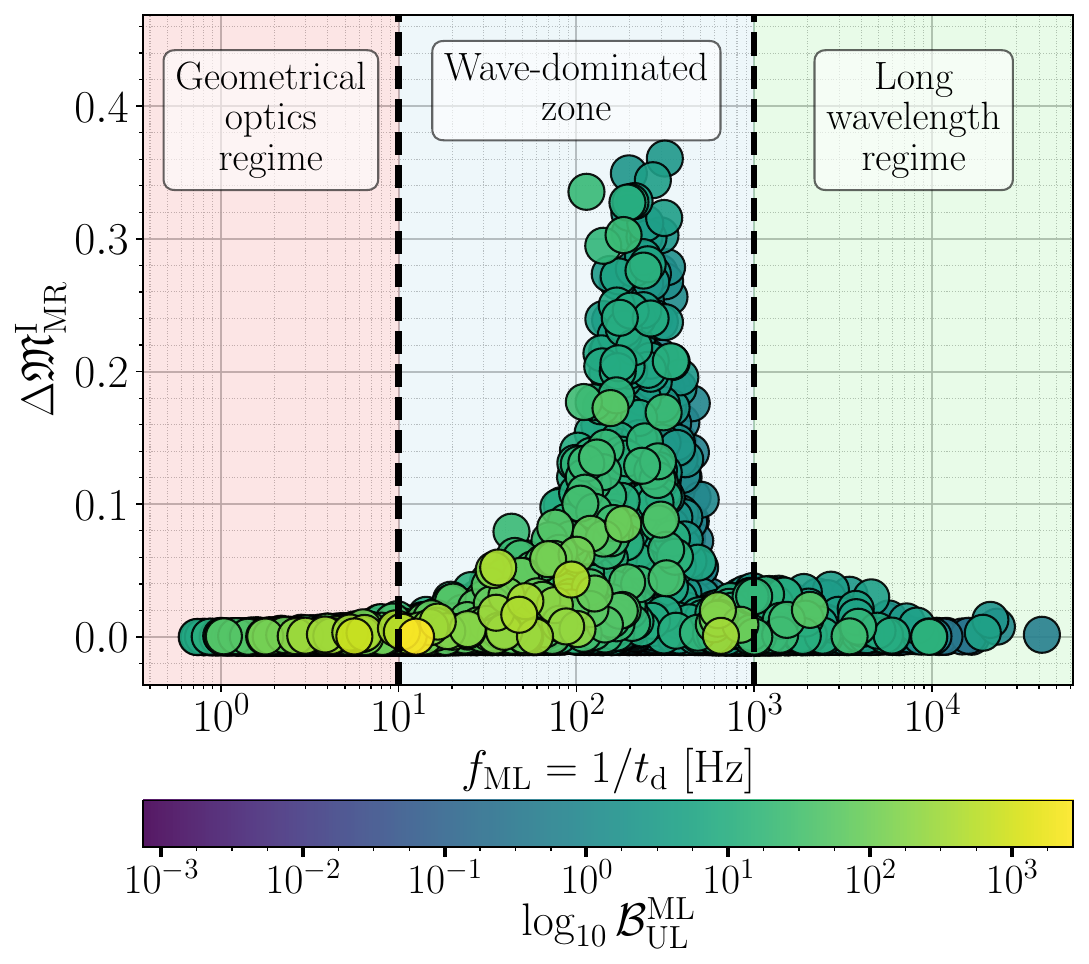}
        \caption*{}
        \label{fig:pop_IMRCT_v2}
     \end{subfigure}     
        \caption{\justifying IMRCT results for a population of simulated microlensed injections using $\Delta\mathfrak{M}^\mathrm{I}_\mathrm{MR}$ (\Eref{eq:delta_M_I-MR}) as a proxy to quantify deviations in IMRCT. \textit{Top panel}: Depicts $\Delta\mathfrak{M}^\mathrm{I}_\mathrm{MR}$ in the $\log_{10}M_\mathrm{Lz}-y$ parameter space. Marker color and size represent the strength of $\Delta\mathfrak{M}^\mathrm{I}_\mathrm{MR}$. \textit{Bottom panel}: Depicts $\Delta\mathfrak{M}^\mathrm{I}_\mathrm{MR}$ as a function of $f_\mathrm{ML}$. The colorbar represents Bayes factor values $\log_{10} \mathcal{B}^\mathrm{ML}_\mathrm{UL}$ estimated using \Eref{eq:bayes_factor_estimate}.}
        \label{fig:pop_IMRCT}
\end{figure}

\begin{figure*}
     \centering
     \begin{subfigure}[b]{0.47\textwidth}
         \centering
             \includegraphics[width=\textwidth]{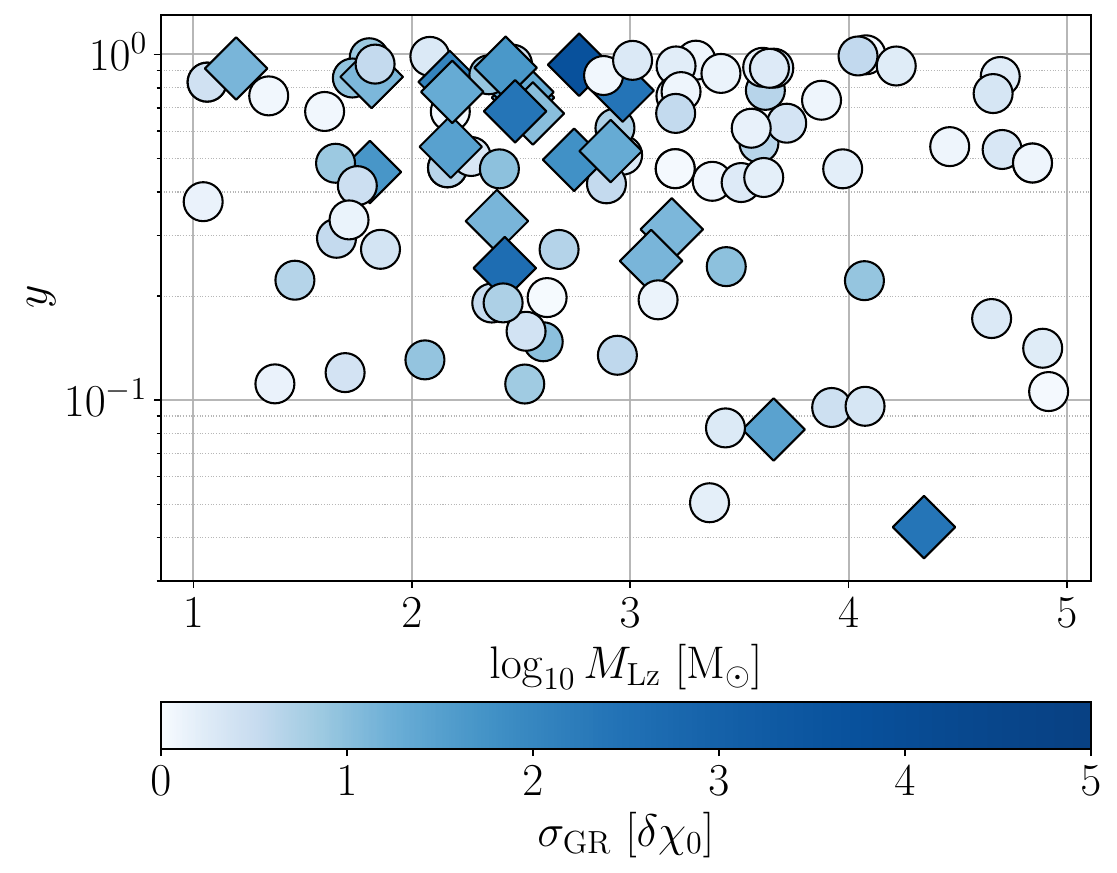}
         \caption*{}
         \label{fig:dchi_0_pop_TIGER}
     \end{subfigure}
     \begin{subfigure}[b]{0.47\textwidth}
         \centering
         \includegraphics[width=\textwidth]{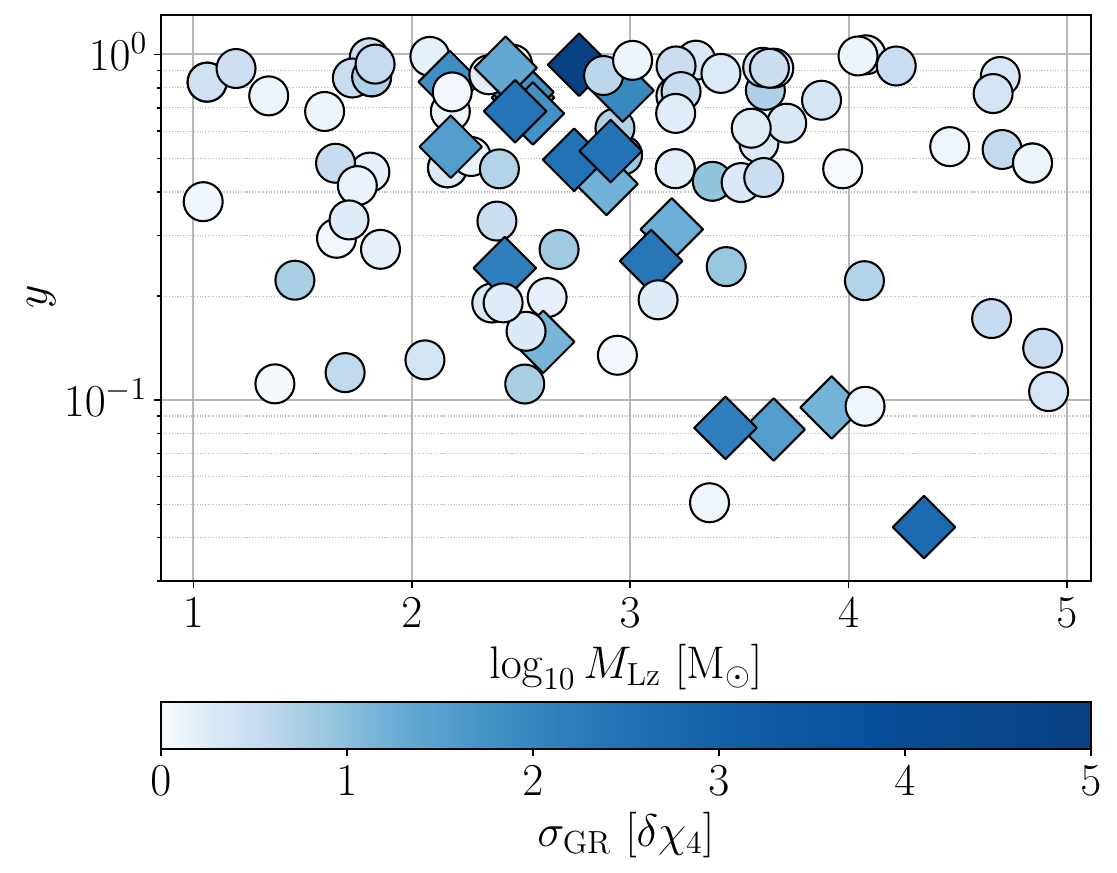}
         \caption*{}
         \label{fig:dchi_4_pop_TIGER}
     \end{subfigure}     
     \begin{subfigure}[b]{0.47\textwidth}
         \centering
         \includegraphics[width=\textwidth]{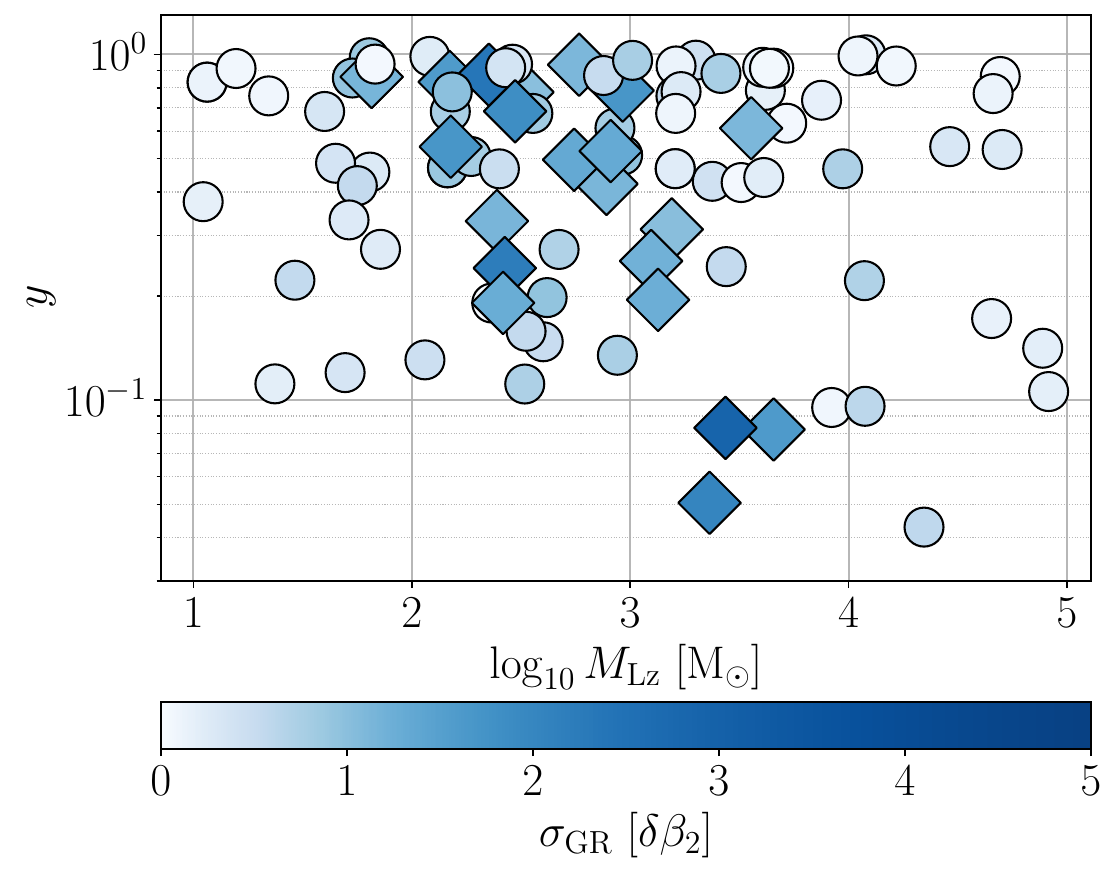}
         \caption*{}
         \label{fig:dalpha_2_pop_TIGER}
     \end{subfigure}
     \begin{subfigure}[b]{0.47\textwidth}
         \centering
         \includegraphics[width=\textwidth]{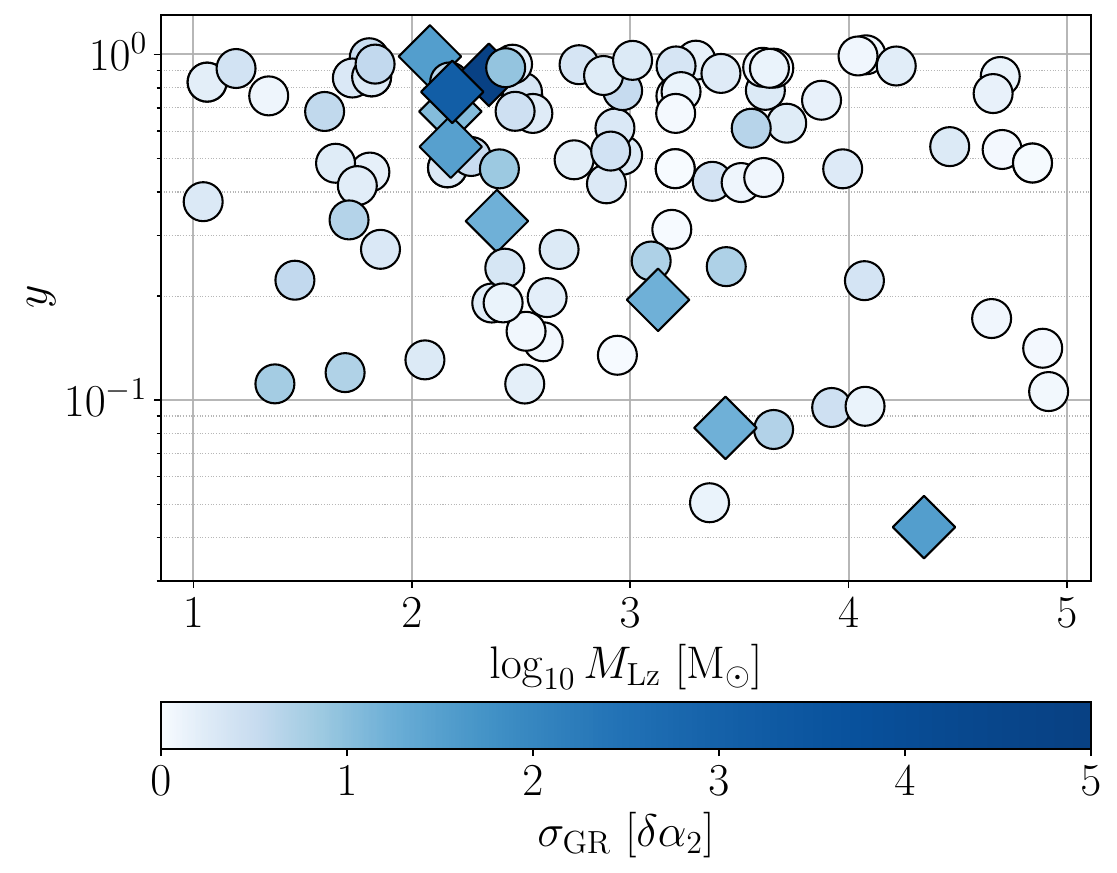}
         \caption*{}
         \label{fig:dbeta_2_pop_TIGER}
     \end{subfigure}     
        \caption{\justifying Same as \Fref{fig:gw150914_TIGER}, but for a population of microlensed injections as studied in \Sref{subsubsec:pop_tiger_study}.}
        \label{fig:pop_TIGER}
\end{figure*}

\begin{figure*}
    \centering
    \includegraphics[width=\textwidth]{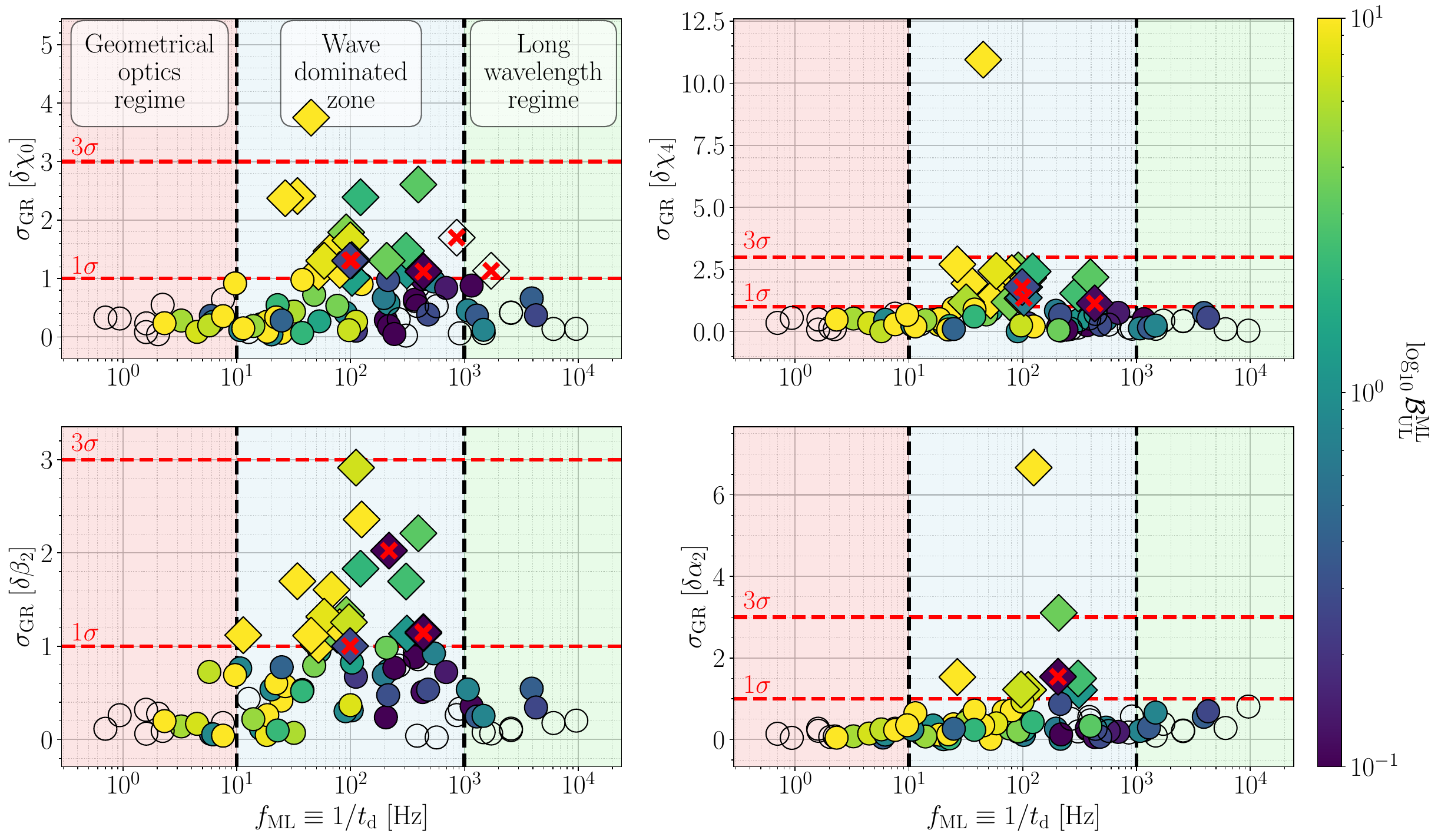}
    \caption{\justifying Same as \Fref{fig:gw150914_TIGER_with_fML}, but for a population of microlensed injections as studied in \Sref{subsubsec:pop_tiger_study}.}
    \label{fig:pop_TIGER_with_fML}
\end{figure*}

\subsection{\label{subsec:population_study}Population of Microlensed Injections}
To robustly analyze the effect of microlensing on tests of GR, we consider a population of simulated microlensed signals (see \Sref{subsubsec:injections} for details).

\subsubsection{\label{subsubsec:pop_imrct_study}IMRCT}
For the study of IMRCT on our microlensed injections, we only consider a subset of those having total mass in the range $50-100~$M$_\odot$, to which our detectors are sensitive from inspiral to ringdown phase of the coalescence (also called ``golden" binaries). This leaves us with a total of $\sim 1.5\times10^4$ injections.

Unlike our strategy in Sect \ref{subsubsec:GW150914_imrct_study}, we do not perform parameter estimation for this study. We save the computational expense of performing parameter estimation runs by instead studying a quantity that closely resembles the deviation parameter $Q_\mathrm{GR}$ in IMRCT. We realize from our study of GW150914-like injections in \Sref{subsubsec:GW150914_imrct_study} that, for IMRCT, the $Q_\mathrm{GR}$ value shows a positive correlation with the difference in the match between the microlensed and unlensed waveforms in the inspiral and post-inspiral phases, respectively. In quantitative terms, we observe that the quantity:
\begin{equation}
    \Delta\mathfrak{M}^\mathrm{I}_\mathrm{MR} = |\mathfrak{M}(h_\mathrm{UL}^\mathrm{I}, h_\mathrm{ML}^\mathrm{I}) - \mathfrak{M}(h_\mathrm{UL}^\mathrm{MR}, h_\mathrm{ML}^\mathrm{MR})|,
    \label{eq:delta_M_I-MR}
\end{equation}
where $\mathfrak{M}$ represents match as defined in \Eref{eq:match_func_def}
, shows a positive Pearson correlation value of around $\sim57\%$ with $Q_\mathrm{GR}$~(see Fig.~\ref{fig:delMatch_vs_QGR}
). 
Hence, we compute this quantity for our population as a computationally more cost-effective alternative for estimating the trend of the $Q_\mathrm{GR}$ value in IMRCT.

The results are shown in \Fref{fig:pop_IMRCT}. In the top panel, we depict the quantity $\Delta\mathfrak{M}^\mathrm{I}_\mathrm{MR}$ in the parameter space of $\log_{10}M_\mathrm{Lz}$ and $y$. Both color and size of the markers are proportional to the strength of $\Delta\mathfrak{M}^\mathrm{I}_\mathrm{MR}$. In other words, markers with dark red color and relatively larger size represent regions where we expect IMRCT to be biased due to microlensing. We again observe that deviations from GR are prominent mainly in the wave-dominated zone. This is more explicitly shown in bottom panel, where we plot the deviation parameter $\Delta\mathfrak{M}^\mathrm{I}_\mathrm{MR}$ as a function of $f_\mathrm{ML}$. We also color the markers based on their Bayes factor values $\log_{10} \mathcal{B}^\mathrm{ML}_\mathrm{UL}$, estimated using \Eref{eq:bayes_factor_estimate}. We clearly see deviations to increase sharply in the wave-dominated zone. In the geometrical-optics regime and the transition region from the wave-dominated to geometrical-optics regime,  we again notice several cases having high $\log_{10} \mathcal{B}^\mathrm{ML}_\mathrm{UL}$ values but with no significant $\Delta\mathfrak{M}^\mathrm{I}_\mathrm{MR}$ value.
These results are consistent with our observations in the previous sections.


\subsubsection{\label{subsubsec:pop_tiger_study}Parameterized test of
GR}
To get a population-wide behavior of deviations in the parameterized test of GR, we choose 
$100$ injections from our population of microlensed signals. To ensure a significant number of events strongly favors microlensing, we restrict the injected impact parameter value to be less than unity (i.e., $y<1$). Furthermore, the selection employs \Eref{eq:bayes_factor_estimate} to determine $\log_{10} \mathcal{B}^\mathrm{ML}_\mathrm{UL}$. Since this estimation is not reliable at lower SNRs, we keep the threshold on $\log_{10} \mathcal{B}^\mathrm{ML}_\mathrm{UL}>9$ for the majority of events $(\sim 50\%)$ \citep{mishra2023exploring}. The rest of the population corresponds to cases where microlensing is too weak to be correctly identified in the model selection process. 
Such a population of weakly microlensed signals help us in investigating the possibility of \textit{stealth bias} \citep{Cornish:2011ys, Vallisneri:2012qq, Vitale:2013bma} in the context of microlensing -  cases where microlensing itself is not large enough to be detectable but the systematic errors due to it remain significant (i.e., larger than the statistical uncertainties in parameter estimation), appearing as deviations from GR\footnote{The term `stealth-bias' was coined by \cite{Cornish:2011ys} in the context that GR templates can be significantly biased even when there is no significant evidence for adopting an alternative theory of gravity. In our work, we use this term to mean when microlensing effects lead to biases in tests of GR without being detectable themselves.}.

As discussed in \Sref{subsubsec:tiger}, we study the deviations in each of the four deviation parameters $(\delta\hat{\chi}_0,~\delta\hat{\chi}_4,~\delta\hat{\alpha}_2,~\delta\hat{\beta}_2)$ separately. The results are plotted in \Fref{fig:pop_TIGER} and \Fref{fig:pop_TIGER_with_fML}, in a similar fashion as in \Sref{subsubsec:GW150914_ppn_study}. Particularly, in \Fref{fig:pop_TIGER}, we plot the Gaussian sigma deviations in the parameter space of $M_\mathrm{Lz}-y$. We again show the cases having $\sigma_\mathrm{GR}>1$ with a diamond marker for better visualization. 
As we can see, there are several cases that give significant deviation from GR $(\sigma_\mathrm{GR}>1)$. These include cases that lie in the modest regime in our parameter space, i.e., having the characteristic impact parameter value of $y\approx 1$ and the lens mass $M_\mathrm{Lz}<100~$M$_\odot$. 

Meanwhile, in \Fref{fig:pop_TIGER_with_fML}, we plot it against the $f_\mathrm{ML}$ with colors representing the Bayes factor $\log_\mathrm{10} \mathcal{B}^\mathrm{ML}_\mathrm{UL}$ values obtained via nested sampling. We also put an upper cap of $10$ on the colorbar to get a better visibility of Bayes factors around unity.
We observed several instances where $\sigma_\mathrm{GR}>1$ for all the parameters, including cases where it is even beyond $3\sigma_\mathrm{GR}$. However, we only saw two cases having $\sigma_\mathrm{GR}>5$, one for $\delta\hat{\chi}_4$ and the other for $\delta\hat{\alpha}_2$.
We clearly see that deviations increase pre-dominantly in the wave-dominated zone and fall off as we go further into the long-wavelength regime. 
We do not see any deviation from GR in the geometrical optics regime.  
These results are consistent with all our previous observations. 

To investigate the possibility of \textit{stealth bias}, we specifically identify cases with significant deviations from GR ($\sigma_\mathrm{GR}>1$) but without a strong preference for the microlensing model ($\log_{10} \mathcal{B}^\mathrm{ML}_\mathrm{UL}<1$), marking them with a red cross.
It is worth noting that although several such cases are found, none reach the 'evidence'-level significance ($\sigma_\mathrm{GR}>3$). This is because all cases with $\sigma_\mathrm{GR}>3$ also exhibit a strong preference for the microlensing hypothesis over the null hypothesis ($\log_{10} \mathcal{B}^\mathrm{ML}_\mathrm{UL}>1$).

\section{\label{sec:conclusion}Conclusion}
In this study, we examined the potential impact of microlensing effects on tests of GR. We adopted an isolated point-lens model for our study, covering a parameter space typically relevant for the ground-based detectors, i.e., $M_\mathrm{Lz}\in(1,10^5)~$M$_\odot$ and $y\in(0.01, 3)$. 
However, it is important to note that our findings and conclusions are expected to apply broadly to any microlensing scenario, regardless of the specific parameter space, as we relate the biases observed to the fundamental characteristic of gravitational lensing.
Our investigation centered on two theory-agnostic tests of GR: the inspiral-merger-ringdown consistency test (IMRCT) and the parameterized tests of GR. These tests allowed us to explore deviations from GR across different evolutionary phases of a GW signal: inspiral, intermediate, and merger-ringdown. We consider both high-SNR GW150914-like systems and a population of microlensed signals similar to GWTC-3 with added microlensing effects.

Our findings lead to the following conclusions:
\begin{enumerate}

\item Microlensing can significantly bias tests of GR, with confidence levels even exceeding $5\sigma$. Fortunately, whenever there is a strong deviation from GR  $(\sigma>3)$, there is also a strong preference for microlensing over the null hypothesis that the signal is unlensed. In other words, we consistently observe  $\log_{10}\mathcal{B}^\mathrm{ML}_\mathrm{UL}>1$ in cases where $\sigma_\mathrm{GR}>3$, preventing us from falsely claiming deviations from GR.
However, it is important to note that we do encounter scenarios in which deviations from GR are still significant $\sigma_\mathrm{GR}\in(1,3)$, but the Bayes factor isn't strong enough to confidently assert microlensing ($\log_{10}\mathcal{B}^\mathrm{ML}_\mathrm{UL}<1$). We refer to these situations as the \textit{stealth bias} of microlensing in tests of GR.

\item In general, we do not find any correlation between the deviations from GR and the Bayes factor $\log_{10}\mathcal{B}^\mathrm{ML}_\mathrm{UL}$. Upon closer inspection, we demonstrate that the deviations from GR occur primarily when interference effects are pronounced, i.e., when $f_\mathrm{GW}\sim 1/t_\mathrm{d}\equiv f_\mathrm{ML}$. 
These deviations intensify within the wave-dominated region of the $M_\mathrm{Lz}-y$ parameter space, where $f_\mathrm{ML}\in (10,~10^3)~$Hz, and diminish as we move further into the long-wavelength regime $(f_\mathrm{ML} > 10^3~{\rm Hz})$.
In geometrical optics regime ($f_\mathrm{ML} < 10~$Hz), where we can consider the resultant signal to be a trivial superposition of two signals which differ only by a constant amplitude, a phase shift of $\pi/2$, and a time-delay value smaller than the chirp time of the signal, we saw the least bias despite noting several instances where the lensing Bayes factor strongly supported microlensing ($\log_{10}\mathcal{B}^\mathrm{ML}_\mathrm{UL}>1$). 
\item We further refine the microlensing condition for a given GW signal, suggesting that it is more precisely defined as when $f_\mathrm{ML} \in \sim(f_\mathrm{low},~f_\mathrm{RD})$, rather than the generic $(10,~1000)~$Hz range used in our study. Upon closer inspection, we observe that parameters quantifying deviations in the inspiral phase, such as $\delta\hat{\chi}_0$ and $\delta\hat{\chi}_4$, exhibit increasing deviations when $f_\mathrm{ML}\sim f_\mathrm{avg.}$ (power-weighted average frequency). Conversely, for parameters assessing deviations in the post-inspiral phase, such as $\delta\hat{\alpha}_2$ and $\delta\hat{\beta}_2$, the deviations tend to increase when $f_\mathrm{ML}\sim f_\mathrm{ISCO}$ (frequency corresponding to the innermost stable circular orbit).
\end{enumerate}

Since the rate of expected microlensing events is significantly lower in comparison to unlensed signals, the potential for microlensed signals to bias tests of GR at a population level is highly improbable.
Nonetheless, as our study demonstrates, there is a possibility that certain individual events in the future could exhibit notable deviations from GR due to microlensing effects. This would necessitate conducting dedicated microlensing analyses, alongside investigations into other potential effects, before making any erroneous claims of deviations from GR. Notably, these deviations in a few events could serve as indicators, helping to prioritize them as potential microlensing candidates, as microlensing searches are computationally expensive.

While our current study primarily focuses on demonstrating how microlensing effects can introduce deviations from GR, future research has the potential to explore additional aspects. For example, one could investigate potential biases in microlensing searches resulting from non-GR effects, addressing the possible degeneracy between non-GR effects and microlensing. Furthermore, future studies might investigate the impact of microlensed signals on population-level tests of GR, considering various compact dark matter fractions and detector sensitivities. Lastly, a crucial avenue for future research involves a detailed examination of cases leading to stealth biases of microlensing on tests of GR.

\begin{acknowledgments}
The authors are grateful for computational resources provided by the LIGO Laboratory and Inter-University Center for Astronomy \& Astrophysics (IUCAA), Pune, India, and supported by the National Science Foundation Grants PHY-0757058 and PHY-0823459. A. Mishra would like to thank the University Grants Commission (UGC), India, for financial support as a research fellow. N.V.K. acknowledges the support from the Science and Education Research Board, Government of India through the National Post Doctoral Fellowship grant (Reg. No. PDF/2022/000379).
The authors would like to thank Sukanta Bose, N.~K.~Johnson-McDaniel, Ania Liu,, and Anupreeta More for their careful reading of the manuscript and for making various valuable suggestions and comments. Additionally, we thank  P.~Ajith, M. Çalışkan, J.~Ezquiaga, and A.~Vijaykumar for their useful suggestions.

The work utilizes the following software packages:
\texttt{Cython}~\citep{behnel2011cython},
\texttt{NumPy}~\citep{harris2020array}, 
\texttt{SciPy}~\citep{2020SciPy-NMeth}, 
\texttt{LALSuite}~\citep{2020ascl.soft12021L},
\texttt{dynesty}~\citep{speagle2020dynesty}, 
\texttt{Bilby}~\citep{Ashton:2018jfp,Romero-Shaw:2020owr}, 
\texttt{PESummary}~\citep{Hoy:2020vys},
\texttt{GWPopulation}~\citep{2019PhRvD.100d3030T}, 
\texttt{Matplotlib}~\citep{Hunter:2007}, and
\texttt{Jupyter notebook}~\citep{Kluyver2016jupyter}.

\end{acknowledgments}

\appendix
\begin{figure*}
     \centering
     \begin{subfigure}[b]{0.47\textwidth}
         \centering
             \includegraphics[width=\textwidth]{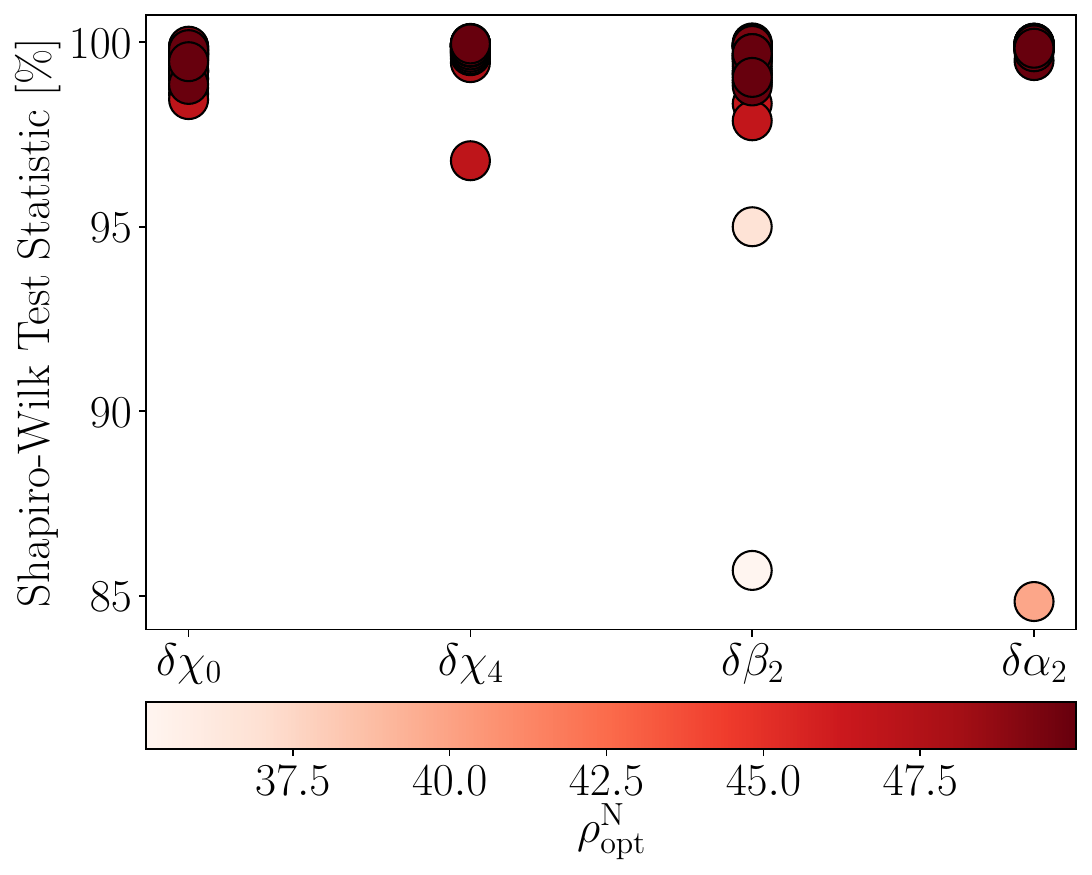}
         \caption*{}
         \label{fig:gw150914_tiger_normality_test}
     \end{subfigure}
     \begin{subfigure}[b]{0.47\textwidth}
         \centering
         \includegraphics[width=\textwidth]{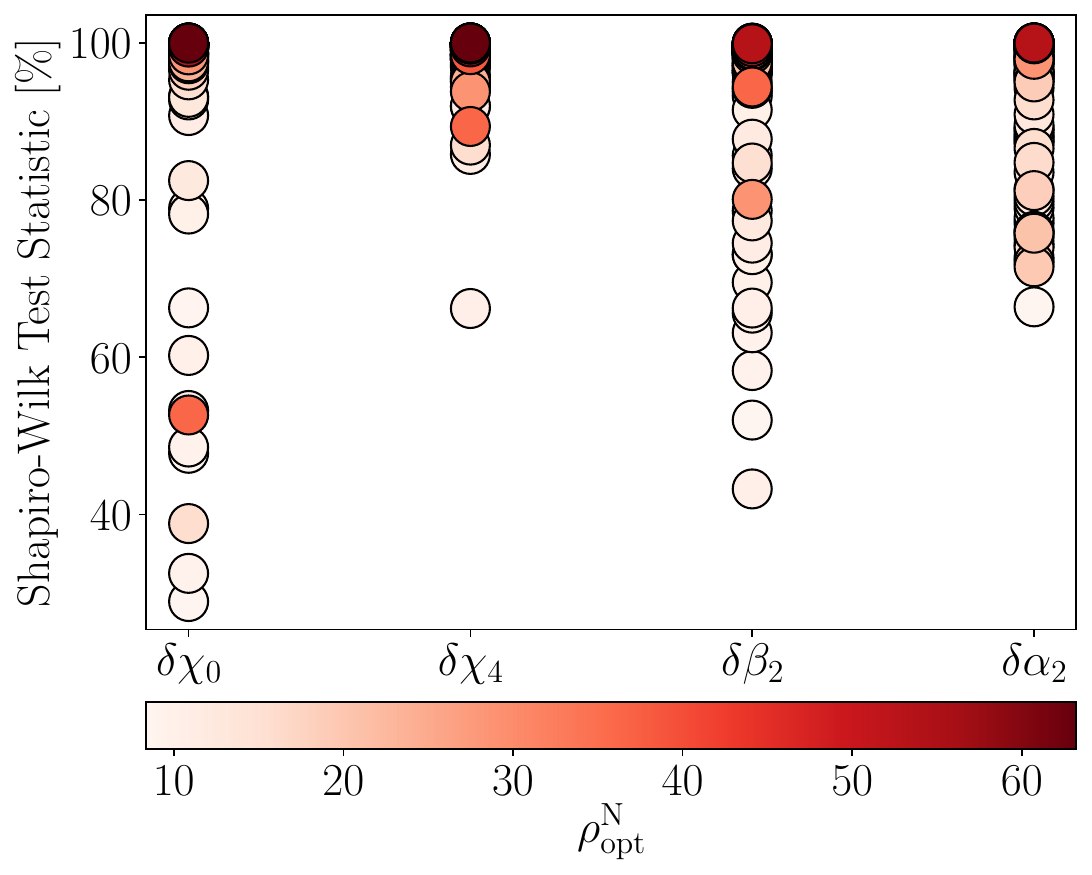}
         \caption*{}
         \label{fig:pop_tiger_normality_test}
     \end{subfigure}     
        \caption{\justifying Shapiro-Wilk normality test for the posterior samples of  $\{\delta\hat{\chi}_0$, $\delta\hat{\chi}_4$, $\delta\hat{\alpha}_2$, $\delta\hat{\beta}_2\}$ in our injections. The colorbar shows the corresponding SNR of the events. \textit{Left-panel:} Shapiro-Wilk test statistic for samples obtained for GW150914-like injections in \Sref{subsubsec:GW150914_ppn_study} (see also Figs. \ref{fig:gw150914_TIGER} and \ref{fig:gw150914_TIGER_with_fML}). \textit{Right-panel:} Shapiro-Wilk  test statistic for the samples obtained for the study of the microlensed population in \Sref{subsubsec:pop_tiger_study} (see also Figs. \ref{fig:pop_TIGER} and \ref{fig:pop_TIGER_with_fML}).}
        \label{fig:shapiro-wilk_normality_test}
\end{figure*}

\section{\label{appendix:normality_test}Normality Test of Posterior Samples}
In Sections \ref{subsubsec:GW150914_ppn_study} and \ref{subsubsec:pop_tiger_study}, we conducted parameterized test analysis using Gaussian sigma values to quantify deviations from General Relativity (GR). However, this approach implicitly assumes that the posterior distributions are Gaussian. To assess the validity of this assumption, we performed the Shapiro-Wilk test of normality \cite{shapiro1965analysis} on our posterior samples of the deviation parameters, namely, $\delta\hat{\chi}_0$, $\delta\hat{\chi}_4$, $\delta\hat{\alpha}_2$, $\delta\hat{\beta}_2$.
The Shapiro-Wilk test is known to be one of the most powerful normality tests \citep{mendes2003type, keskin2006comparison, razali2011power}. The test statistic, denoted as $\mathcal{SW}$, tends to be higher as the samples more closely resemble a Gaussian distribution and should approach $100\%$ for a true Gaussian distribution.

In \Fref{fig:shapiro-wilk_normality_test}, we present the results of the Shapiro-Wilk test. The corresponding SNR values of the events are displayed in the colorbar. The samples obtained for GW150914-like injections in \Sref{subsubsec:GW150914_ppn_study} exhibit a high degree of confidence in being Gaussian distributions. With the exception of two cases, all other cases yield a statistic value above $95\%$ (i.e., $\mathcal{SW}>95\%$ for the majority of cases).

Building on this observation, we introduce a new metric \textit{p-index}, denoted as $\mathcal{P}$, to quantify the nature of these distributions more robustly. It is defined as the value such that $\mathcal{P}~\%$ of the samples in the distribution have values above $\mathcal{P}~\%$.
In mathematical terms, for a distribution $\mathbb{D}=\{d_i\}$ of percentages, it can be expressed as:
\begin{equation}
\mathcal{P} = \argmin_{x \in (0,100)} \left| \frac{\text{count}(\{d_i: d_i > x\})}{\text{count}(\{d_i\})} - \frac{x}{100} \right|,
\end{equation}
where $x$ denotes the possible percentage values, and `count' represents the number of elements in the set inside the parentheses.
This additional metric further reinforces our confidence in the Gaussian nature of the data. Specifically, the Shapiro-Wilk test statistic values for our deviation parameters, including $\delta\hat{\chi}_0$, $\delta\hat{\chi}_4$, $\delta\hat{\alpha}_2$, and $\delta\hat{\beta}_2$, yield $\mathcal{P}$ values of $98.5\%,~96.8\%,~95.0\%,$ and $96.7\%$, respectively. For instance, this means that the statistic values for the posterior distribution of $\delta\hat{\chi}_0$ are above $98.5\%$ for approximately $98.5\%$ of the cases, and so on. 

In a similar fashion, the \textit{right-panel} of \Fref{fig:shapiro-wilk_normality_test} displays the test statistic results for the samples obtained during the study of the microlensed population in \Sref{subsubsec:pop_tiger_study}. 
Here, $\mathcal{P}$ values for our deviation parameters, $\delta\hat{\chi}_0$, $\delta\hat{\chi}_4$, $\delta\hat{\alpha}_2$, and $\delta\hat{\beta}_2$ are  $88.0\%,~93.8\%,~84.7\%,$ and $82.7\%$, respectively.

In conclusion, the assumption of Gaussianity holds well in our case, especially for high SNR events. For GW150914-like injections, approximately $95\%$ of the cases yielded a statistic value exceeding $95\%$, while in the case of the population study, approximately $85\%$ of the cases yielded a statistic value above $85\%$.

\begin{figure*}
    \centering
    \includegraphics[width=\textwidth]{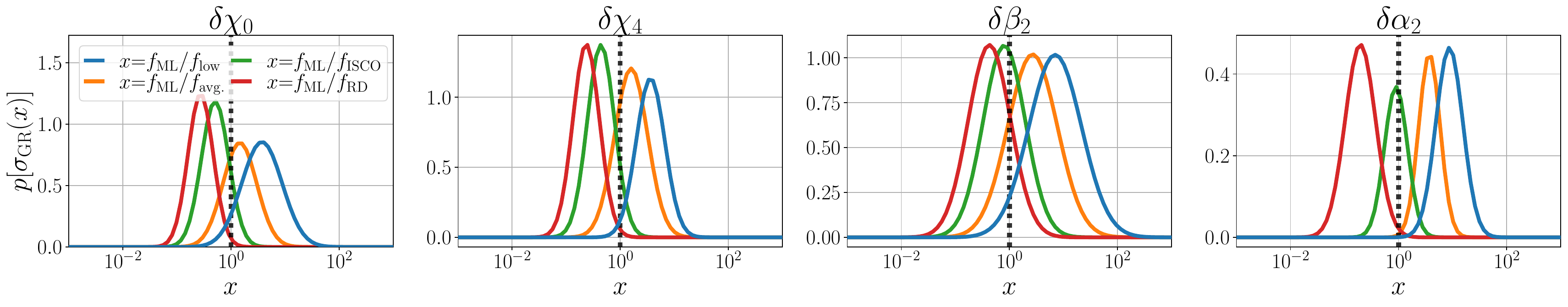}
    \caption{\justifying Gaussian fits to the distribution of $\sigma_\mathrm{GR}$ for the microlensed population injections plotted against the variable $x \in \{f_\mathrm{ML}/f_\mathrm{low},~f_\mathrm{ML}/f_\mathrm{avg.},~f_\mathrm{ML}/f_\mathrm{ISCO},~f_\mathrm{ML}/f_\mathrm{RD}\}$ for all the deviation parameters $\{\delta\hat{\chi}_0,~\delta\hat{\chi}_4,~\delta\hat{\alpha}_2,~\delta\hat{\beta}_2\}$ (indicated on top of each panel). A dashed-black line is included at $x=1$ for reference, representing the characteristic frequency of GW $f_{\mathrm{GW}}^{\mathrm{c}}$ which when comparable to $f_{\mathrm{ML}}$ gives rise to wave effects.}
    \label{fig:pop_TIGER_fML_pdf}
\end{figure*}

\begin{figure*}
    \centering
    \includegraphics[width=\textwidth]{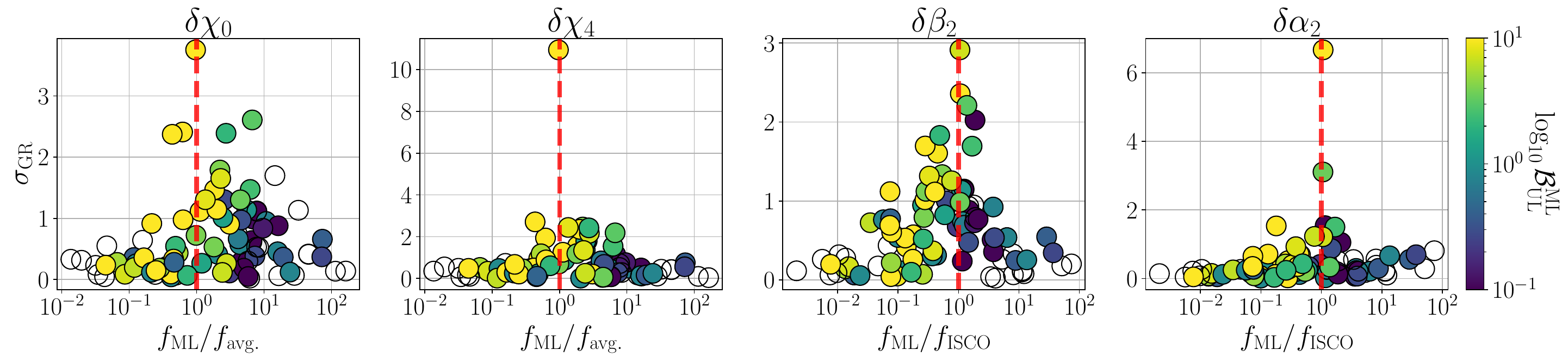}
    \caption{\justifying The distribution of $\sigma_\mathrm{GR}$ is analyzed in the context of some selected values of $x$ for different deviation parameters. \textit{Leftmost two panels}: $x=f_\mathrm{ML}/f_\mathrm{avg.}$ is considered with deviation parameters $\{\delta\hat{\chi}_0,\delta\hat{\chi}_4\}$. \textit{Rightmost two panels}: For parameters $\{\delta\hat{\alpha}_2,\delta\hat{\beta}_2\}$, we choose $x=f_\mathrm{ML}/f_\mathrm{ISCO}$. A dashed-red line at $x=1$ is included for reference.}
    \label{fig:pop_TIGER_fML_selected}
\end{figure*}

\section{\label{appendix:fML_closer_look}Investigating the correspondence between $f_\mathrm{ML}$ and $f_\mathrm{GW}$: A closer perspective}
As discussed in Section \ref{subsec:basics_microlensing}, the condition for wave effects is that $f_{\mathrm{ML}} \sim f_{\mathrm{GW}}$. In the sections leading up to this point, we assumed that $f_{\mathrm{GW}}\in (10, 1000)$ Hz, based on the sensitivity of ground-based detectors. However, we can be more precise in defining the characteristic frequency of GW, denoted as $f_{\mathrm{GW}}^{\mathrm{c}}$, which when comparable to $f_{\mathrm{ML}}$ gives rise to wave effects. This approach is possible because, for a given GW signal, we can explicitly determine the frequency range where it exhibits significant power in the detectors. Therefore, in this section, we leverage this knowledge to establish a closer correspondence between $f_{\mathrm{ML}}$ and $f_{\mathrm{GW}}$.

The lower frequency cutoff typically used for GW data analysis is $f_\mathrm{low} = 20~$Hz, below which the noise dramatically increases, especially due to seismic and thermal noise sources \citep{Abbott:2016xvh, Valdes:2022pnm}.
Next, the inspiral and post-inspiral sections are typically demarcated using the frequency at the inner-most stable circular orbit $f_\mathrm{ISCO}$ \citep{Hanna:2008um, GWTC-3-TGR-LVK}, which is the same cutoff frequency we used for IMRCT (see \Sref{subsubsec:imrct}). 
We note that this demarcation is not the same as the demarcation used for the parameterized test.
Moreover, a rough estimate of the upper frequency of a GW signal can be determined based on the ringdown frequency, $f_\mathrm{RD}$, associated with the dominant quasi-normal mode $(l=2,~m=2,~n=0)$ \citep{Hanna:2008um, Berti:2005ys}.
Additionally, one can define a signal's power-weighted average frequency, denoted as $f_\mathrm{avg.}$, using the following equation:
\begin{equation}
f_\mathrm{avg.} = \frac{\int_{-\infty}^{\infty}f\cdot|\widetilde{h}(f)|^2df}{\int_{-\infty}^{\infty}|\widetilde{h}(f)|^2df}.
\end{equation}
This power-weighted average frequency reflects where most of the signal power is concentrated. In the case of GW signals, the inspiral phase typically contributes the most power. Consequently, we usually expect $f_\mathrm{avg.} < f_\mathrm{ISCO}$. Thus, a signal that spans a sufficient number of cycles in the inspiral phase within the current detectors can be effectively divided into different phases of its evolution using these frequency markers: $f_\mathrm{low} < f_\mathrm{avg.} < f_\mathrm{ISCO} < f_\mathrm{RD}$.

We consider posterior distributions of the deviation parameters, namely, $\delta\hat{\chi}_0$, $\delta\hat{\chi}_4$, $\delta\hat{\alpha}_2$, and $\delta\hat{\beta}_2$, for the different injections studied in \ref{subsubsec:pop_tiger_study}.
Since the characteristic frequencies mentioned above differ for each injection, we study the ratio of $f_\mathrm{ML}$ with these frequencies, i.e., we examine the quantity:
$x:x \in \{f_\mathrm{ML}/f_\mathrm{low},~f_\mathrm{ML}/f_\mathrm{avg.},~f_\mathrm{ML}/f_\mathrm{ISCO},~f_\mathrm{ML}/f_\mathrm{RD}\}$.
Given that the condition for pronounced wave effects is $f_\mathrm{ML}\sim f_\mathrm{GW}^{\mathrm{c}}$, we analyze the distribution of $\sigma_\mathrm{GR}$ as a function of $x$ to determine which ratio results in a peak near $1$.

In \Fref{fig:pop_TIGER_fML_pdf}, we present the distribution of $\sigma_\mathrm{GR}$ for the microlensed population injections against the variable $x$ for all the deviation parameters (indicated on top of each panel). To enhance clarity, we employ Gaussian fits to combine all four $\sigma_\mathrm{GR}-x$ distributions of a given deviation parameter into a single panel. An example illustrating the actual $\sigma_\mathrm{GR}-x$ distribution is provided in \Fref{fig:pop_TIGER_fML_selected} for specific cases. We also include a dashed-black reference line at $x=1$, representing the frequency where $f_{\mathrm{ML}} \sim f_{\mathrm{GW}}^{\mathrm{c}}$.
Given that, for our injections, $1/f_\mathrm{RD} < 1/f_\mathrm{ISCO} < 1/f_\mathrm{avg.} < 1/f_\mathrm{low}$, the Gaussian curves follow a consistent trend. Specifically, the $\sigma_\mathrm{GR}-x$ distribution for $x=f_\mathrm{ML}/f_\mathrm{RD}$ occupies the leftmost position with its peak significantly below 1 (indicated by the red curves), while for $x=f_\mathrm{ML}/f_\mathrm{low}$, it occupies the rightmost position in all panels with the peak significantly above 1 (indicated by the blue curves). In other words, assuming $f_\mathrm{GW}^\mathrm{c}=f_\mathrm{low}$ would result in an underestimation, whereas assuming $f_\mathrm{GW}^\mathrm{c}=f_\mathrm{RD}$ would lead to an overestimation. Therefore, the fact that the curves of these ratios lie on different sides of unity implies that the condition for microlensing in the case of a given GW signal can be more precisely defined as when $f_\mathrm{ML} \in (f_\mathrm{low},~f_\mathrm{RD})$, as opposed to the generic range of $(10,~1000)$Hz we used earlier.

We also note that for the deviation parameters $\delta\hat{\chi}_0$ and $\delta\hat{\chi}_4$, the peak corresponding to $x=f_\mathrm{ML}/f_\mathrm{avg.}$ is closest to unity. This implies that these deviation parameters exhibit the highest bias when $f_\mathrm{ML} \sim f_\mathrm{GW}^\mathrm{c} \sim f_\mathrm{avg.}$.
This outcome is expected because these parameters primarily measure deviations from GR during the inspiral phase of the signal. In \Fref{fig:pop_TIGER_fML_selected}, we explicitly depict the distribution of $\sigma_\mathrm{GR}$ for the $x=f_\mathrm{ML}/f_\mathrm{avg.}$ ratio for these two parameters in the leftmost two columns, where we observe an increase in deviations around $x\sim 1$.
In contrast, for the deviation parameters $\delta\hat{\alpha}_2$ and $\delta\hat{\beta}_2$, the peak corresponding to $x=f_\mathrm{ML}/f_\mathrm{ISCO}$ is closest to unity, indicating that these deviation parameters exhibit the highest bias when $f_\mathrm{ML} \sim f_\mathrm{GW}^\mathrm{c} \sim f_\mathrm{ISCO}$.
This can be explained by the fact that these parameters primarily measure deviations from GR during the post-inspiral phase of the signal. In \Fref{fig:pop_TIGER_fML_selected}, we explicitly display the distribution of $\sigma_\mathrm{GR}$ for the $x=f_\mathrm{ML}/f_\mathrm{ISCO}$ ratio for these two parameters in the rightmost two columns, where, once again, we notice an increase in deviations around $x\sim 1$.

In conclusion, we find that the condition for microlensing in the case of a given GW signal can be more precisely defined as when $f_\mathrm{ML} \in (f_\mathrm{low},~f_\mathrm{RD})$, as opposed to the generic range of $(10,~1000)~$Hz we used earlier. We went further and tried to find even specific conditions. For example, for parameters that measure deviations in the inspiral regime, such as $\delta\hat{\chi}_0$ and $\delta\hat{\chi}_4$, the deviations increase when  $f_\mathrm{ML}\sim f_\mathrm{avg.}$. While for parameters that measure deviations in the post-inspiral regime, such as $\delta\hat{\alpha}_2$ and $\delta\hat{\beta}_2$, the deviations increase when  $f_\mathrm{ML}\sim f_\mathrm{ISCO}$.

\section{Additional Figures}
\begin{figure}
    \centering
    \includegraphics[width=\linewidth]{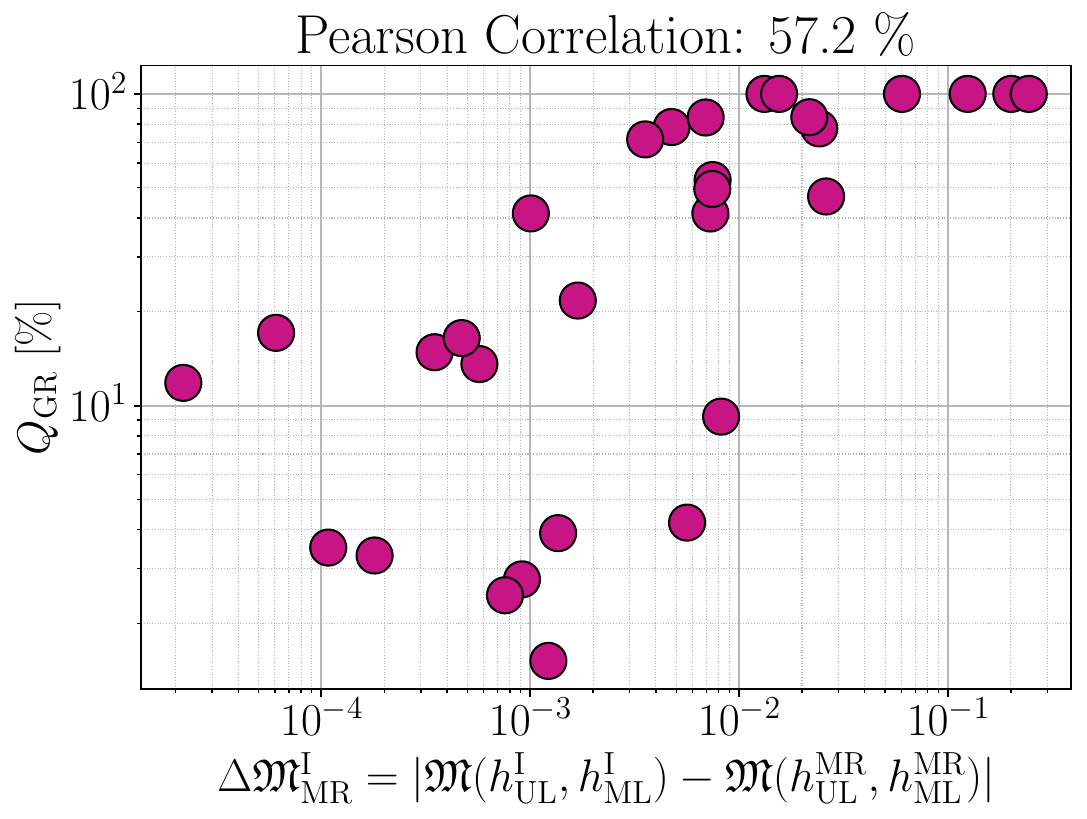}
    \caption{\justifying Illustration of correlation between $\mathcal{Q}_\mathrm{GR}$ and $\Delta\mathfrak{M}^\mathrm{I}_\mathrm{MR}$.}
    \label{fig:delMatch_vs_QGR}
\end{figure}

\bibliography{bibliography}

\begin{thebibliography}{120}%
\makeatletter
\providecommand \@ifxundefined [1]{%
 \@ifx{#1\undefined}
}%
\providecommand \@ifnum [1]{%
 \ifnum #1\expandafter \@firstoftwo
 \else \expandafter \@secondoftwo
 \fi
}%
\providecommand \@ifx [1]{%
 \ifx #1\expandafter \@firstoftwo
 \else \expandafter \@secondoftwo
 \fi
}%
\providecommand \natexlab [1]{#1}%
\providecommand \enquote  [1]{``#1''}%
\providecommand \bibnamefont  [1]{#1}%
\providecommand \bibfnamefont [1]{#1}%
\providecommand \citenamefont [1]{#1}%
\providecommand \href@noop [0]{\@secondoftwo}%
\providecommand \href [0]{\begingroup \@sanitize@url \@href}%
\providecommand \@href[1]{\@@startlink{#1}\@@href}%
\providecommand \@@href[1]{\endgroup#1\@@endlink}%
\providecommand \@sanitize@url [0]{\catcode `\\12\catcode `\$12\catcode
  `\&12\catcode `\#12\catcode `\^12\catcode `\_12\catcode `\%12\relax}%
\providecommand \@@startlink[1]{}%
\providecommand \@@endlink[0]{}%
\providecommand \url  [0]{\begingroup\@sanitize@url \@url }%
\providecommand \@url [1]{\endgroup\@href {#1}{\urlprefix }}%
\providecommand \urlprefix  [0]{URL }%
\providecommand \Eprint [0]{\href }%
\providecommand \doibase [0]{http://dx.doi.org/}%
\providecommand \selectlanguage [0]{\@gobble}%
\providecommand \bibinfo  [0]{\@secondoftwo}%
\providecommand \bibfield  [0]{\@secondoftwo}%
\providecommand \translation [1]{[#1]}%
\providecommand \BibitemOpen [0]{}%
\providecommand \bibitemStop [0]{}%
\providecommand \bibitemNoStop [0]{.\EOS\space}%
\providecommand \EOS [0]{\spacefactor3000\relax}%
\providecommand \BibitemShut  [1]{\csname bibitem#1\endcsname}%
\let\auto@bib@innerbib\@empty
\bibitem [{\citenamefont {Will}(2014)}]{Will:2014kxa}%
  \BibitemOpen
  \bibfield  {author} {\bibinfo {author} {\bibfnamefont {Clifford~M.}\
  \bibnamefont {Will}},\ }\bibfield  {title} {\enquote {\bibinfo {title} {{The
  Confrontation between General Relativity and Experiment}},}\ }\href {\doibase
  10.12942/lrr-2014-4} {\bibfield  {journal} {\bibinfo  {journal} {Living Rev.
  Rel.}\ }\textbf {\bibinfo {volume} {17}},\ \bibinfo {pages} {4} (\bibinfo
  {year} {2014})},\ \Eprint {http://arxiv.org/abs/1403.7377} {arXiv:1403.7377
  [gr-qc]} \BibitemShut {NoStop}%
\bibitem [{\citenamefont {Wex}(2014)}]{Wex:2014nva}%
  \BibitemOpen
  \bibfield  {author} {\bibinfo {author} {\bibfnamefont {Norbert}\ \bibnamefont
  {Wex}},\ }\bibfield  {title} {\enquote {\bibinfo {title} {{Testing
  Relativistic Gravity with Radio Pulsars}},}\ }\href@noop {} {\  (\bibinfo
  {year} {2014})},\ \Eprint {http://arxiv.org/abs/1402.5594} {arXiv:1402.5594
  [gr-qc]} \BibitemShut {NoStop}%
\bibitem [{\citenamefont {Aasi}\ \emph {et~al.}(2015)\citenamefont {Aasi} \emph
  {et~al.}}]{LIGOScientific:2014pky}%
  \BibitemOpen
  \bibfield  {author} {\bibinfo {author} {\bibfnamefont {J.}~\bibnamefont
  {Aasi}} \emph {et~al.} (\bibinfo {collaboration} {LIGO Scientific}),\
  }\bibfield  {title} {\enquote {\bibinfo {title} {{Advanced LIGO}},}\ }\href
  {\doibase 10.1088/0264-9381/32/7/074001} {\bibfield  {journal} {\bibinfo
  {journal} {Class. Quant. Grav.}\ }\textbf {\bibinfo {volume} {32}},\ \bibinfo
  {pages} {074001} (\bibinfo {year} {2015})},\ \Eprint
  {http://arxiv.org/abs/1411.4547} {arXiv:1411.4547 [gr-qc]} \BibitemShut
  {NoStop}%
\bibitem [{\citenamefont {Acernese}\ \emph {et~al.}(2015)\citenamefont
  {Acernese} \emph {et~al.}}]{VIRGO:2014yos}%
  \BibitemOpen
  \bibfield  {author} {\bibinfo {author} {\bibfnamefont {F.}~\bibnamefont
  {Acernese}} \emph {et~al.} (\bibinfo {collaboration} {VIRGO}),\ }\bibfield
  {title} {\enquote {\bibinfo {title} {{Advanced Virgo: a second-generation
  interferometric gravitational wave detector}},}\ }\href {\doibase
  10.1088/0264-9381/32/2/024001} {\bibfield  {journal} {\bibinfo  {journal}
  {Class. Quant. Grav.}\ }\textbf {\bibinfo {volume} {32}},\ \bibinfo {pages}
  {024001} (\bibinfo {year} {2015})},\ \Eprint {http://arxiv.org/abs/1408.3978}
  {arXiv:1408.3978 [gr-qc]} \BibitemShut {NoStop}%
\bibitem [{\citenamefont {Abbott}\ \emph
  {et~al.}(2019{\natexlab{a}})\citenamefont {Abbott} \emph
  {et~al.}}]{LIGOScientific:2018mvr}%
  \BibitemOpen
  \bibfield  {author} {\bibinfo {author} {\bibfnamefont {B.~P.}\ \bibnamefont
  {Abbott}} \emph {et~al.} (\bibinfo {collaboration} {LIGO Scientific,
  Virgo}),\ }\bibfield  {title} {\enquote {\bibinfo {title} {{GWTC-1: A
  Gravitational-Wave Transient Catalog of Compact Binary Mergers Observed by
  LIGO and Virgo during the First and Second Observing Runs}},}\ }\href
  {\doibase 10.1103/PhysRevX.9.031040} {\bibfield  {journal} {\bibinfo
  {journal} {Phys. Rev. X}\ }\textbf {\bibinfo {volume} {9}},\ \bibinfo {pages}
  {031040} (\bibinfo {year} {2019}{\natexlab{a}})},\ \Eprint
  {http://arxiv.org/abs/1811.12907} {arXiv:1811.12907 [astro-ph.HE]}
  \BibitemShut {NoStop}%
\bibitem [{\citenamefont {Abbott}\ \emph
  {et~al.}(2021{\natexlab{a}})\citenamefont {Abbott} \emph
  {et~al.}}]{LIGOScientific:2020ibl}%
  \BibitemOpen
  \bibfield  {author} {\bibinfo {author} {\bibfnamefont {R.}~\bibnamefont
  {Abbott}} \emph {et~al.} (\bibinfo {collaboration} {LIGO Scientific,
  Virgo}),\ }\bibfield  {title} {\enquote {\bibinfo {title} {{GWTC-2: Compact
  Binary Coalescences Observed by LIGO and Virgo During the First Half of the
  Third Observing Run}},}\ }\href {\doibase 10.1103/PhysRevX.11.021053}
  {\bibfield  {journal} {\bibinfo  {journal} {Phys. Rev. X}\ }\textbf {\bibinfo
  {volume} {11}},\ \bibinfo {pages} {021053} (\bibinfo {year}
  {2021}{\natexlab{a}})},\ \Eprint {http://arxiv.org/abs/2010.14527}
  {arXiv:2010.14527 [gr-qc]} \BibitemShut {NoStop}%
\bibitem [{\citenamefont {Abbott}\ \emph
  {et~al.}(2021{\natexlab{b}})\citenamefont {Abbott} \emph
  {et~al.}}]{LIGOScientific:2021usb}%
  \BibitemOpen
  \bibfield  {author} {\bibinfo {author} {\bibfnamefont {R.}~\bibnamefont
  {Abbott}} \emph {et~al.} (\bibinfo {collaboration} {LIGO Scientific,
  VIRGO}),\ }\bibfield  {title} {\enquote {\bibinfo {title} {{GWTC-2.1: Deep
  Extended Catalog of Compact Binary Coalescences Observed by LIGO and Virgo
  During the First Half of the Third Observing Run}},}\ }\href@noop {} {\
  (\bibinfo {year} {2021}{\natexlab{b}})},\ \Eprint
  {http://arxiv.org/abs/2108.01045} {arXiv:2108.01045 [gr-qc]} \BibitemShut
  {NoStop}%
\bibitem [{\citenamefont {Abbott}\ \emph
  {et~al.}(2021{\natexlab{c}})\citenamefont {Abbott} \emph
  {et~al.}}]{LIGOScientific:2021djp}%
  \BibitemOpen
  \bibfield  {author} {\bibinfo {author} {\bibfnamefont {R.}~\bibnamefont
  {Abbott}} \emph {et~al.} (\bibinfo {collaboration} {LIGO Scientific, VIRGO,
  KAGRA}),\ }\bibfield  {title} {\enquote {\bibinfo {title} {{GWTC-3: Compact
  Binary Coalescences Observed by LIGO and Virgo During the Second Part of the
  Third Observing Run}},}\ }\href@noop {} {\  (\bibinfo {year}
  {2021}{\natexlab{c}})},\ \Eprint {http://arxiv.org/abs/2111.03606}
  {arXiv:2111.03606 [gr-qc]} \BibitemShut {NoStop}%
\bibitem [{\citenamefont {Abbott}\ \emph
  {et~al.}(2016{\natexlab{a}})\citenamefont {Abbott} \emph
  {et~al.}}]{GW150914-TGR-LVC}%
  \BibitemOpen
  \bibfield  {author} {\bibinfo {author} {\bibfnamefont {B.~P.}\ \bibnamefont
  {Abbott}} \emph {et~al.} (\bibinfo {collaboration} {LIGO Scientific,
  Virgo}),\ }\bibfield  {title} {\enquote {\bibinfo {title} {{Tests of general
  relativity with GW150914}},}\ }\href {\doibase
  10.1103/PhysRevLett.116.221101} {\bibfield  {journal} {\bibinfo  {journal}
  {Phys. Rev. Lett.}\ }\textbf {\bibinfo {volume} {116}},\ \bibinfo {pages}
  {221101} (\bibinfo {year} {2016}{\natexlab{a}})},\ \bibinfo {note} {[Erratum:
  Phys.Rev.Lett. 121, 129902 (2018)]},\ \Eprint
  {http://arxiv.org/abs/1602.03841} {arXiv:1602.03841 [gr-qc]} \BibitemShut
  {NoStop}%
\bibitem [{\citenamefont {Abbott}\ \emph
  {et~al.}(2019{\natexlab{b}})\citenamefont {Abbott} \emph
  {et~al.}}]{GW170817_TGR_LVC}%
  \BibitemOpen
  \bibfield  {author} {\bibinfo {author} {\bibfnamefont {B.~P.}\ \bibnamefont
  {Abbott}} \emph {et~al.} (\bibinfo {collaboration} {LIGO Scientific,
  Virgo}),\ }\bibfield  {title} {\enquote {\bibinfo {title} {{Tests of General
  Relativity with GW170817}},}\ }\href {\doibase
  10.1103/PhysRevLett.123.011102} {\bibfield  {journal} {\bibinfo  {journal}
  {Phys. Rev. Lett.}\ }\textbf {\bibinfo {volume} {123}},\ \bibinfo {pages}
  {011102} (\bibinfo {year} {2019}{\natexlab{b}})},\ \Eprint
  {http://arxiv.org/abs/1811.00364} {arXiv:1811.00364 [gr-qc]} \BibitemShut
  {NoStop}%
\bibitem [{\citenamefont {Abbott}\ \emph
  {et~al.}(2019{\natexlab{c}})\citenamefont {Abbott} \emph
  {et~al.}}]{GWTC-1-TGR-LVC}%
  \BibitemOpen
  \bibfield  {author} {\bibinfo {author} {\bibfnamefont {B.~P.}\ \bibnamefont
  {Abbott}} \emph {et~al.} (\bibinfo {collaboration} {LIGO Scientific,
  Virgo}),\ }\bibfield  {title} {\enquote {\bibinfo {title} {{Tests of General
  Relativity with the Binary Black Hole Signals from the LIGO-Virgo Catalog
  GWTC-1}},}\ }\href {\doibase 10.1103/PhysRevD.100.104036} {\bibfield
  {journal} {\bibinfo  {journal} {Phys. Rev. D}\ }\textbf {\bibinfo {volume}
  {100}},\ \bibinfo {pages} {104036} (\bibinfo {year} {2019}{\natexlab{c}})},\
  \Eprint {http://arxiv.org/abs/1903.04467} {arXiv:1903.04467 [gr-qc]}
  \BibitemShut {NoStop}%
\bibitem [{\citenamefont {Abbott}\ \emph
  {et~al.}(2021{\natexlab{d}})\citenamefont {Abbott} \emph
  {et~al.}}]{GWTC-2-TGR-LVC}%
  \BibitemOpen
  \bibfield  {author} {\bibinfo {author} {\bibfnamefont {R.}~\bibnamefont
  {Abbott}} \emph {et~al.} (\bibinfo {collaboration} {LIGO Scientific,
  Virgo}),\ }\bibfield  {title} {\enquote {\bibinfo {title} {{Tests of general
  relativity with binary black holes from the second LIGO-Virgo
  gravitational-wave transient catalog}},}\ }\href {\doibase
  10.1103/PhysRevD.103.122002} {\bibfield  {journal} {\bibinfo  {journal}
  {Phys. Rev. D}\ }\textbf {\bibinfo {volume} {103}},\ \bibinfo {pages}
  {122002} (\bibinfo {year} {2021}{\natexlab{d}})},\ \Eprint
  {http://arxiv.org/abs/2010.14529} {arXiv:2010.14529 [gr-qc]} \BibitemShut
  {NoStop}%
\bibitem [{\citenamefont {Abbott}\ \emph
  {et~al.}(2021{\natexlab{e}})\citenamefont {Abbott} \emph
  {et~al.}}]{GWTC-3-TGR-LVK}%
  \BibitemOpen
  \bibfield  {author} {\bibinfo {author} {\bibfnamefont {R.}~\bibnamefont
  {Abbott}} \emph {et~al.} (\bibinfo {collaboration} {LIGO Scientific, VIRGO,
  KAGRA}),\ }\bibfield  {title} {\enquote {\bibinfo {title} {{Tests of General
  Relativity with GWTC-3}},}\ }\href@noop {} {\  (\bibinfo {year}
  {2021}{\natexlab{e}})},\ \Eprint {http://arxiv.org/abs/2112.06861}
  {arXiv:2112.06861 [gr-qc]} \BibitemShut {NoStop}%
\bibitem [{\citenamefont {Collett}\ and\ \citenamefont
  {Bacon}(2017)}]{Collett:2016dey}%
  \BibitemOpen
  \bibfield  {author} {\bibinfo {author} {\bibfnamefont {Thomas~E.}\
  \bibnamefont {Collett}}\ and\ \bibinfo {author} {\bibfnamefont {David}\
  \bibnamefont {Bacon}},\ }\bibfield  {title} {\enquote {\bibinfo {title}
  {{Testing the speed of gravitational waves over cosmological distances with
  strong gravitational lensing}},}\ }\href {\doibase
  10.1103/PhysRevLett.118.091101} {\bibfield  {journal} {\bibinfo  {journal}
  {Phys. Rev. Lett.}\ }\textbf {\bibinfo {volume} {118}},\ \bibinfo {pages}
  {091101} (\bibinfo {year} {2017})},\ \Eprint
  {http://arxiv.org/abs/1602.05882} {arXiv:1602.05882 [astro-ph.HE]}
  \BibitemShut {NoStop}%
\bibitem [{\citenamefont {Fan}\ \emph {et~al.}(2017)\citenamefont {Fan},
  \citenamefont {Liao}, \citenamefont {Biesiada}, \citenamefont
  {Piorkowska-Kurpas},\ and\ \citenamefont {Zhu}}]{Fan:2016swi}%
  \BibitemOpen
  \bibfield  {author} {\bibinfo {author} {\bibfnamefont {Xi-Long}\ \bibnamefont
  {Fan}}, \bibinfo {author} {\bibfnamefont {Kai}\ \bibnamefont {Liao}},
  \bibinfo {author} {\bibfnamefont {Marek}\ \bibnamefont {Biesiada}}, \bibinfo
  {author} {\bibfnamefont {Aleksandra}\ \bibnamefont {Piorkowska-Kurpas}}, \
  and\ \bibinfo {author} {\bibfnamefont {Zong-Hong}\ \bibnamefont {Zhu}},\
  }\bibfield  {title} {\enquote {\bibinfo {title} {{Speed of Gravitational
  Waves from Strongly Lensed Gravitational Waves and Electromagnetic
  Signals}},}\ }\href {\doibase 10.1103/PhysRevLett.118.091102} {\bibfield
  {journal} {\bibinfo  {journal} {Phys. Rev. Lett.}\ }\textbf {\bibinfo
  {volume} {118}},\ \bibinfo {pages} {091102} (\bibinfo {year} {2017})},\
  \Eprint {http://arxiv.org/abs/1612.04095} {arXiv:1612.04095 [gr-qc]}
  \BibitemShut {NoStop}%
\bibitem [{\citenamefont {Yang}\ \emph {et~al.}(2019)\citenamefont {Yang},
  \citenamefont {Hu}, \citenamefont {Cai},\ and\ \citenamefont
  {Wang}}]{Yang:2018bdf}%
  \BibitemOpen
  \bibfield  {author} {\bibinfo {author} {\bibfnamefont {Tao}\ \bibnamefont
  {Yang}}, \bibinfo {author} {\bibfnamefont {Bin}\ \bibnamefont {Hu}}, \bibinfo
  {author} {\bibfnamefont {Rong-Gen}\ \bibnamefont {Cai}}, \ and\ \bibinfo
  {author} {\bibfnamefont {Bin}\ \bibnamefont {Wang}},\ }\bibfield  {title}
  {\enquote {\bibinfo {title} {{New probe of gravity: strongly lensed
  gravitational wave multi-messenger approach}},}\ }\href {\doibase
  10.3847/1538-4357/ab271e} {\bibfield  {journal} {\bibinfo  {journal}
  {Astrophys. J.}\ }\textbf {\bibinfo {volume} {880}},\ \bibinfo {pages} {50}
  (\bibinfo {year} {2019})},\ \Eprint {http://arxiv.org/abs/1810.00164}
  {arXiv:1810.00164 [astro-ph.CO]} \BibitemShut {NoStop}%
\bibitem [{\citenamefont {Chung}\ and\ \citenamefont
  {Li}(2021)}]{Chung:2021rcu}%
  \BibitemOpen
  \bibfield  {author} {\bibinfo {author} {\bibfnamefont {Adrian Ka-Wai}\
  \bibnamefont {Chung}}\ and\ \bibinfo {author} {\bibfnamefont {Tjonnie
  Guang~Feng}\ \bibnamefont {Li}},\ }\bibfield  {title} {\enquote {\bibinfo
  {title} {{Lensing of gravitational waves as a novel probe of graviton
  mass}},}\ }\href {\doibase 10.1103/PhysRevD.104.124060} {\bibfield  {journal}
  {\bibinfo  {journal} {Phys. Rev. D}\ }\textbf {\bibinfo {volume} {104}},\
  \bibinfo {pages} {124060} (\bibinfo {year} {2021})},\ \Eprint
  {http://arxiv.org/abs/2106.09630} {arXiv:2106.09630 [gr-qc]} \BibitemShut
  {NoStop}%
\bibitem [{\citenamefont {Mukherjee}\ \emph
  {et~al.}(2020{\natexlab{a}})\citenamefont {Mukherjee}, \citenamefont
  {Wandelt},\ and\ \citenamefont {Silk}}]{Mukherjee:2019wcg}%
  \BibitemOpen
  \bibfield  {author} {\bibinfo {author} {\bibfnamefont {Suvodip}\ \bibnamefont
  {Mukherjee}}, \bibinfo {author} {\bibfnamefont {Benjamin~D.}\ \bibnamefont
  {Wandelt}}, \ and\ \bibinfo {author} {\bibfnamefont {Joseph}\ \bibnamefont
  {Silk}},\ }\bibfield  {title} {\enquote {\bibinfo {title} {{Probing the
  theory of gravity with gravitational lensing of gravitational waves and
  galaxy surveys}},}\ }\href {\doibase 10.1093/mnras/staa827} {\bibfield
  {journal} {\bibinfo  {journal} {Mon. Not. Roy. Astron. Soc.}\ }\textbf
  {\bibinfo {volume} {494}},\ \bibinfo {pages} {1956--1970} (\bibinfo {year}
  {2020}{\natexlab{a}})},\ \Eprint {http://arxiv.org/abs/1908.08951}
  {arXiv:1908.08951 [astro-ph.CO]} \BibitemShut {NoStop}%
\bibitem [{\citenamefont {Mukherjee}\ \emph
  {et~al.}(2020{\natexlab{b}})\citenamefont {Mukherjee}, \citenamefont
  {Wandelt},\ and\ \citenamefont {Silk}}]{Mukherjee:2019wfw}%
  \BibitemOpen
  \bibfield  {author} {\bibinfo {author} {\bibfnamefont {Suvodip}\ \bibnamefont
  {Mukherjee}}, \bibinfo {author} {\bibfnamefont {Benjamin~D.}\ \bibnamefont
  {Wandelt}}, \ and\ \bibinfo {author} {\bibfnamefont {Joseph}\ \bibnamefont
  {Silk}},\ }\bibfield  {title} {\enquote {\bibinfo {title} {{Multimessenger
  tests of gravity with weakly lensed gravitational waves}},}\ }\href {\doibase
  10.1103/PhysRevD.101.103509} {\bibfield  {journal} {\bibinfo  {journal}
  {Phys. Rev. D}\ }\textbf {\bibinfo {volume} {101}},\ \bibinfo {pages}
  {103509} (\bibinfo {year} {2020}{\natexlab{b}})},\ \Eprint
  {http://arxiv.org/abs/1908.08950} {arXiv:1908.08950 [astro-ph.CO]}
  \BibitemShut {NoStop}%
\bibitem [{\citenamefont {Finke}\ \emph {et~al.}(2021)\citenamefont {Finke},
  \citenamefont {Foffa}, \citenamefont {Iacovelli}, \citenamefont {Maggiore},\
  and\ \citenamefont {Mancarella}}]{Finke:2021znb}%
  \BibitemOpen
  \bibfield  {author} {\bibinfo {author} {\bibfnamefont {Andreas}\ \bibnamefont
  {Finke}}, \bibinfo {author} {\bibfnamefont {Stefano}\ \bibnamefont {Foffa}},
  \bibinfo {author} {\bibfnamefont {Francesco}\ \bibnamefont {Iacovelli}},
  \bibinfo {author} {\bibfnamefont {Michele}\ \bibnamefont {Maggiore}}, \ and\
  \bibinfo {author} {\bibfnamefont {Michele}\ \bibnamefont {Mancarella}},\
  }\bibfield  {title} {\enquote {\bibinfo {title} {{Probing modified
  gravitational wave propagation with strongly lensed coalescing binaries}},}\
  }\href {\doibase 10.1103/PhysRevD.104.084057} {\bibfield  {journal} {\bibinfo
   {journal} {Phys. Rev. D}\ }\textbf {\bibinfo {volume} {104}},\ \bibinfo
  {pages} {084057} (\bibinfo {year} {2021})},\ \Eprint
  {http://arxiv.org/abs/2107.05046} {arXiv:2107.05046 [gr-qc]} \BibitemShut
  {NoStop}%
\bibitem [{\citenamefont {Iacovelli}\ \emph {et~al.}(2022)\citenamefont
  {Iacovelli}, \citenamefont {Finke}, \citenamefont {Foffa}, \citenamefont
  {Maggiore},\ and\ \citenamefont {Mancarella}}]{Iacovelli:2022tlw}%
  \BibitemOpen
  \bibfield  {author} {\bibinfo {author} {\bibfnamefont {Francesco}\
  \bibnamefont {Iacovelli}}, \bibinfo {author} {\bibfnamefont {Andreas}\
  \bibnamefont {Finke}}, \bibinfo {author} {\bibfnamefont {Stefano}\
  \bibnamefont {Foffa}}, \bibinfo {author} {\bibfnamefont {Michele}\
  \bibnamefont {Maggiore}}, \ and\ \bibinfo {author} {\bibfnamefont {Michele}\
  \bibnamefont {Mancarella}},\ }\bibfield  {title} {\enquote {\bibinfo {title}
  {{Modified gravitational wave propagation: information from strongly lensed
  binaries and the BNS mass function}},}\ }in\ \href@noop {} {\emph {\bibinfo
  {booktitle} {{56th Rencontres de Moriond on Gravitation}}}}\ (\bibinfo {year}
  {2022})\ \Eprint {http://arxiv.org/abs/2203.09237} {arXiv:2203.09237 [gr-qc]}
  \BibitemShut {NoStop}%
\bibitem [{\citenamefont {Goyal}\ \emph {et~al.}(2021)\citenamefont {Goyal},
  \citenamefont {Haris}, \citenamefont {Mehta},\ and\ \citenamefont
  {Ajith}}]{Goyal:2020bkm}%
  \BibitemOpen
  \bibfield  {author} {\bibinfo {author} {\bibfnamefont {Srashti}\ \bibnamefont
  {Goyal}}, \bibinfo {author} {\bibfnamefont {K.}~\bibnamefont {Haris}},
  \bibinfo {author} {\bibfnamefont {Ajit~Kumar}\ \bibnamefont {Mehta}}, \ and\
  \bibinfo {author} {\bibfnamefont {Parameswaran}\ \bibnamefont {Ajith}},\
  }\bibfield  {title} {\enquote {\bibinfo {title} {{Testing the nature of
  gravitational-wave polarizations using strongly lensed signals}},}\ }\href
  {\doibase 10.1103/PhysRevD.103.024038} {\bibfield  {journal} {\bibinfo
  {journal} {Phys. Rev. D}\ }\textbf {\bibinfo {volume} {103}},\ \bibinfo
  {pages} {024038} (\bibinfo {year} {2021})},\ \Eprint
  {http://arxiv.org/abs/2008.07060} {arXiv:2008.07060 [gr-qc]} \BibitemShut
  {NoStop}%
\bibitem [{\citenamefont {Maga\~na Hernandez}(2022)}]{MaganaHernandez:2022ayv}%
  \BibitemOpen
  \bibfield  {author} {\bibinfo {author} {\bibfnamefont {Ignacio}\ \bibnamefont
  {Maga\~na Hernandez}},\ }\bibfield  {title} {\enquote {\bibinfo {title}
  {{Measuring the polarization content of gravitational waves with strongly
  lensed binary black hole mergers}},}\ }\href@noop {} {\  (\bibinfo {year}
  {2022})},\ \Eprint {http://arxiv.org/abs/2211.01272} {arXiv:2211.01272
  [gr-qc]} \BibitemShut {NoStop}%
\bibitem [{\citenamefont {Goyal}\ \emph {et~al.}(2023)\citenamefont {Goyal},
  \citenamefont {Vijaykumar}, \citenamefont {Ezquiaga},\ and\ \citenamefont
  {Zumalacarregui}}]{Goyal:2023uvm}%
  \BibitemOpen
  \bibfield  {author} {\bibinfo {author} {\bibfnamefont {Srashti}\ \bibnamefont
  {Goyal}}, \bibinfo {author} {\bibfnamefont {Aditya}\ \bibnamefont
  {Vijaykumar}}, \bibinfo {author} {\bibfnamefont {Jose~Maria}\ \bibnamefont
  {Ezquiaga}}, \ and\ \bibinfo {author} {\bibfnamefont {Miguel}\ \bibnamefont
  {Zumalacarregui}},\ }\bibfield  {title} {\enquote {\bibinfo {title} {{Probing
  lens-induced gravitational-wave birefringence as a test of general
  relativity}},}\ }\href {\doibase 10.1103/PhysRevD.108.024052} {\bibfield
  {journal} {\bibinfo  {journal} {Phys. Rev. D}\ }\textbf {\bibinfo {volume}
  {108}},\ \bibinfo {pages} {024052} (\bibinfo {year} {2023})},\ \Eprint
  {http://arxiv.org/abs/2301.04826} {arXiv:2301.04826 [gr-qc]} \BibitemShut
  {NoStop}%
\bibitem [{\citenamefont {Liu}\ \emph {et~al.}(2023)\citenamefont {Liu},
  \citenamefont {Wong}, \citenamefont {Leong}, \citenamefont {More},
  \citenamefont {Hannuksela},\ and\ \citenamefont {Li}}]{Liu:2023ikc}%
  \BibitemOpen
  \bibfield  {author} {\bibinfo {author} {\bibfnamefont {Anna}\ \bibnamefont
  {Liu}}, \bibinfo {author} {\bibfnamefont {Isaac C.~F.}\ \bibnamefont {Wong}},
  \bibinfo {author} {\bibfnamefont {Samson H.~W.}\ \bibnamefont {Leong}},
  \bibinfo {author} {\bibfnamefont {Anupreeta}\ \bibnamefont {More}}, \bibinfo
  {author} {\bibfnamefont {Otto~A.}\ \bibnamefont {Hannuksela}}, \ and\
  \bibinfo {author} {\bibfnamefont {Tjonnie G.~F.}\ \bibnamefont {Li}},\
  }\bibfield  {title} {\enquote {\bibinfo {title} {{Exploring the hidden
  Universe: a novel phenomenological approach for recovering arbitrary
  gravitational-wave millilensing configurations}},}\ }\href {\doibase
  10.1093/mnras/stad1302} {\bibfield  {journal} {\bibinfo  {journal} {Mon. Not.
  Roy. Astron. Soc.}\ }\textbf {\bibinfo {volume} {525}},\ \bibinfo {pages}
  {4149--4160} (\bibinfo {year} {2023})},\ \Eprint
  {http://arxiv.org/abs/2302.09870} {arXiv:2302.09870 [gr-qc]} \BibitemShut
  {NoStop}%
\bibitem [{\citenamefont {Janquart}\ \emph {et~al.}(2023)\citenamefont
  {Janquart} \emph {et~al.}}]{Janquart:2023mvf}%
  \BibitemOpen
  \bibfield  {author} {\bibinfo {author} {\bibfnamefont {Justin}\ \bibnamefont
  {Janquart}} \emph {et~al.},\ }\bibfield  {title} {\enquote {\bibinfo {title}
  {{Follow-up Analyses to the O3 LIGO-Virgo-KAGRA Lensing Searches}},}\ }\href
  {\doibase 10.1093/mnras/stad2909} {\  (\bibinfo {year} {2023}),\
  10.1093/mnras/stad2909},\ \Eprint {http://arxiv.org/abs/2306.03827}
  {arXiv:2306.03827 [gr-qc]} \BibitemShut {NoStop}%
\bibitem [{\citenamefont {{Deguchi}}\ and\ \citenamefont
  {{Watson}}(1986)}]{1986ApJ...307...30D}%
  \BibitemOpen
  \bibfield  {author} {\bibinfo {author} {\bibfnamefont {S.}~\bibnamefont
  {{Deguchi}}}\ and\ \bibinfo {author} {\bibfnamefont {W.~D.}\ \bibnamefont
  {{Watson}}},\ }\bibfield  {title} {\enquote {\bibinfo {title} {{Diffraction
  in Gravitational Lensing for Compact Objects of Low Mass}},}\ }\href
  {\doibase 10.1086/164389} {\bibfield  {journal} {\bibinfo  {journal} {\apj}\
  }\textbf {\bibinfo {volume} {307}},\ \bibinfo {pages} {30} (\bibinfo {year}
  {1986})}\BibitemShut {NoStop}%
\bibitem [{\citenamefont {Nakamura}(1998)}]{Nakamura:1997sw}%
  \BibitemOpen
  \bibfield  {author} {\bibinfo {author} {\bibfnamefont {Takahiro~T.}\
  \bibnamefont {Nakamura}},\ }\bibfield  {title} {\enquote {\bibinfo {title}
  {{Gravitational lensing of gravitational waves from inspiraling binaries by a
  point mass lens}},}\ }\href {\doibase 10.1103/PhysRevLett.80.1138} {\bibfield
   {journal} {\bibinfo  {journal} {Phys. Rev. Lett.}\ }\textbf {\bibinfo
  {volume} {80}},\ \bibinfo {pages} {1138--1141} (\bibinfo {year}
  {1998})}\BibitemShut {NoStop}%
\bibitem [{\citenamefont {Baraldo}\ \emph {et~al.}(1999)\citenamefont
  {Baraldo}, \citenamefont {Hosoya},\ and\ \citenamefont
  {Nakamura}}]{Baraldo:1999ny}%
  \BibitemOpen
  \bibfield  {author} {\bibinfo {author} {\bibfnamefont {Christian}\
  \bibnamefont {Baraldo}}, \bibinfo {author} {\bibfnamefont {Akio}\
  \bibnamefont {Hosoya}}, \ and\ \bibinfo {author} {\bibfnamefont
  {Takahiro~T.}\ \bibnamefont {Nakamura}},\ }\bibfield  {title} {\enquote
  {\bibinfo {title} {{Gravitationally induced interference of gravitational
  waves by a rotating massive object}},}\ }\href {\doibase
  10.1103/PhysRevD.59.083001} {\bibfield  {journal} {\bibinfo  {journal} {Phys.
  Rev. D}\ }\textbf {\bibinfo {volume} {59}},\ \bibinfo {pages} {083001}
  (\bibinfo {year} {1999})}\BibitemShut {NoStop}%
\bibitem [{\citenamefont {Nakamura}\ and\ \citenamefont
  {Deguchi}(1999)}]{Nakamura:1999uwi}%
  \BibitemOpen
  \bibfield  {author} {\bibinfo {author} {\bibfnamefont {Takahiro~T.}\
  \bibnamefont {Nakamura}}\ and\ \bibinfo {author} {\bibfnamefont {Shuji}\
  \bibnamefont {Deguchi}},\ }\bibfield  {title} {\enquote {\bibinfo {title}
  {{Wave Optics in Gravitational Lensing}},}\ }\href {\doibase
  10.1143/ptps.133.137} {\bibfield  {journal} {\bibinfo  {journal} {Prog.
  Theor. Phys. Suppl.}\ }\textbf {\bibinfo {volume} {133}},\ \bibinfo {pages}
  {137--153} (\bibinfo {year} {1999})}\BibitemShut {NoStop}%
\bibitem [{\citenamefont {Bulashenko}\ and\ \citenamefont
  {Ubach}(2022)}]{Bulashenko:2021fes}%
  \BibitemOpen
  \bibfield  {author} {\bibinfo {author} {\bibfnamefont {Oleg}\ \bibnamefont
  {Bulashenko}}\ and\ \bibinfo {author} {\bibfnamefont {Helena}\ \bibnamefont
  {Ubach}},\ }\bibfield  {title} {\enquote {\bibinfo {title} {{Lensing of
  gravitational waves: universal signatures in the beating pattern}},}\ }\href
  {\doibase 10.1088/1475-7516/2022/07/022} {\bibfield  {journal} {\bibinfo
  {journal} {JCAP}\ }\textbf {\bibinfo {volume} {07}},\ \bibinfo {pages} {022}
  (\bibinfo {year} {2022})},\ \Eprint {http://arxiv.org/abs/2112.10773}
  {arXiv:2112.10773 [gr-qc]} \BibitemShut {NoStop}%
\bibitem [{\citenamefont {Leung}\ \emph {et~al.}(2023)\citenamefont {Leung},
  \citenamefont {Jow}, \citenamefont {Saha}, \citenamefont {Dai}, \citenamefont
  {Oguri},\ and\ \citenamefont {Koopmans}}]{Leung:2023lmq}%
  \BibitemOpen
  \bibfield  {author} {\bibinfo {author} {\bibfnamefont {Calvin}\ \bibnamefont
  {Leung}}, \bibinfo {author} {\bibfnamefont {Dylan}\ \bibnamefont {Jow}},
  \bibinfo {author} {\bibfnamefont {Prasenjit}\ \bibnamefont {Saha}}, \bibinfo
  {author} {\bibfnamefont {Liang}\ \bibnamefont {Dai}}, \bibinfo {author}
  {\bibfnamefont {Masamune}\ \bibnamefont {Oguri}}, \ and\ \bibinfo {author}
  {\bibfnamefont {L\'eon V.~E.}\ \bibnamefont {Koopmans}},\ }\bibfield  {title}
  {\enquote {\bibinfo {title} {{Wave Mechanics, Interference, and Decoherence
  in Strong Gravitational Lensing}},}\ }\href@noop {} {\  (\bibinfo {year}
  {2023})},\ \Eprint {http://arxiv.org/abs/2304.01202} {arXiv:2304.01202
  [astro-ph.HE]} \BibitemShut {NoStop}%
\bibitem [{\citenamefont {Mishra}\ \emph {et~al.}(2023)\citenamefont {Mishra},
  \citenamefont {Meena}, \citenamefont {More},\ and\ \citenamefont
  {Bose}}]{mishra2023exploring}%
  \BibitemOpen
  \bibfield  {author} {\bibinfo {author} {\bibfnamefont {Anuj}\ \bibnamefont
  {Mishra}}, \bibinfo {author} {\bibfnamefont {Ashish~Kumar}\ \bibnamefont
  {Meena}}, \bibinfo {author} {\bibfnamefont {Anupreeta}\ \bibnamefont {More}},
  \ and\ \bibinfo {author} {\bibfnamefont {Sukanta}\ \bibnamefont {Bose}},\
  }\bibfield  {title} {\enquote {\bibinfo {title} {Exploring the impact of
  microlensing on gravitational wave signals: Biases, population
  characteristics, and prospects for detection},}\ }\href@noop {} {\bibfield
  {journal} {\bibinfo  {journal} {arXiv preprint arXiv:2306.11479}\ } (\bibinfo
  {year} {2023})}\BibitemShut {NoStop}%
\bibitem [{\citenamefont {Abbott}\ \emph
  {et~al.}(2021{\natexlab{f}})\citenamefont {Abbott} \emph
  {et~al.}}]{LIGOScientific:2021izm}%
  \BibitemOpen
  \bibfield  {author} {\bibinfo {author} {\bibfnamefont {R.}~\bibnamefont
  {Abbott}} \emph {et~al.} (\bibinfo {collaboration} {LIGO Scientific,
  VIRGO}),\ }\bibfield  {title} {\enquote {\bibinfo {title} {{Search for
  Lensing Signatures in the Gravitational-Wave Observations from the First Half
  of LIGO\textendash{}Virgo\textquoteright{}s Third Observing Run}},}\ }\href
  {\doibase 10.3847/1538-4357/ac23db} {\bibfield  {journal} {\bibinfo
  {journal} {Astrophys. J.}\ }\textbf {\bibinfo {volume} {923}},\ \bibinfo
  {pages} {14} (\bibinfo {year} {2021}{\natexlab{f}})},\ \Eprint
  {http://arxiv.org/abs/2105.06384} {arXiv:2105.06384 [gr-qc]} \BibitemShut
  {NoStop}%
\bibitem [{\citenamefont {Abbott}\ \emph {et~al.}(2023)\citenamefont {Abbott}
  \emph {et~al.}}]{LIGOScientific:2023bwz}%
  \BibitemOpen
  \bibfield  {author} {\bibinfo {author} {\bibfnamefont {R.}~\bibnamefont
  {Abbott}} \emph {et~al.} (\bibinfo {collaboration} {LIGO Scientific, VIRGO,
  KAGRA}),\ }\bibfield  {title} {\enquote {\bibinfo {title} {{Search for
  gravitational-lensing signatures in the full third observing run of the
  LIGO-Virgo network}},}\ }\href@noop {} {\  (\bibinfo {year} {2023})},\
  \Eprint {http://arxiv.org/abs/2304.08393} {arXiv:2304.08393 [gr-qc]}
  \BibitemShut {NoStop}%
\bibitem [{\citenamefont {Jung}\ and\ \citenamefont
  {Shin}(2019)}]{Jung:2017flg}%
  \BibitemOpen
  \bibfield  {author} {\bibinfo {author} {\bibfnamefont {Sunghoon}\
  \bibnamefont {Jung}}\ and\ \bibinfo {author} {\bibfnamefont {Chang~Sub}\
  \bibnamefont {Shin}},\ }\bibfield  {title} {\enquote {\bibinfo {title}
  {{Gravitational-Wave Fringes at LIGO: Detecting Compact Dark Matter by
  Gravitational Lensing}},}\ }\href {\doibase 10.1103/PhysRevLett.122.041103}
  {\bibfield  {journal} {\bibinfo  {journal} {Phys. Rev. Lett.}\ }\textbf
  {\bibinfo {volume} {122}},\ \bibinfo {pages} {041103} (\bibinfo {year}
  {2019})},\ \Eprint {http://arxiv.org/abs/1712.01396} {arXiv:1712.01396
  [astro-ph.CO]} \BibitemShut {NoStop}%
\bibitem [{\citenamefont {Basak}\ \emph {et~al.}(2022)\citenamefont {Basak},
  \citenamefont {Ganguly}, \citenamefont {Haris}, \citenamefont {Kapadia},
  \citenamefont {Mehta},\ and\ \citenamefont {Ajith}}]{Basak:2021ten}%
  \BibitemOpen
  \bibfield  {author} {\bibinfo {author} {\bibfnamefont {S.}~\bibnamefont
  {Basak}}, \bibinfo {author} {\bibfnamefont {A.}~\bibnamefont {Ganguly}},
  \bibinfo {author} {\bibfnamefont {K.}~\bibnamefont {Haris}}, \bibinfo
  {author} {\bibfnamefont {S.}~\bibnamefont {Kapadia}}, \bibinfo {author}
  {\bibfnamefont {A.~K.}\ \bibnamefont {Mehta}}, \ and\ \bibinfo {author}
  {\bibfnamefont {P.}~\bibnamefont {Ajith}},\ }\bibfield  {title} {\enquote
  {\bibinfo {title} {{Constraints on Compact Dark Matter from Gravitational
  Wave Microlensing}},}\ }\href {\doibase 10.3847/2041-8213/ac4dfa} {\bibfield
  {journal} {\bibinfo  {journal} {Astrophys. J.}\ }\textbf {\bibinfo {volume}
  {926}},\ \bibinfo {pages} {L28} (\bibinfo {year} {2022})},\ \Eprint
  {http://arxiv.org/abs/2109.06456} {arXiv:2109.06456 [gr-qc]} \BibitemShut
  {NoStop}%
\bibitem [{\citenamefont {Diego}(2020)}]{Diego:2019rzc}%
  \BibitemOpen
  \bibfield  {author} {\bibinfo {author} {\bibfnamefont {Jose~M.}\ \bibnamefont
  {Diego}},\ }\bibfield  {title} {\enquote {\bibinfo {title} {{Constraining the
  abundance of primordial black holes with gravitational lensing of
  gravitational waves at LIGO frequencies}},}\ }\href {\doibase
  10.1103/PhysRevD.101.123512} {\bibfield  {journal} {\bibinfo  {journal}
  {Phys. Rev. D}\ }\textbf {\bibinfo {volume} {101}},\ \bibinfo {pages}
  {123512} (\bibinfo {year} {2020})},\ \Eprint
  {http://arxiv.org/abs/1911.05736} {arXiv:1911.05736 [astro-ph.CO]}
  \BibitemShut {NoStop}%
\bibitem [{\citenamefont {Diego}\ \emph {et~al.}(2019)\citenamefont {Diego},
  \citenamefont {Hannuksela}, \citenamefont {Kelly}, \citenamefont
  {Broadhurst}, \citenamefont {Kim}, \citenamefont {Li}, \citenamefont
  {Smoot},\ and\ \citenamefont {Pagano}}]{Diego:2019lcd}%
  \BibitemOpen
  \bibfield  {author} {\bibinfo {author} {\bibfnamefont {J.~M.}\ \bibnamefont
  {Diego}}, \bibinfo {author} {\bibfnamefont {O.~A.}\ \bibnamefont
  {Hannuksela}}, \bibinfo {author} {\bibfnamefont {P.~L.}\ \bibnamefont
  {Kelly}}, \bibinfo {author} {\bibfnamefont {T.}~\bibnamefont {Broadhurst}},
  \bibinfo {author} {\bibfnamefont {K.}~\bibnamefont {Kim}}, \bibinfo {author}
  {\bibfnamefont {T.~G.~F.}\ \bibnamefont {Li}}, \bibinfo {author}
  {\bibfnamefont {G.~F.}\ \bibnamefont {Smoot}}, \ and\ \bibinfo {author}
  {\bibfnamefont {G.}~\bibnamefont {Pagano}},\ }\bibfield  {title} {\enquote
  {\bibinfo {title} {{Observational signatures of microlensing in gravitational
  waves at LIGO/Virgo frequencies}},}\ }\href {\doibase
  10.1051/0004-6361/201935490} {\bibfield  {journal} {\bibinfo  {journal}
  {Astron. Astrophys.}\ }\textbf {\bibinfo {volume} {627}},\ \bibinfo {pages}
  {A130} (\bibinfo {year} {2019})},\ \Eprint {http://arxiv.org/abs/1903.04513}
  {arXiv:1903.04513 [astro-ph.CO]} \BibitemShut {NoStop}%
\bibitem [{\citenamefont {Seo}\ \emph {et~al.}(2021)\citenamefont {Seo},
  \citenamefont {Hannuksela},\ and\ \citenamefont {Li}}]{Seo:2021ucd}%
  \BibitemOpen
  \bibfield  {author} {\bibinfo {author} {\bibfnamefont {Eungwang}\
  \bibnamefont {Seo}}, \bibinfo {author} {\bibfnamefont {Otto~A.}\ \bibnamefont
  {Hannuksela}}, \ and\ \bibinfo {author} {\bibfnamefont {Tjonnie G.~F.}\
  \bibnamefont {Li}},\ }\bibfield  {title} {\enquote {\bibinfo {title} {{Strong
  lensing: A magnifying glass to detect gravitational-wave microlensing}},}\
  }in\ \href@noop {} {\emph {\bibinfo {booktitle} {{17th International
  Conference on Topics in Astroparticle and Underground Physics}}}}\ (\bibinfo
  {year} {2021})\ \Eprint {http://arxiv.org/abs/2110.03308} {arXiv:2110.03308
  [astro-ph.HE]} \BibitemShut {NoStop}%
\bibitem [{\citenamefont {Mishra}\ \emph {et~al.}(2021)\citenamefont {Mishra},
  \citenamefont {Meena}, \citenamefont {More}, \citenamefont {Bose},\ and\
  \citenamefont {Bagla}}]{Mishra:2021xzz}%
  \BibitemOpen
  \bibfield  {author} {\bibinfo {author} {\bibfnamefont {Anuj}\ \bibnamefont
  {Mishra}}, \bibinfo {author} {\bibfnamefont {Ashish~Kumar}\ \bibnamefont
  {Meena}}, \bibinfo {author} {\bibfnamefont {Anupreeta}\ \bibnamefont {More}},
  \bibinfo {author} {\bibfnamefont {Sukanta}\ \bibnamefont {Bose}}, \ and\
  \bibinfo {author} {\bibfnamefont {Jasjeet~Singh}\ \bibnamefont {Bagla}},\
  }\bibfield  {title} {\enquote {\bibinfo {title} {{Gravitational lensing of
  gravitational waves: effect of microlens population in lensing galaxies}},}\
  }\href {\doibase 10.1093/mnras/stab2875} {\bibfield  {journal} {\bibinfo
  {journal} {Mon. Not. Roy. Astron. Soc.}\ }\textbf {\bibinfo {volume} {508}},\
  \bibinfo {pages} {4869--4886} (\bibinfo {year} {2021})},\ \Eprint
  {http://arxiv.org/abs/2102.03946} {arXiv:2102.03946 [astro-ph.CO]}
  \BibitemShut {NoStop}%
\bibitem [{\citenamefont {Meena}\ \emph {et~al.}(2022)\citenamefont {Meena},
  \citenamefont {Mishra}, \citenamefont {More}, \citenamefont {Bose},\ and\
  \citenamefont {Bagla}}]{Meena:2022unp}%
  \BibitemOpen
  \bibfield  {author} {\bibinfo {author} {\bibfnamefont {Ashish~Kumar}\
  \bibnamefont {Meena}}, \bibinfo {author} {\bibfnamefont {Anuj}\ \bibnamefont
  {Mishra}}, \bibinfo {author} {\bibfnamefont {Anupreeta}\ \bibnamefont
  {More}}, \bibinfo {author} {\bibfnamefont {Sukanta}\ \bibnamefont {Bose}}, \
  and\ \bibinfo {author} {\bibfnamefont {Jasjeet~Singh}\ \bibnamefont
  {Bagla}},\ }\bibfield  {title} {\enquote {\bibinfo {title} {{Gravitational
  lensing of gravitational waves: Probability of microlensing in galaxy-scale
  lens population}},}\ }\href {\doibase 10.1093/mnras/stac2721} {\bibfield
  {journal} {\bibinfo  {journal} {Mon. Not. Roy. Astron. Soc.}\ }\textbf
  {\bibinfo {volume} {517}},\ \bibinfo {pages} {872--884} (\bibinfo {year}
  {2022})},\ \Eprint {http://arxiv.org/abs/2205.05409} {arXiv:2205.05409
  [astro-ph.GA]} \BibitemShut {NoStop}%
\bibitem [{\citenamefont {Meena}(2023)}]{Meena:2023qdq}%
  \BibitemOpen
  \bibfield  {author} {\bibinfo {author} {\bibfnamefont {Ashish~Kumar}\
  \bibnamefont {Meena}},\ }\bibfield  {title} {\enquote {\bibinfo {title}
  {{Gravitational Lensing of Gravitational Waves: Probing Intermediate Mass
  Black Holes in Galaxy Lenses with Global Minima}},}\ }\href@noop {} {\
  (\bibinfo {year} {2023})},\ \Eprint {http://arxiv.org/abs/2305.02880}
  {arXiv:2305.02880 [astro-ph.CO]} \BibitemShut {NoStop}%
\bibitem [{\citenamefont {\c{C}al\i{}\c{s}kan}\ \emph
  {et~al.}(2023{\natexlab{a}})\citenamefont {\c{C}al\i{}\c{s}kan},
  \citenamefont {Ji}, \citenamefont {Cotesta}, \citenamefont {Berti},
  \citenamefont {Kamionkowski},\ and\ \citenamefont
  {Marsat}}]{Caliskan:2022hbu}%
  \BibitemOpen
  \bibfield  {author} {\bibinfo {author} {\bibfnamefont {Mesut}\ \bibnamefont
  {\c{C}al\i{}\c{s}kan}}, \bibinfo {author} {\bibfnamefont {Lingyuan}\
  \bibnamefont {Ji}}, \bibinfo {author} {\bibfnamefont {Roberto}\ \bibnamefont
  {Cotesta}}, \bibinfo {author} {\bibfnamefont {Emanuele}\ \bibnamefont
  {Berti}}, \bibinfo {author} {\bibfnamefont {Marc}\ \bibnamefont
  {Kamionkowski}}, \ and\ \bibinfo {author} {\bibfnamefont {Sylvain}\
  \bibnamefont {Marsat}},\ }\bibfield  {title} {\enquote {\bibinfo {title}
  {{Observability of lensing of gravitational waves from massive black hole
  binaries with LISA}},}\ }\href {\doibase 10.1103/PhysRevD.107.043029}
  {\bibfield  {journal} {\bibinfo  {journal} {Phys. Rev. D}\ }\textbf {\bibinfo
  {volume} {107}},\ \bibinfo {pages} {043029} (\bibinfo {year}
  {2023}{\natexlab{a}})},\ \Eprint {http://arxiv.org/abs/2206.02803}
  {arXiv:2206.02803 [astro-ph.CO]} \BibitemShut {NoStop}%
\bibitem [{\citenamefont {\c{C}al\i{}\c{s}kan}\ \emph
  {et~al.}(2023{\natexlab{b}})\citenamefont {\c{C}al\i{}\c{s}kan},
  \citenamefont {Anil~Kumar}, \citenamefont {Ji}, \citenamefont {Ezquiaga},
  \citenamefont {Cotesta}, \citenamefont {Berti},\ and\ \citenamefont
  {Kamionkowski}}]{Caliskan:2023zqm}%
  \BibitemOpen
  \bibfield  {author} {\bibinfo {author} {\bibfnamefont {Mesut}\ \bibnamefont
  {\c{C}al\i{}\c{s}kan}}, \bibinfo {author} {\bibfnamefont {Neha}\ \bibnamefont
  {Anil~Kumar}}, \bibinfo {author} {\bibfnamefont {Lingyuan}\ \bibnamefont
  {Ji}}, \bibinfo {author} {\bibfnamefont {Jose~M.}\ \bibnamefont {Ezquiaga}},
  \bibinfo {author} {\bibfnamefont {Roberto}\ \bibnamefont {Cotesta}}, \bibinfo
  {author} {\bibfnamefont {Emanuele}\ \bibnamefont {Berti}}, \ and\ \bibinfo
  {author} {\bibfnamefont {Marc}\ \bibnamefont {Kamionkowski}},\ }\bibfield
  {title} {\enquote {\bibinfo {title} {{Probing wave-optics effects and
  dark-matter subhalos with lensing of gravitational waves from massive black
  holes}},}\ }\href@noop {} {\  (\bibinfo {year} {2023}{\natexlab{b}})},\
  \Eprint {http://arxiv.org/abs/2307.06990} {arXiv:2307.06990 [astro-ph.CO]}
  \BibitemShut {NoStop}%
\bibitem [{\citenamefont {Ghosh}\ \emph {et~al.}(2016)\citenamefont {Ghosh}
  \emph {et~al.}}]{Ghosh2016}%
  \BibitemOpen
  \bibfield  {author} {\bibinfo {author} {\bibfnamefont {Abhirup}\ \bibnamefont
  {Ghosh}} \emph {et~al.},\ }\bibfield  {title} {\enquote {\bibinfo {title}
  {{Testing general relativity using golden black-hole binaries}},}\ }\href
  {\doibase 10.1103/PhysRevD.94.021101} {\bibfield  {journal} {\bibinfo
  {journal} {Phys. Rev. D}\ }\textbf {\bibinfo {volume} {94}},\ \bibinfo
  {pages} {021101} (\bibinfo {year} {2016})},\ \Eprint
  {http://arxiv.org/abs/1602.02453} {arXiv:1602.02453 [gr-qc]} \BibitemShut
  {NoStop}%
\bibitem [{\citenamefont {Ghosh}\ \emph {et~al.}(2018)\citenamefont {Ghosh},
  \citenamefont {Johnson-Mcdaniel}, \citenamefont {Ghosh}, \citenamefont
  {Mishra}, \citenamefont {Ajith}, \citenamefont {Del~Pozzo}, \citenamefont
  {Berry}, \citenamefont {Nielsen},\ and\ \citenamefont {London}}]{Ghosh2017}%
  \BibitemOpen
  \bibfield  {author} {\bibinfo {author} {\bibfnamefont {Abhirup}\ \bibnamefont
  {Ghosh}}, \bibinfo {author} {\bibfnamefont {Nathan~K.}\ \bibnamefont
  {Johnson-Mcdaniel}}, \bibinfo {author} {\bibfnamefont {Archisman}\
  \bibnamefont {Ghosh}}, \bibinfo {author} {\bibfnamefont {Chandra~Kant}\
  \bibnamefont {Mishra}}, \bibinfo {author} {\bibfnamefont {Parameswaran}\
  \bibnamefont {Ajith}}, \bibinfo {author} {\bibfnamefont {Walter}\
  \bibnamefont {Del~Pozzo}}, \bibinfo {author} {\bibfnamefont {Christopher
  P.~L.}\ \bibnamefont {Berry}}, \bibinfo {author} {\bibfnamefont {Alex~B.}\
  \bibnamefont {Nielsen}}, \ and\ \bibinfo {author} {\bibfnamefont {Lionel}\
  \bibnamefont {London}},\ }\bibfield  {title} {\enquote {\bibinfo {title}
  {{Testing general relativity using gravitational wave signals from the
  inspiral, merger and ringdown of binary black holes}},}\ }\href {\doibase
  10.1088/1361-6382/aa972e} {\bibfield  {journal} {\bibinfo  {journal} {Class.
  Quant. Grav.}\ }\textbf {\bibinfo {volume} {35}},\ \bibinfo {pages} {014002}
  (\bibinfo {year} {2018})},\ \Eprint {http://arxiv.org/abs/1704.06784}
  {arXiv:1704.06784 [gr-qc]} \BibitemShut {NoStop}%
\bibitem [{\citenamefont {Arun}\ \emph
  {et~al.}(2006{\natexlab{a}})\citenamefont {Arun}, \citenamefont {Iyer},
  \citenamefont {Qusailah},\ and\ \citenamefont
  {Sathyaprakash}}]{ParametrizedTests_Arun2006a}%
  \BibitemOpen
  \bibfield  {author} {\bibinfo {author} {\bibfnamefont {K.~G.}\ \bibnamefont
  {Arun}}, \bibinfo {author} {\bibfnamefont {Bala~R.}\ \bibnamefont {Iyer}},
  \bibinfo {author} {\bibfnamefont {M.~S.~S.}\ \bibnamefont {Qusailah}}, \ and\
  \bibinfo {author} {\bibfnamefont {B.~S.}\ \bibnamefont {Sathyaprakash}},\
  }\bibfield  {title} {\enquote {\bibinfo {title} {{Testing post-Newtonian
  theory with gravitational wave observations}},}\ }\href {\doibase
  10.1088/0264-9381/23/9/L01} {\bibfield  {journal} {\bibinfo  {journal}
  {Class. Quant. Grav.}\ }\textbf {\bibinfo {volume} {23}},\ \bibinfo {pages}
  {L37--L43} (\bibinfo {year} {2006}{\natexlab{a}})},\ \Eprint
  {http://arxiv.org/abs/gr-qc/0604018} {arXiv:gr-qc/0604018} \BibitemShut
  {NoStop}%
\bibitem [{\citenamefont {Arun}\ \emph
  {et~al.}(2006{\natexlab{b}})\citenamefont {Arun}, \citenamefont {Iyer},
  \citenamefont {Qusailah},\ and\ \citenamefont
  {Sathyaprakash}}]{ParametrizedTests_Arun2006b}%
  \BibitemOpen
  \bibfield  {author} {\bibinfo {author} {\bibfnamefont {K.~G.}\ \bibnamefont
  {Arun}}, \bibinfo {author} {\bibfnamefont {Bala~R.}\ \bibnamefont {Iyer}},
  \bibinfo {author} {\bibfnamefont {M.~S.~S.}\ \bibnamefont {Qusailah}}, \ and\
  \bibinfo {author} {\bibfnamefont {B.~S.}\ \bibnamefont {Sathyaprakash}},\
  }\bibfield  {title} {\enquote {\bibinfo {title} {{Probing the non-linear
  structure of general relativity with black hole binaries}},}\ }\href
  {\doibase 10.1103/PhysRevD.74.024006} {\bibfield  {journal} {\bibinfo
  {journal} {Phys. Rev. D}\ }\textbf {\bibinfo {volume} {74}},\ \bibinfo
  {pages} {024006} (\bibinfo {year} {2006}{\natexlab{b}})},\ \Eprint
  {http://arxiv.org/abs/gr-qc/0604067} {arXiv:gr-qc/0604067} \BibitemShut
  {NoStop}%
\bibitem [{\citenamefont {Yunes}\ and\ \citenamefont
  {Pretorius}(2009)}]{PPE_Yunes2009}%
  \BibitemOpen
  \bibfield  {author} {\bibinfo {author} {\bibfnamefont {Nicolas}\ \bibnamefont
  {Yunes}}\ and\ \bibinfo {author} {\bibfnamefont {Frans}\ \bibnamefont
  {Pretorius}},\ }\bibfield  {title} {\enquote {\bibinfo {title} {{Fundamental
  Theoretical Bias in Gravitational Wave Astrophysics and the Parameterized
  Post-Einsteinian Framework}},}\ }\href {\doibase 10.1103/PhysRevD.80.122003}
  {\bibfield  {journal} {\bibinfo  {journal} {Phys. Rev. D}\ }\textbf {\bibinfo
  {volume} {80}},\ \bibinfo {pages} {122003} (\bibinfo {year} {2009})},\
  \Eprint {http://arxiv.org/abs/0909.3328} {arXiv:0909.3328 [gr-qc]}
  \BibitemShut {NoStop}%
\bibitem [{\citenamefont {Li}\ \emph {et~al.}(2012)\citenamefont {Li},
  \citenamefont {Del~Pozzo}, \citenamefont {Vitale}, \citenamefont {Van
  Den~Broeck}, \citenamefont {Agathos}, \citenamefont {Veitch}, \citenamefont
  {Grover}, \citenamefont {Sidery}, \citenamefont {Sturani},\ and\
  \citenamefont {Vecchio}}]{TIGER_2011}%
  \BibitemOpen
  \bibfield  {author} {\bibinfo {author} {\bibfnamefont {T.~G.~F.}\
  \bibnamefont {Li}}, \bibinfo {author} {\bibfnamefont {W.}~\bibnamefont
  {Del~Pozzo}}, \bibinfo {author} {\bibfnamefont {S.}~\bibnamefont {Vitale}},
  \bibinfo {author} {\bibfnamefont {C.}~\bibnamefont {Van Den~Broeck}},
  \bibinfo {author} {\bibfnamefont {M.}~\bibnamefont {Agathos}}, \bibinfo
  {author} {\bibfnamefont {J.}~\bibnamefont {Veitch}}, \bibinfo {author}
  {\bibfnamefont {K.}~\bibnamefont {Grover}}, \bibinfo {author} {\bibfnamefont
  {T.}~\bibnamefont {Sidery}}, \bibinfo {author} {\bibfnamefont
  {R.}~\bibnamefont {Sturani}}, \ and\ \bibinfo {author} {\bibfnamefont
  {A.}~\bibnamefont {Vecchio}},\ }\bibfield  {title} {\enquote {\bibinfo
  {title} {{Towards a generic test of the strong field dynamics of general
  relativity using compact binary coalescence}},}\ }\href {\doibase
  10.1103/PhysRevD.85.082003} {\bibfield  {journal} {\bibinfo  {journal} {Phys.
  Rev. D}\ }\textbf {\bibinfo {volume} {85}},\ \bibinfo {pages} {082003}
  (\bibinfo {year} {2012})},\ \Eprint {http://arxiv.org/abs/1110.0530}
  {arXiv:1110.0530 [gr-qc]} \BibitemShut {NoStop}%
\bibitem [{\citenamefont {Yunes}\ \emph {et~al.}(2016)\citenamefont {Yunes},
  \citenamefont {Yagi},\ and\ \citenamefont {Pretorius}}]{Yunes:2016jcc}%
  \BibitemOpen
  \bibfield  {author} {\bibinfo {author} {\bibfnamefont {Nicolas}\ \bibnamefont
  {Yunes}}, \bibinfo {author} {\bibfnamefont {Kent}\ \bibnamefont {Yagi}}, \
  and\ \bibinfo {author} {\bibfnamefont {Frans}\ \bibnamefont {Pretorius}},\
  }\bibfield  {title} {\enquote {\bibinfo {title} {{Theoretical Physics
  Implications of the Binary Black-Hole Mergers GW150914 and GW151226}},}\
  }\href {\doibase 10.1103/PhysRevD.94.084002} {\bibfield  {journal} {\bibinfo
  {journal} {Phys. Rev. D}\ }\textbf {\bibinfo {volume} {94}},\ \bibinfo
  {pages} {084002} (\bibinfo {year} {2016})},\ \Eprint
  {http://arxiv.org/abs/1603.08955} {arXiv:1603.08955 [gr-qc]} \BibitemShut
  {NoStop}%
\bibitem [{\citenamefont {Ezquiaga}\ \emph {et~al.}(2022)\citenamefont
  {Ezquiaga}, \citenamefont {Hu}, \citenamefont {Lagos}, \citenamefont {Lin},\
  and\ \citenamefont {Xu}}]{Ezquiaga:2022nak}%
  \BibitemOpen
  \bibfield  {author} {\bibinfo {author} {\bibfnamefont {Jose~Maria}\
  \bibnamefont {Ezquiaga}}, \bibinfo {author} {\bibfnamefont {Wayne}\
  \bibnamefont {Hu}}, \bibinfo {author} {\bibfnamefont {Macarena}\ \bibnamefont
  {Lagos}}, \bibinfo {author} {\bibfnamefont {Meng-Xiang}\ \bibnamefont {Lin}},
  \ and\ \bibinfo {author} {\bibfnamefont {Fei}\ \bibnamefont {Xu}},\
  }\bibfield  {title} {\enquote {\bibinfo {title} {{Modified gravitational wave
  propagation with higher modes and its degeneracies with lensing}},}\ }\href
  {\doibase 10.1088/1475-7516/2022/08/016} {\bibfield  {journal} {\bibinfo
  {journal} {JCAP}\ }\textbf {\bibinfo {volume} {08}},\ \bibinfo {pages} {016}
  (\bibinfo {year} {2022})},\ \Eprint {http://arxiv.org/abs/2203.13252}
  {arXiv:2203.13252 [gr-qc]} \BibitemShut {NoStop}%
\bibitem [{\citenamefont {Janquart}\ \emph {et~al.}(2021)\citenamefont
  {Janquart}, \citenamefont {Seo}, \citenamefont {Hannuksela}, \citenamefont
  {Li},\ and\ \citenamefont {Broeck}}]{Janquart:2021nus}%
  \BibitemOpen
  \bibfield  {author} {\bibinfo {author} {\bibfnamefont {Justin}\ \bibnamefont
  {Janquart}}, \bibinfo {author} {\bibfnamefont {Eungwang}\ \bibnamefont
  {Seo}}, \bibinfo {author} {\bibfnamefont {Otto~A.}\ \bibnamefont
  {Hannuksela}}, \bibinfo {author} {\bibfnamefont {Tjonnie G.~F.}\ \bibnamefont
  {Li}}, \ and\ \bibinfo {author} {\bibfnamefont {Chris Van~Den}\ \bibnamefont
  {Broeck}},\ }\bibfield  {title} {\enquote {\bibinfo {title} {{On the
  Identification of Individual Gravitational-wave Image Types of a Lensed
  System Using Higher-order Modes}},}\ }\href {\doibase
  10.3847/2041-8213/ac3bcf} {\bibfield  {journal} {\bibinfo  {journal}
  {Astrophys. J. Lett.}\ }\textbf {\bibinfo {volume} {923}},\ \bibinfo {pages}
  {L1} (\bibinfo {year} {2021})},\ \Eprint {http://arxiv.org/abs/2110.06873}
  {arXiv:2110.06873 [gr-qc]} \BibitemShut {NoStop}%
\bibitem [{\citenamefont {Vijaykumar}\ \emph {et~al.}(2023)\citenamefont
  {Vijaykumar}, \citenamefont {Mehta},\ and\ \citenamefont
  {Ganguly}}]{Vijaykumar:2022dlp}%
  \BibitemOpen
  \bibfield  {author} {\bibinfo {author} {\bibfnamefont {Aditya}\ \bibnamefont
  {Vijaykumar}}, \bibinfo {author} {\bibfnamefont {Ajit~Kumar}\ \bibnamefont
  {Mehta}}, \ and\ \bibinfo {author} {\bibfnamefont {Apratim}\ \bibnamefont
  {Ganguly}},\ }\bibfield  {title} {\enquote {\bibinfo {title} {{Detection and
  parameter estimation challenges of type-II lensed binary black hole
  signals}},}\ }\href {\doibase 10.1103/PhysRevD.108.043036} {\bibfield
  {journal} {\bibinfo  {journal} {Phys. Rev. D}\ }\textbf {\bibinfo {volume}
  {108}},\ \bibinfo {pages} {043036} (\bibinfo {year} {2023})},\ \Eprint
  {http://arxiv.org/abs/2202.06334} {arXiv:2202.06334 [gr-qc]} \BibitemShut
  {NoStop}%
\bibitem [{\citenamefont {Mollerach}\ and\ \citenamefont
  {Roulet}(2002)}]{mollerach2002gravitational}%
  \BibitemOpen
  \bibfield  {author} {\bibinfo {author} {\bibfnamefont {Silvia}\ \bibnamefont
  {Mollerach}}\ and\ \bibinfo {author} {\bibfnamefont {Esteban}\ \bibnamefont
  {Roulet}},\ }\href@noop {} {\emph {\bibinfo {title} {Gravitational lensing
  and microlensing}}}\ (\bibinfo  {publisher} {World Scientific},\ \bibinfo
  {year} {2002})\BibitemShut {NoStop}%
\bibitem [{\citenamefont {{Schneider}}\ \emph {et~al.}(1992)\citenamefont
  {{Schneider}}, \citenamefont {{Ehlers}},\ and\ \citenamefont
  {{Falco}}}]{1992grle.book.....S}%
  \BibitemOpen
  \bibfield  {author} {\bibinfo {author} {\bibfnamefont {Peter}\ \bibnamefont
  {{Schneider}}}, \bibinfo {author} {\bibfnamefont {J{\"u}rgen}\ \bibnamefont
  {{Ehlers}}}, \ and\ \bibinfo {author} {\bibfnamefont {Emilio~E.}\
  \bibnamefont {{Falco}}},\ }\href@noop {} {\emph {\bibinfo {title}
  {{Gravitational Lenses}}}}\ (\bibinfo  {publisher} {Springer, Berlin,
  Heidelberg},\ \bibinfo {year} {1992})\BibitemShut {NoStop}%
\bibitem [{\citenamefont {Oguri}(2018)}]{Oguri:2018muv}%
  \BibitemOpen
  \bibfield  {author} {\bibinfo {author} {\bibfnamefont {Masamune}\
  \bibnamefont {Oguri}},\ }\bibfield  {title} {\enquote {\bibinfo {title}
  {{Effect of gravitational lensing on the distribution of gravitational waves
  from distant binary black hole mergers}},}\ }\href {\doibase
  10.1093/mnras/sty2145} {\bibfield  {journal} {\bibinfo  {journal} {Mon. Not.
  Roy. Astron. Soc.}\ }\textbf {\bibinfo {volume} {480}},\ \bibinfo {pages}
  {3842--3855} (\bibinfo {year} {2018})},\ \Eprint
  {http://arxiv.org/abs/1807.02584} {arXiv:1807.02584 [astro-ph.CO]}
  \BibitemShut {NoStop}%
\bibitem [{\citenamefont {More}\ and\ \citenamefont
  {More}(2022)}]{More:2021kpb}%
  \BibitemOpen
  \bibfield  {author} {\bibinfo {author} {\bibfnamefont {Anupreeta}\
  \bibnamefont {More}}\ and\ \bibinfo {author} {\bibfnamefont {Surhud}\
  \bibnamefont {More}},\ }\bibfield  {title} {\enquote {\bibinfo {title}
  {{Improved statistic to identify strongly lensed gravitational wave
  events}},}\ }\href {\doibase 10.1093/mnras/stac1704} {\bibfield  {journal}
  {\bibinfo  {journal} {Mon. Not. Roy. Astron. Soc.}\ }\textbf {\bibinfo
  {volume} {515}},\ \bibinfo {pages} {1044--1051} (\bibinfo {year} {2022})},\
  \Eprint {http://arxiv.org/abs/2111.03091} {arXiv:2111.03091 [astro-ph.CO]}
  \BibitemShut {NoStop}%
\bibitem [{\citenamefont {Hannuksela}\ \emph {et~al.}(2019)\citenamefont
  {Hannuksela}, \citenamefont {Haris}, \citenamefont {Ng}, \citenamefont
  {Kumar}, \citenamefont {Mehta}, \citenamefont {Keitel}, \citenamefont {Li},\
  and\ \citenamefont {Ajith}}]{Hannuksela:2019kle}%
  \BibitemOpen
  \bibfield  {author} {\bibinfo {author} {\bibfnamefont {O.~A.}\ \bibnamefont
  {Hannuksela}}, \bibinfo {author} {\bibfnamefont {K.}~\bibnamefont {Haris}},
  \bibinfo {author} {\bibfnamefont {K.~K.~Y.}\ \bibnamefont {Ng}}, \bibinfo
  {author} {\bibfnamefont {S.}~\bibnamefont {Kumar}}, \bibinfo {author}
  {\bibfnamefont {A.~K.}\ \bibnamefont {Mehta}}, \bibinfo {author}
  {\bibfnamefont {D.}~\bibnamefont {Keitel}}, \bibinfo {author} {\bibfnamefont
  {T.~G.~F.}\ \bibnamefont {Li}}, \ and\ \bibinfo {author} {\bibfnamefont
  {P.}~\bibnamefont {Ajith}},\ }\bibfield  {title} {\enquote {\bibinfo {title}
  {{Search for gravitational lensing signatures in LIGO-Virgo binary black hole
  events}},}\ }\href {\doibase 10.3847/2041-8213/ab0c0f} {\bibfield  {journal}
  {\bibinfo  {journal} {Astrophys. J. Lett.}\ }\textbf {\bibinfo {volume}
  {874}},\ \bibinfo {pages} {L2} (\bibinfo {year} {2019})},\ \Eprint
  {http://arxiv.org/abs/1901.02674} {arXiv:1901.02674 [gr-qc]} \BibitemShut
  {NoStop}%
\bibitem [{\citenamefont {Dai}\ \emph {et~al.}(2020)\citenamefont {Dai},
  \citenamefont {Zackay}, \citenamefont {Venumadhav}, \citenamefont {Roulet},\
  and\ \citenamefont {Zaldarriaga}}]{Dai:2020tpj}%
  \BibitemOpen
  \bibfield  {author} {\bibinfo {author} {\bibfnamefont {Liang}\ \bibnamefont
  {Dai}}, \bibinfo {author} {\bibfnamefont {Barak}\ \bibnamefont {Zackay}},
  \bibinfo {author} {\bibfnamefont {Tejaswi}\ \bibnamefont {Venumadhav}},
  \bibinfo {author} {\bibfnamefont {Javier}\ \bibnamefont {Roulet}}, \ and\
  \bibinfo {author} {\bibfnamefont {Matias}\ \bibnamefont {Zaldarriaga}},\
  }\bibfield  {title} {\enquote {\bibinfo {title} {{Search for Lensed
  Gravitational Waves Including Morse Phase Information: An Intriguing
  Candidate in O2}},}\ }\href@noop {} {\  (\bibinfo {year} {2020})},\ \Eprint
  {http://arxiv.org/abs/2007.12709} {arXiv:2007.12709 [astro-ph.HE]}
  \BibitemShut {NoStop}%
\bibitem [{\citenamefont {Dai}\ and\ \citenamefont
  {Venumadhav}(2017)}]{Dai:2017huk}%
  \BibitemOpen
  \bibfield  {author} {\bibinfo {author} {\bibfnamefont {Liang}\ \bibnamefont
  {Dai}}\ and\ \bibinfo {author} {\bibfnamefont {Tejaswi}\ \bibnamefont
  {Venumadhav}},\ }\bibfield  {title} {\enquote {\bibinfo {title} {{On the
  waveforms of gravitationally lensed gravitational waves}},}\ }\href@noop {}
  {\  (\bibinfo {year} {2017})},\ \Eprint {http://arxiv.org/abs/1702.04724}
  {arXiv:1702.04724 [gr-qc]} \BibitemShut {NoStop}%
\bibitem [{\citenamefont {Takahashi}\ and\ \citenamefont
  {Nakamura}(2003)}]{Takahashi:2003ix}%
  \BibitemOpen
  \bibfield  {author} {\bibinfo {author} {\bibfnamefont {Ryuichi}\ \bibnamefont
  {Takahashi}}\ and\ \bibinfo {author} {\bibfnamefont {Takashi}\ \bibnamefont
  {Nakamura}},\ }\bibfield  {title} {\enquote {\bibinfo {title} {{Wave effects
  in gravitational lensing of gravitational waves from chirping binaries}},}\
  }\href {\doibase 10.1086/377430} {\bibfield  {journal} {\bibinfo  {journal}
  {Astrophys. J.}\ }\textbf {\bibinfo {volume} {595}},\ \bibinfo {pages}
  {1039--1051} (\bibinfo {year} {2003})},\ \Eprint
  {http://arxiv.org/abs/astro-ph/0305055} {arXiv:astro-ph/0305055} \BibitemShut
  {NoStop}%
\bibitem [{\citenamefont {Goodman}(2005)}]{goodman2005introduction}%
  \BibitemOpen
  \bibfield  {author} {\bibinfo {author} {\bibfnamefont {Joseph~W}\
  \bibnamefont {Goodman}},\ }\href@noop {} {\emph {\bibinfo {title}
  {Introduction to Fourier optics}}}\ (\bibinfo  {publisher} {Roberts and
  Company Publishers},\ \bibinfo {year} {2005})\BibitemShut {NoStop}%
\bibitem [{\citenamefont {Schneider}\ and\ \citenamefont
  {Weiss}(1986)}]{schneider1986two}%
  \BibitemOpen
  \bibfield  {author} {\bibinfo {author} {\bibfnamefont {Peter}\ \bibnamefont
  {Schneider}}\ and\ \bibinfo {author} {\bibfnamefont {Achim}\ \bibnamefont
  {Weiss}},\ }\bibfield  {title} {\enquote {\bibinfo {title} {The
  two-point-mass lens-detailed investigation of a special asymmetric
  gravitational lens},}\ }\href@noop {} {\bibfield  {journal} {\bibinfo
  {journal} {Astronomy and Astrophysics}\ }\textbf {\bibinfo {volume} {164}},\
  \bibinfo {pages} {237--259} (\bibinfo {year} {1986})}\BibitemShut {NoStop}%
\bibitem [{\citenamefont {Bondarescu}\ \emph {et~al.}(2023)\citenamefont
  {Bondarescu}, \citenamefont {Ubach}, \citenamefont {Bulashenko},\ and\
  \citenamefont {Lundgren}}]{Bondarescu:2022srx}%
  \BibitemOpen
  \bibfield  {author} {\bibinfo {author} {\bibfnamefont {Ruxandra}\
  \bibnamefont {Bondarescu}}, \bibinfo {author} {\bibfnamefont {Helena}\
  \bibnamefont {Ubach}}, \bibinfo {author} {\bibfnamefont {Oleg}\ \bibnamefont
  {Bulashenko}}, \ and\ \bibinfo {author} {\bibfnamefont {Andrew~P.}\
  \bibnamefont {Lundgren}},\ }\bibfield  {title} {\enquote {\bibinfo {title}
  {{Compact binaries through a lens: Silent versus detectable microlensing for
  the LIGO-Virgo-KAGRA gravitational wave observatories}},}\ }\href {\doibase
  10.1103/PhysRevD.108.084033} {\bibfield  {journal} {\bibinfo  {journal}
  {Phys. Rev. D}\ }\textbf {\bibinfo {volume} {108}},\ \bibinfo {pages}
  {084033} (\bibinfo {year} {2023})},\ \Eprint
  {http://arxiv.org/abs/2211.13604} {arXiv:2211.13604 [gr-qc]} \BibitemShut
  {NoStop}%
\bibitem [{\citenamefont {Tambalo}\ \emph {et~al.}(2023)\citenamefont
  {Tambalo}, \citenamefont {Zumalac\'arregui}, \citenamefont {Dai},\ and\
  \citenamefont {Cheung}}]{Tambalo:2022plm}%
  \BibitemOpen
  \bibfield  {author} {\bibinfo {author} {\bibfnamefont {Giovanni}\
  \bibnamefont {Tambalo}}, \bibinfo {author} {\bibfnamefont {Miguel}\
  \bibnamefont {Zumalac\'arregui}}, \bibinfo {author} {\bibfnamefont {Liang}\
  \bibnamefont {Dai}}, \ and\ \bibinfo {author} {\bibfnamefont {Mark Ho-Yeuk}\
  \bibnamefont {Cheung}},\ }\bibfield  {title} {\enquote {\bibinfo {title}
  {{Lensing of gravitational waves: Efficient wave-optics methods and
  validation with symmetric lenses}},}\ }\href {\doibase
  10.1103/PhysRevD.108.043527} {\bibfield  {journal} {\bibinfo  {journal}
  {Phys. Rev. D}\ }\textbf {\bibinfo {volume} {108}},\ \bibinfo {pages}
  {043527} (\bibinfo {year} {2023})},\ \Eprint
  {http://arxiv.org/abs/2210.05658} {arXiv:2210.05658 [gr-qc]} \BibitemShut
  {NoStop}%
\bibitem [{\citenamefont {Narayan}\ \emph {et~al.}(2023)\citenamefont
  {Narayan}, \citenamefont {Johnson-McDaniel},\ and\ \citenamefont
  {Gupta}}]{Narayan:2023vhm}%
  \BibitemOpen
  \bibfield  {author} {\bibinfo {author} {\bibfnamefont {Purnima}\ \bibnamefont
  {Narayan}}, \bibinfo {author} {\bibfnamefont {Nathan~K.}\ \bibnamefont
  {Johnson-McDaniel}}, \ and\ \bibinfo {author} {\bibfnamefont {Anuradha}\
  \bibnamefont {Gupta}},\ }\bibfield  {title} {\enquote {\bibinfo {title}
  {{Effect of ignoring eccentricity in testing general relativity with
  gravitational waves}},}\ }\href {\doibase 10.1103/PhysRevD.108.064003}
  {\bibfield  {journal} {\bibinfo  {journal} {Phys. Rev. D}\ }\textbf {\bibinfo
  {volume} {108}},\ \bibinfo {pages} {064003} (\bibinfo {year} {2023})},\
  \Eprint {http://arxiv.org/abs/2306.04068} {arXiv:2306.04068 [gr-qc]}
  \BibitemShut {NoStop}%
\bibitem [{\citenamefont {Saini}\ \emph {et~al.}(2022)\citenamefont {Saini},
  \citenamefont {Favata},\ and\ \citenamefont {Arun}}]{Saini:2022igm}%
  \BibitemOpen
  \bibfield  {author} {\bibinfo {author} {\bibfnamefont {Pankaj}\ \bibnamefont
  {Saini}}, \bibinfo {author} {\bibfnamefont {Marc}\ \bibnamefont {Favata}}, \
  and\ \bibinfo {author} {\bibfnamefont {K.~G.}\ \bibnamefont {Arun}},\
  }\bibfield  {title} {\enquote {\bibinfo {title} {{Systematic bias on
  parametrized tests of general relativity due to neglect of orbital
  eccentricity}},}\ }\href {\doibase 10.1103/PhysRevD.106.084031} {\bibfield
  {journal} {\bibinfo  {journal} {Phys. Rev. D}\ }\textbf {\bibinfo {volume}
  {106}},\ \bibinfo {pages} {084031} (\bibinfo {year} {2022})},\ \Eprint
  {http://arxiv.org/abs/2203.04634} {arXiv:2203.04634 [gr-qc]} \BibitemShut
  {NoStop}%
\bibitem [{\citenamefont {Bhat}\ \emph {et~al.}(2023)\citenamefont {Bhat},
  \citenamefont {Saini}, \citenamefont {Favata},\ and\ \citenamefont
  {Arun}}]{Bhat:2022amc}%
  \BibitemOpen
  \bibfield  {author} {\bibinfo {author} {\bibfnamefont {Sajad~A.}\
  \bibnamefont {Bhat}}, \bibinfo {author} {\bibfnamefont {Pankaj}\ \bibnamefont
  {Saini}}, \bibinfo {author} {\bibfnamefont {Marc}\ \bibnamefont {Favata}}, \
  and\ \bibinfo {author} {\bibfnamefont {K.~G.}\ \bibnamefont {Arun}},\
  }\bibfield  {title} {\enquote {\bibinfo {title} {{Systematic bias on the
  inspiral-merger-ringdown consistency test due to neglect of orbital
  eccentricity}},}\ }\href {\doibase 10.1103/PhysRevD.107.024009} {\bibfield
  {journal} {\bibinfo  {journal} {Phys. Rev. D}\ }\textbf {\bibinfo {volume}
  {107}},\ \bibinfo {pages} {024009} (\bibinfo {year} {2023})},\ \Eprint
  {http://arxiv.org/abs/2207.13761} {arXiv:2207.13761 [gr-qc]} \BibitemShut
  {NoStop}%
\bibitem [{\citenamefont {Abbott}\ \emph {et~al.}(2017)\citenamefont {Abbott}
  \emph {et~al.}}]{LIGOScientific:2017bnn}%
  \BibitemOpen
  \bibfield  {author} {\bibinfo {author} {\bibfnamefont {Benjamin~P.}\
  \bibnamefont {Abbott}} \emph {et~al.} (\bibinfo {collaboration} {LIGO
  Scientific, VIRGO}),\ }\bibfield  {title} {\enquote {\bibinfo {title}
  {{GW170104: Observation of a 50-Solar-Mass Binary Black Hole Coalescence at
  Redshift 0.2}},}\ }\href {\doibase 10.1103/PhysRevLett.118.221101} {\bibfield
   {journal} {\bibinfo  {journal} {Phys. Rev. Lett.}\ }\textbf {\bibinfo
  {volume} {118}},\ \bibinfo {pages} {221101} (\bibinfo {year} {2017})},\
  \bibinfo {note} {[Erratum: Phys.Rev.Lett. 121, 129901 (2018)]},\ \Eprint
  {http://arxiv.org/abs/1706.01812} {arXiv:1706.01812 [gr-qc]} \BibitemShut
  {NoStop}%
\bibitem [{\citenamefont {Johnson-McDaniel}\ \emph {et~al.}()\citenamefont
  {Johnson-McDaniel} \emph {et~al.}}]{Final_mass_and_spin_public}%
  \BibitemOpen
  \bibfield  {author} {\bibinfo {author} {\bibfnamefont {Nathan~K.}\
  \bibnamefont {Johnson-McDaniel}} \emph {et~al.},\ }\href
  {https://dcc.ligo.org/public/0126/T1600168/006/Final_mass_and_spin_public.pdf}
  {\enquote {\bibinfo {title} {{Determining the final spin of a binary black
  hole system including in-plane spins: Method and checks of accuracy}},}\
  }\BibitemShut {NoStop}%
\bibitem [{\citenamefont {Hofmann}\ \emph {et~al.}(2016)\citenamefont
  {Hofmann}, \citenamefont {Barausse},\ and\ \citenamefont
  {Rezzolla}}]{Hofmann:2016yih}%
  \BibitemOpen
  \bibfield  {author} {\bibinfo {author} {\bibfnamefont {Fabian}\ \bibnamefont
  {Hofmann}}, \bibinfo {author} {\bibfnamefont {Enrico}\ \bibnamefont
  {Barausse}}, \ and\ \bibinfo {author} {\bibfnamefont {Luciano}\ \bibnamefont
  {Rezzolla}},\ }\bibfield  {title} {\enquote {\bibinfo {title} {{The final
  spin from binary black holes in quasi-circular orbits}},}\ }\href {\doibase
  10.3847/2041-8205/825/2/L19} {\bibfield  {journal} {\bibinfo  {journal}
  {Astrophys. J. Lett.}\ }\textbf {\bibinfo {volume} {825}},\ \bibinfo {pages}
  {L19} (\bibinfo {year} {2016})},\ \Eprint {http://arxiv.org/abs/1605.01938}
  {arXiv:1605.01938 [gr-qc]} \BibitemShut {NoStop}%
\bibitem [{\citenamefont {Jim\'enez-Forteza}\ \emph {et~al.}(2017)\citenamefont
  {Jim\'enez-Forteza}, \citenamefont {Keitel}, \citenamefont {Husa},
  \citenamefont {Hannam}, \citenamefont {Khan},\ and\ \citenamefont
  {P\"urrer}}]{Jimenez-Forteza:2016oae}%
  \BibitemOpen
  \bibfield  {author} {\bibinfo {author} {\bibfnamefont {Xisco}\ \bibnamefont
  {Jim\'enez-Forteza}}, \bibinfo {author} {\bibfnamefont {David}\ \bibnamefont
  {Keitel}}, \bibinfo {author} {\bibfnamefont {Sascha}\ \bibnamefont {Husa}},
  \bibinfo {author} {\bibfnamefont {Mark}\ \bibnamefont {Hannam}}, \bibinfo
  {author} {\bibfnamefont {Sebastian}\ \bibnamefont {Khan}}, \ and\ \bibinfo
  {author} {\bibfnamefont {Michael}\ \bibnamefont {P\"urrer}},\ }\bibfield
  {title} {\enquote {\bibinfo {title} {{Hierarchical data-driven approach to
  fitting numerical relativity data for nonprecessing binary black holes with
  an application to final spin and radiated energy}},}\ }\href {\doibase
  10.1103/PhysRevD.95.064024} {\bibfield  {journal} {\bibinfo  {journal} {Phys.
  Rev. D}\ }\textbf {\bibinfo {volume} {95}},\ \bibinfo {pages} {064024}
  (\bibinfo {year} {2017})},\ \Eprint {http://arxiv.org/abs/1611.00332}
  {arXiv:1611.00332 [gr-qc]} \BibitemShut {NoStop}%
\bibitem [{\citenamefont {Healy}\ and\ \citenamefont
  {Lousto}(2017)}]{Healy:2016lce}%
  \BibitemOpen
  \bibfield  {author} {\bibinfo {author} {\bibfnamefont {James}\ \bibnamefont
  {Healy}}\ and\ \bibinfo {author} {\bibfnamefont {Carlos~O.}\ \bibnamefont
  {Lousto}},\ }\bibfield  {title} {\enquote {\bibinfo {title} {{Remnant of
  binary black-hole mergers: New simulations and peak luminosity studies}},}\
  }\href {\doibase 10.1103/PhysRevD.95.024037} {\bibfield  {journal} {\bibinfo
  {journal} {Phys. Rev. D}\ }\textbf {\bibinfo {volume} {95}},\ \bibinfo
  {pages} {024037} (\bibinfo {year} {2017})},\ \Eprint
  {http://arxiv.org/abs/1610.09713} {arXiv:1610.09713 [gr-qc]} \BibitemShut
  {NoStop}%
\bibitem [{\citenamefont {Hoy}\ and\ \citenamefont
  {Raymond}(2021)}]{Hoy:2020vys}%
  \BibitemOpen
  \bibfield  {author} {\bibinfo {author} {\bibfnamefont {Charlie}\ \bibnamefont
  {Hoy}}\ and\ \bibinfo {author} {\bibfnamefont {Vivien}\ \bibnamefont
  {Raymond}},\ }\bibfield  {title} {\enquote {\bibinfo {title} {{PESummary: the
  code agnostic Parameter Estimation Summary page builder}},}\ }\href {\doibase
  10.1016/j.softx.2021.100765} {\bibfield  {journal} {\bibinfo  {journal}
  {SoftwareX}\ }\textbf {\bibinfo {volume} {15}},\ \bibinfo {pages} {100765}
  (\bibinfo {year} {2021})},\ \Eprint {http://arxiv.org/abs/2006.06639}
  {arXiv:2006.06639 [astro-ph.IM]} \BibitemShut {NoStop}%
\bibitem [{\citenamefont {Pratten}\ \emph {et~al.}(2021)\citenamefont {Pratten}
  \emph {et~al.}}]{PhenomXPHMPratten2020}%
  \BibitemOpen
  \bibfield  {author} {\bibinfo {author} {\bibfnamefont {Geraint}\ \bibnamefont
  {Pratten}} \emph {et~al.},\ }\bibfield  {title} {\enquote {\bibinfo {title}
  {{Computationally efficient models for the dominant and subdominant harmonic
  modes of precessing binary black holes}},}\ }\href {\doibase
  10.1103/PhysRevD.103.104056} {\bibfield  {journal} {\bibinfo  {journal}
  {Phys. Rev. D}\ }\textbf {\bibinfo {volume} {103}},\ \bibinfo {pages}
  {104056} (\bibinfo {year} {2021})},\ \Eprint
  {http://arxiv.org/abs/2004.06503} {arXiv:2004.06503 [gr-qc]} \BibitemShut
  {NoStop}%
\bibitem [{\citenamefont {Blanchet}(2014)}]{Blanchet2013_LivingReview}%
  \BibitemOpen
  \bibfield  {author} {\bibinfo {author} {\bibfnamefont {Luc}\ \bibnamefont
  {Blanchet}},\ }\bibfield  {title} {\enquote {\bibinfo {title} {{Gravitational
  Radiation from Post-Newtonian Sources and Inspiralling Compact Binaries}},}\
  }\href {\doibase 10.12942/lrr-2014-2} {\bibfield  {journal} {\bibinfo
  {journal} {Living Rev. Rel.}\ }\textbf {\bibinfo {volume} {17}},\ \bibinfo
  {pages} {2} (\bibinfo {year} {2014})},\ \Eprint
  {http://arxiv.org/abs/1310.1528} {arXiv:1310.1528 [gr-qc]} \BibitemShut
  {NoStop}%
\bibitem [{\citenamefont {Pretorius}(2007)}]{Pretorius:2007nq}%
  \BibitemOpen
  \bibfield  {author} {\bibinfo {author} {\bibfnamefont {Frans}\ \bibnamefont
  {Pretorius}},\ }\bibfield  {title} {\enquote {\bibinfo {title} {{Binary Black
  Hole Coalescence}},}\ }\href@noop {} {\  (\bibinfo {year} {2007})},\ \Eprint
  {http://arxiv.org/abs/0710.1338} {arXiv:0710.1338 [gr-qc]} \BibitemShut
  {NoStop}%
\bibitem [{\citenamefont {Duez}\ and\ \citenamefont
  {Zlochower}(2019)}]{Duez:2018jaf}%
  \BibitemOpen
  \bibfield  {author} {\bibinfo {author} {\bibfnamefont {Matthew~D.}\
  \bibnamefont {Duez}}\ and\ \bibinfo {author} {\bibfnamefont {Yosef}\
  \bibnamefont {Zlochower}},\ }\bibfield  {title} {\enquote {\bibinfo {title}
  {{Numerical Relativity of Compact Binaries in the 21st Century}},}\ }\href
  {\doibase 10.1088/1361-6633/aadb16} {\bibfield  {journal} {\bibinfo
  {journal} {Rept. Prog. Phys.}\ }\textbf {\bibinfo {volume} {82}},\ \bibinfo
  {pages} {016902} (\bibinfo {year} {2019})},\ \Eprint
  {http://arxiv.org/abs/1808.06011} {arXiv:1808.06011 [gr-qc]} \BibitemShut
  {NoStop}%
\bibitem [{\citenamefont {Berti}\ \emph {et~al.}(2006)\citenamefont {Berti},
  \citenamefont {Cardoso},\ and\ \citenamefont {Will}}]{Berti:2005ys}%
  \BibitemOpen
  \bibfield  {author} {\bibinfo {author} {\bibfnamefont {Emanuele}\
  \bibnamefont {Berti}}, \bibinfo {author} {\bibfnamefont {Vitor}\ \bibnamefont
  {Cardoso}}, \ and\ \bibinfo {author} {\bibfnamefont {Clifford~M.}\
  \bibnamefont {Will}},\ }\bibfield  {title} {\enquote {\bibinfo {title} {{On
  gravitational-wave spectroscopy of massive black holes with the space
  interferometer LISA}},}\ }\href {\doibase 10.1103/PhysRevD.73.064030}
  {\bibfield  {journal} {\bibinfo  {journal} {Phys. Rev. D}\ }\textbf {\bibinfo
  {volume} {73}},\ \bibinfo {pages} {064030} (\bibinfo {year} {2006})},\
  \Eprint {http://arxiv.org/abs/gr-qc/0512160} {arXiv:gr-qc/0512160}
  \BibitemShut {NoStop}%
\bibitem [{\citenamefont {Agathos}\ \emph {et~al.}(2014)\citenamefont
  {Agathos}, \citenamefont {Del~Pozzo}, \citenamefont {Li}, \citenamefont {Van
  Den~Broeck}, \citenamefont {Veitch},\ and\ \citenamefont
  {Vitale}}]{Agathos:2013upa}%
  \BibitemOpen
  \bibfield  {author} {\bibinfo {author} {\bibfnamefont {Michalis}\
  \bibnamefont {Agathos}}, \bibinfo {author} {\bibfnamefont {Walter}\
  \bibnamefont {Del~Pozzo}}, \bibinfo {author} {\bibfnamefont {Tjonnie G.~F.}\
  \bibnamefont {Li}}, \bibinfo {author} {\bibfnamefont {Chris}\ \bibnamefont
  {Van Den~Broeck}}, \bibinfo {author} {\bibfnamefont {John}\ \bibnamefont
  {Veitch}}, \ and\ \bibinfo {author} {\bibfnamefont {Salvatore}\ \bibnamefont
  {Vitale}},\ }\bibfield  {title} {\enquote {\bibinfo {title} {{TIGER: A data
  analysis pipeline for testing the strong-field dynamics of general relativity
  with gravitational wave signals from coalescing compact binaries}},}\ }\href
  {\doibase 10.1103/PhysRevD.89.082001} {\bibfield  {journal} {\bibinfo
  {journal} {Phys. Rev. D}\ }\textbf {\bibinfo {volume} {89}},\ \bibinfo
  {pages} {082001} (\bibinfo {year} {2014})},\ \Eprint
  {http://arxiv.org/abs/1311.0420} {arXiv:1311.0420 [gr-qc]} \BibitemShut
  {NoStop}%
\bibitem [{\citenamefont {Meidam}\ \emph {et~al.}(2018)\citenamefont {Meidam}
  \emph {et~al.}}]{Meidam:2017dgf}%
  \BibitemOpen
  \bibfield  {author} {\bibinfo {author} {\bibfnamefont {Jeroen}\ \bibnamefont
  {Meidam}} \emph {et~al.},\ }\bibfield  {title} {\enquote {\bibinfo {title}
  {{Parametrized tests of the strong-field dynamics of general relativity using
  gravitational wave signals from coalescing binary black holes: Fast
  likelihood calculations and sensitivity of the method}},}\ }\href {\doibase
  10.1103/PhysRevD.97.044033} {\bibfield  {journal} {\bibinfo  {journal} {Phys.
  Rev. D}\ }\textbf {\bibinfo {volume} {97}},\ \bibinfo {pages} {044033}
  (\bibinfo {year} {2018})},\ \Eprint {http://arxiv.org/abs/1712.08772}
  {arXiv:1712.08772 [gr-qc]} \BibitemShut {NoStop}%
\bibitem [{\citenamefont {Ajith}\ \emph {et~al.}(2008)\citenamefont {Ajith}
  \emph {et~al.}}]{Ajith:2007kx}%
  \BibitemOpen
  \bibfield  {author} {\bibinfo {author} {\bibfnamefont {P.}~\bibnamefont
  {Ajith}} \emph {et~al.},\ }\bibfield  {title} {\enquote {\bibinfo {title} {{A
  Template bank for gravitational waveforms from coalescing binary black holes.
  I. Non-spinning binaries}},}\ }\href {\doibase 10.1103/PhysRevD.77.104017}
  {\bibfield  {journal} {\bibinfo  {journal} {Phys. Rev. D}\ }\textbf {\bibinfo
  {volume} {77}},\ \bibinfo {pages} {104017} (\bibinfo {year} {2008})},\
  \bibinfo {note} {[Erratum: Phys.Rev.D 79, 129901 (2009)]},\ \Eprint
  {http://arxiv.org/abs/0710.2335} {arXiv:0710.2335 [gr-qc]} \BibitemShut
  {NoStop}%
\bibitem [{\citenamefont {Ajith}\ \emph {et~al.}(2011)\citenamefont {Ajith}
  \emph {et~al.}}]{Ajith:2009bn}%
  \BibitemOpen
  \bibfield  {author} {\bibinfo {author} {\bibfnamefont {P.}~\bibnamefont
  {Ajith}} \emph {et~al.},\ }\bibfield  {title} {\enquote {\bibinfo {title}
  {{Inspiral-merger-ringdown waveforms for black-hole binaries with
  non-precessing spins}},}\ }\href {\doibase 10.1103/PhysRevLett.106.241101}
  {\bibfield  {journal} {\bibinfo  {journal} {Phys. Rev. Lett.}\ }\textbf
  {\bibinfo {volume} {106}},\ \bibinfo {pages} {241101} (\bibinfo {year}
  {2011})},\ \Eprint {http://arxiv.org/abs/0909.2867} {arXiv:0909.2867 [gr-qc]}
  \BibitemShut {NoStop}%
\bibitem [{\citenamefont {Santamaria}\ \emph {et~al.}(2010)\citenamefont
  {Santamaria} \emph {et~al.}}]{Santamaria:2010yb}%
  \BibitemOpen
  \bibfield  {author} {\bibinfo {author} {\bibfnamefont {L.}~\bibnamefont
  {Santamaria}} \emph {et~al.},\ }\bibfield  {title} {\enquote {\bibinfo
  {title} {{Matching post-Newtonian and numerical relativity waveforms:
  systematic errors and a new phenomenological model for non-precessing black
  hole binaries}},}\ }\href {\doibase 10.1103/PhysRevD.82.064016} {\bibfield
  {journal} {\bibinfo  {journal} {Phys. Rev. D}\ }\textbf {\bibinfo {volume}
  {82}},\ \bibinfo {pages} {064016} (\bibinfo {year} {2010})},\ \Eprint
  {http://arxiv.org/abs/1005.3306} {arXiv:1005.3306 [gr-qc]} \BibitemShut
  {NoStop}%
\bibitem [{\citenamefont {Husa}\ \emph {et~al.}(2016)\citenamefont {Husa},
  \citenamefont {Khan}, \citenamefont {Hannam}, \citenamefont {P\"urrer},
  \citenamefont {Ohme}, \citenamefont {Jim\'enez~Forteza},\ and\ \citenamefont
  {Boh\'e}}]{Husa:2015iqa}%
  \BibitemOpen
  \bibfield  {author} {\bibinfo {author} {\bibfnamefont {Sascha}\ \bibnamefont
  {Husa}}, \bibinfo {author} {\bibfnamefont {Sebastian}\ \bibnamefont {Khan}},
  \bibinfo {author} {\bibfnamefont {Mark}\ \bibnamefont {Hannam}}, \bibinfo
  {author} {\bibfnamefont {Michael}\ \bibnamefont {P\"urrer}}, \bibinfo
  {author} {\bibfnamefont {Frank}\ \bibnamefont {Ohme}}, \bibinfo {author}
  {\bibfnamefont {Xisco}\ \bibnamefont {Jim\'enez~Forteza}}, \ and\ \bibinfo
  {author} {\bibfnamefont {Alejandro}\ \bibnamefont {Boh\'e}},\ }\bibfield
  {title} {\enquote {\bibinfo {title} {{Frequency-domain gravitational waves
  from nonprecessing black-hole binaries. I. New numerical waveforms and
  anatomy of the signal}},}\ }\href {\doibase 10.1103/PhysRevD.93.044006}
  {\bibfield  {journal} {\bibinfo  {journal} {Phys. Rev. D}\ }\textbf {\bibinfo
  {volume} {93}},\ \bibinfo {pages} {044006} (\bibinfo {year} {2016})},\
  \Eprint {http://arxiv.org/abs/1508.07250} {arXiv:1508.07250 [gr-qc]}
  \BibitemShut {NoStop}%
\bibitem [{\citenamefont {Khan}\ \emph {et~al.}(2016)\citenamefont {Khan},
  \citenamefont {Husa}, \citenamefont {Hannam}, \citenamefont {Ohme},
  \citenamefont {P\"urrer}, \citenamefont {Jim\'enez~Forteza},\ and\
  \citenamefont {Boh\'e}}]{Khan:2015jqa}%
  \BibitemOpen
  \bibfield  {author} {\bibinfo {author} {\bibfnamefont {Sebastian}\
  \bibnamefont {Khan}}, \bibinfo {author} {\bibfnamefont {Sascha}\ \bibnamefont
  {Husa}}, \bibinfo {author} {\bibfnamefont {Mark}\ \bibnamefont {Hannam}},
  \bibinfo {author} {\bibfnamefont {Frank}\ \bibnamefont {Ohme}}, \bibinfo
  {author} {\bibfnamefont {Michael}\ \bibnamefont {P\"urrer}}, \bibinfo
  {author} {\bibfnamefont {Xisco}\ \bibnamefont {Jim\'enez~Forteza}}, \ and\
  \bibinfo {author} {\bibfnamefont {Alejandro}\ \bibnamefont {Boh\'e}},\
  }\bibfield  {title} {\enquote {\bibinfo {title} {{Frequency-domain
  gravitational waves from nonprecessing black-hole binaries. II. A
  phenomenological model for the advanced detector era}},}\ }\href {\doibase
  10.1103/PhysRevD.93.044007} {\bibfield  {journal} {\bibinfo  {journal} {Phys.
  Rev. D}\ }\textbf {\bibinfo {volume} {93}},\ \bibinfo {pages} {044007}
  (\bibinfo {year} {2016})},\ \Eprint {http://arxiv.org/abs/1508.07253}
  {arXiv:1508.07253 [gr-qc]} \BibitemShut {NoStop}%
\bibitem [{\citenamefont {Garc\'\i{}a-Quir\'os}\ \emph
  {et~al.}(2020)\citenamefont {Garc\'\i{}a-Quir\'os}, \citenamefont {Colleoni},
  \citenamefont {Husa}, \citenamefont {Estell\'es}, \citenamefont {Pratten},
  \citenamefont {Ramos-Buades}, \citenamefont {Mateu-Lucena},\ and\
  \citenamefont {Jaume}}]{Garcia-Quiros:2020qpx}%
  \BibitemOpen
  \bibfield  {author} {\bibinfo {author} {\bibfnamefont {Cecilio}\ \bibnamefont
  {Garc\'\i{}a-Quir\'os}}, \bibinfo {author} {\bibfnamefont {Marta}\
  \bibnamefont {Colleoni}}, \bibinfo {author} {\bibfnamefont {Sascha}\
  \bibnamefont {Husa}}, \bibinfo {author} {\bibfnamefont {H\'ector}\
  \bibnamefont {Estell\'es}}, \bibinfo {author} {\bibfnamefont {Geraint}\
  \bibnamefont {Pratten}}, \bibinfo {author} {\bibfnamefont {Antoni}\
  \bibnamefont {Ramos-Buades}}, \bibinfo {author} {\bibfnamefont {Maite}\
  \bibnamefont {Mateu-Lucena}}, \ and\ \bibinfo {author} {\bibfnamefont
  {Rafel}\ \bibnamefont {Jaume}},\ }\bibfield  {title} {\enquote {\bibinfo
  {title} {{Multimode frequency-domain model for the gravitational wave signal
  from nonprecessing black-hole binaries}},}\ }\href {\doibase
  10.1103/PhysRevD.102.064002} {\bibfield  {journal} {\bibinfo  {journal}
  {Phys. Rev. D}\ }\textbf {\bibinfo {volume} {102}},\ \bibinfo {pages}
  {064002} (\bibinfo {year} {2020})},\ \Eprint
  {http://arxiv.org/abs/2001.10914} {arXiv:2001.10914 [gr-qc]} \BibitemShut
  {NoStop}%
\bibitem [{\citenamefont {Pratten}\ \emph {et~al.}(2020)\citenamefont
  {Pratten}, \citenamefont {Husa}, \citenamefont {Garcia-Quiros}, \citenamefont
  {Colleoni}, \citenamefont {Ramos-Buades}, \citenamefont {Estelles},\ and\
  \citenamefont {Jaume}}]{Pratten:2020fqn}%
  \BibitemOpen
  \bibfield  {author} {\bibinfo {author} {\bibfnamefont {Geraint}\ \bibnamefont
  {Pratten}}, \bibinfo {author} {\bibfnamefont {Sascha}\ \bibnamefont {Husa}},
  \bibinfo {author} {\bibfnamefont {Cecilio}\ \bibnamefont {Garcia-Quiros}},
  \bibinfo {author} {\bibfnamefont {Marta}\ \bibnamefont {Colleoni}}, \bibinfo
  {author} {\bibfnamefont {Antoni}\ \bibnamefont {Ramos-Buades}}, \bibinfo
  {author} {\bibfnamefont {Hector}\ \bibnamefont {Estelles}}, \ and\ \bibinfo
  {author} {\bibfnamefont {Rafel}\ \bibnamefont {Jaume}},\ }\bibfield  {title}
  {\enquote {\bibinfo {title} {{Setting the cornerstone for a family of models
  for gravitational waves from compact binaries: The dominant harmonic for
  nonprecessing quasicircular black holes}},}\ }\href {\doibase
  10.1103/PhysRevD.102.064001} {\bibfield  {journal} {\bibinfo  {journal}
  {Phys. Rev. D}\ }\textbf {\bibinfo {volume} {102}},\ \bibinfo {pages}
  {064001} (\bibinfo {year} {2020})},\ \Eprint
  {http://arxiv.org/abs/2001.11412} {arXiv:2001.11412 [gr-qc]} \BibitemShut
  {NoStop}%
\bibitem [{\citenamefont {Hannam}\ \emph {et~al.}(2014)\citenamefont {Hannam},
  \citenamefont {Schmidt}, \citenamefont {Boh\'e}, \citenamefont {Haegel},
  \citenamefont {Husa}, \citenamefont {Ohme}, \citenamefont {Pratten},\ and\
  \citenamefont {P\"urrer}}]{Hannam:2013oca}%
  \BibitemOpen
  \bibfield  {author} {\bibinfo {author} {\bibfnamefont {Mark}\ \bibnamefont
  {Hannam}}, \bibinfo {author} {\bibfnamefont {Patricia}\ \bibnamefont
  {Schmidt}}, \bibinfo {author} {\bibfnamefont {Alejandro}\ \bibnamefont
  {Boh\'e}}, \bibinfo {author} {\bibfnamefont {Le\"\i{}la}\ \bibnamefont
  {Haegel}}, \bibinfo {author} {\bibfnamefont {Sascha}\ \bibnamefont {Husa}},
  \bibinfo {author} {\bibfnamefont {Frank}\ \bibnamefont {Ohme}}, \bibinfo
  {author} {\bibfnamefont {Geraint}\ \bibnamefont {Pratten}}, \ and\ \bibinfo
  {author} {\bibfnamefont {Michael}\ \bibnamefont {P\"urrer}},\ }\bibfield
  {title} {\enquote {\bibinfo {title} {{Simple Model of Complete Precessing
  Black-Hole-Binary Gravitational Waveforms}},}\ }\href {\doibase
  10.1103/PhysRevLett.113.151101} {\bibfield  {journal} {\bibinfo  {journal}
  {Phys. Rev. Lett.}\ }\textbf {\bibinfo {volume} {113}},\ \bibinfo {pages}
  {151101} (\bibinfo {year} {2014})},\ \Eprint {http://arxiv.org/abs/1308.3271}
  {arXiv:1308.3271 [gr-qc]} \BibitemShut {NoStop}%
\bibitem [{\citenamefont {Alejandro}\ \emph {et~al.}()\citenamefont {Alejandro}
  \emph {et~al.}}]{PhenomPv2_Technical_document}%
  \BibitemOpen
  \bibfield  {author} {\bibinfo {author} {\bibfnamefont {Bohe}\ \bibnamefont
  {Alejandro}} \emph {et~al.},\ }\href
  {https://dcc.ligo.org/LIGO-T1500602/public} {\enquote {\bibinfo {title}
  {{PhenomPv2 -- technical notes for the LAL implementation}},}\ }\BibitemShut
  {NoStop}%
\bibitem [{\citenamefont {Abbott}\ \emph {et~al.}(2020)\citenamefont {Abbott},
  \citenamefont {Abbott}, \citenamefont {Abbott}, \citenamefont {Abraham},
  \citenamefont {Acernese}, \citenamefont {Ackley}, \citenamefont {Adams},
  \citenamefont {Adya}, \citenamefont {Affeldt}, \citenamefont {Agathos} \emph
  {et~al.}}]{Abbott:2020search}%
  \BibitemOpen
  \bibfield  {author} {\bibinfo {author} {\bibfnamefont {Benjamin~P}\
  \bibnamefont {Abbott}}, \bibinfo {author} {\bibfnamefont {R}~\bibnamefont
  {Abbott}}, \bibinfo {author} {\bibfnamefont {TD}~\bibnamefont {Abbott}},
  \bibinfo {author} {\bibfnamefont {S}~\bibnamefont {Abraham}}, \bibinfo
  {author} {\bibfnamefont {F}~\bibnamefont {Acernese}}, \bibinfo {author}
  {\bibfnamefont {K}~\bibnamefont {Ackley}}, \bibinfo {author} {\bibfnamefont
  {C}~\bibnamefont {Adams}}, \bibinfo {author} {\bibfnamefont {VB}~\bibnamefont
  {Adya}}, \bibinfo {author} {\bibfnamefont {C}~\bibnamefont {Affeldt}},
  \bibinfo {author} {\bibfnamefont {M}~\bibnamefont {Agathos}},  \emph
  {et~al.},\ }\bibfield  {title} {\enquote {\bibinfo {title} {Prospects for
  observing and localizing gravitational-wave transients with advanced ligo,
  advanced virgo and kagra},}\ }\href@noop {} {\bibfield  {journal} {\bibinfo
  {journal} {Living reviews in relativity}\ }\textbf {\bibinfo {volume} {23}},\
  \bibinfo {pages} {1--69} (\bibinfo {year} {2020})}\BibitemShut {NoStop}%
\bibitem [{\citenamefont {P\"urrer}\ \emph {et~al.}(2017)\citenamefont
  {P\"urrer}, \citenamefont {Hannam},\ and\ \citenamefont
  {Ohme}}]{Purrer:2017uch}%
  \BibitemOpen
  \bibfield  {author} {\bibinfo {author} {\bibfnamefont {Michael}\ \bibnamefont
  {P\"urrer}}, \bibinfo {author} {\bibfnamefont {Mark}\ \bibnamefont {Hannam}},
  \ and\ \bibinfo {author} {\bibfnamefont {Frank}\ \bibnamefont {Ohme}},\
  }\bibfield  {title} {\enquote {\bibinfo {title} {{Can we measure individual
  black-hole spins from gravitational-wave observations?}}}\ }in\ \href
  {\doibase 10.1142/9789813226609_0399} {\emph {\bibinfo {booktitle} {{14th
  Marcel Grossmann Meeting on Recent Developments in Theoretical and
  Experimental General Relativity, Astrophysics, and Relativistic Field
  Theories}}}},\ Vol.~\bibinfo {volume} {3}\ (\bibinfo {year} {2017})\ pp.\
  \bibinfo {pages} {3144--3148}\BibitemShut {NoStop}%
\bibitem [{\citenamefont {Madau}\ and\ \citenamefont
  {Dickinson}(2014)}]{Madau:2014bja}%
  \BibitemOpen
  \bibfield  {author} {\bibinfo {author} {\bibfnamefont {Piero}\ \bibnamefont
  {Madau}}\ and\ \bibinfo {author} {\bibfnamefont {Mark}\ \bibnamefont
  {Dickinson}},\ }\bibfield  {title} {\enquote {\bibinfo {title} {{Cosmic Star
  Formation History}},}\ }\href {\doibase 10.1146/annurev-astro-081811-125615}
  {\bibfield  {journal} {\bibinfo  {journal} {Ann. Rev. Astron. Astrophys.}\
  }\textbf {\bibinfo {volume} {52}},\ \bibinfo {pages} {415--486} (\bibinfo
  {year} {2014})},\ \Eprint {http://arxiv.org/abs/1403.0007} {arXiv:1403.0007
  [astro-ph.CO]} \BibitemShut {NoStop}%
\bibitem [{\citenamefont {Fishbach}\ \emph {et~al.}(2018)\citenamefont
  {Fishbach}, \citenamefont {Holz},\ and\ \citenamefont
  {Farr}}]{Fishbach:2018edt}%
  \BibitemOpen
  \bibfield  {author} {\bibinfo {author} {\bibfnamefont {Maya}\ \bibnamefont
  {Fishbach}}, \bibinfo {author} {\bibfnamefont {Daniel~E.}\ \bibnamefont
  {Holz}}, \ and\ \bibinfo {author} {\bibfnamefont {Will~M.}\ \bibnamefont
  {Farr}},\ }\bibfield  {title} {\enquote {\bibinfo {title} {{Does the Black
  Hole Merger Rate Evolve with Redshift?}}}\ }\href {\doibase
  10.3847/2041-8213/aad800} {\bibfield  {journal} {\bibinfo  {journal}
  {Astrophys. J. Lett.}\ }\textbf {\bibinfo {volume} {863}},\ \bibinfo {pages}
  {L41} (\bibinfo {year} {2018})},\ \Eprint {http://arxiv.org/abs/1805.10270}
  {arXiv:1805.10270 [astro-ph.HE]} \BibitemShut {NoStop}%
\bibitem [{\citenamefont {Lai}\ \emph {et~al.}(2018)\citenamefont {Lai},
  \citenamefont {Hannuksela}, \citenamefont {Herrera-Mart\'\i{}n},
  \citenamefont {Diego}, \citenamefont {Broadhurst},\ and\ \citenamefont
  {Li}}]{Lai:2018rto}%
  \BibitemOpen
  \bibfield  {author} {\bibinfo {author} {\bibfnamefont {Kwun-Hang}\
  \bibnamefont {Lai}}, \bibinfo {author} {\bibfnamefont {Otto~A.}\ \bibnamefont
  {Hannuksela}}, \bibinfo {author} {\bibfnamefont {Antonio}\ \bibnamefont
  {Herrera-Mart\'\i{}n}}, \bibinfo {author} {\bibfnamefont {Jose~M.}\
  \bibnamefont {Diego}}, \bibinfo {author} {\bibfnamefont {Tom}\ \bibnamefont
  {Broadhurst}}, \ and\ \bibinfo {author} {\bibfnamefont {Tjonnie G.~F.}\
  \bibnamefont {Li}},\ }\bibfield  {title} {\enquote {\bibinfo {title}
  {{Discovering intermediate-mass black hole lenses through gravitational wave
  lensing}},}\ }\href {\doibase 10.1103/PhysRevD.98.083005} {\bibfield
  {journal} {\bibinfo  {journal} {Phys. Rev. D}\ }\textbf {\bibinfo {volume}
  {98}},\ \bibinfo {pages} {083005} (\bibinfo {year} {2018})},\ \Eprint
  {http://arxiv.org/abs/1801.07840} {arXiv:1801.07840 [gr-qc]} \BibitemShut
  {NoStop}%
\bibitem [{\citenamefont {Ashton}\ \emph {et~al.}(2019)\citenamefont {Ashton}
  \emph {et~al.}}]{Ashton:2018jfp}%
  \BibitemOpen
  \bibfield  {author} {\bibinfo {author} {\bibfnamefont {Gregory}\ \bibnamefont
  {Ashton}} \emph {et~al.},\ }\bibfield  {title} {\enquote {\bibinfo {title}
  {{BILBY: A user-friendly Bayesian inference library for gravitational-wave
  astronomy}},}\ }\href {\doibase 10.3847/1538-4365/ab06fc} {\bibfield
  {journal} {\bibinfo  {journal} {Astrophys. J. Suppl.}\ }\textbf {\bibinfo
  {volume} {241}},\ \bibinfo {pages} {27} (\bibinfo {year} {2019})},\ \Eprint
  {http://arxiv.org/abs/1811.02042} {arXiv:1811.02042 [astro-ph.IM]}
  \BibitemShut {NoStop}%
\bibitem [{\citenamefont {Speagle}(2020)}]{speagle2020dynesty}%
  \BibitemOpen
  \bibfield  {author} {\bibinfo {author} {\bibfnamefont {Joshua~S}\
  \bibnamefont {Speagle}},\ }\bibfield  {title} {\enquote {\bibinfo {title}
  {dynesty: a dynamic nested sampling package for estimating bayesian
  posteriors and evidences},}\ }\href@noop {} {\bibfield  {journal} {\bibinfo
  {journal} {Monthly Notices of the Royal Astronomical Society}\ }\textbf
  {\bibinfo {volume} {493}},\ \bibinfo {pages} {3132--3158} (\bibinfo {year}
  {2020})}\BibitemShut {NoStop}%
\bibitem [{\citenamefont {Johnson-McDaniel}\ \emph {et~al.}(2022)\citenamefont
  {Johnson-McDaniel}, \citenamefont {Ghosh}, \citenamefont {Ghonge},
  \citenamefont {Saleem}, \citenamefont {Krishnendu},\ and\ \citenamefont
  {Clark}}]{Johnson-McDaniel:2021yge}%
  \BibitemOpen
  \bibfield  {author} {\bibinfo {author} {\bibfnamefont {Nathan~K.}\
  \bibnamefont {Johnson-McDaniel}}, \bibinfo {author} {\bibfnamefont {Abhirup}\
  \bibnamefont {Ghosh}}, \bibinfo {author} {\bibfnamefont {Sudarshan}\
  \bibnamefont {Ghonge}}, \bibinfo {author} {\bibfnamefont {Muhammed}\
  \bibnamefont {Saleem}}, \bibinfo {author} {\bibfnamefont {N.~V.}\
  \bibnamefont {Krishnendu}}, \ and\ \bibinfo {author} {\bibfnamefont
  {James~A.}\ \bibnamefont {Clark}},\ }\bibfield  {title} {\enquote {\bibinfo
  {title} {{Investigating the relation between gravitational wave tests of
  general relativity}},}\ }\href {\doibase 10.1103/PhysRevD.105.044020}
  {\bibfield  {journal} {\bibinfo  {journal} {Phys. Rev. D}\ }\textbf {\bibinfo
  {volume} {105}},\ \bibinfo {pages} {044020} (\bibinfo {year} {2022})},\
  \Eprint {http://arxiv.org/abs/2109.06988} {arXiv:2109.06988 [gr-qc]}
  \BibitemShut {NoStop}%
\bibitem [{\citenamefont {Cornish}\ \emph {et~al.}(2011)\citenamefont
  {Cornish}, \citenamefont {Sampson}, \citenamefont {Yunes},\ and\
  \citenamefont {Pretorius}}]{Cornish:2011ys}%
  \BibitemOpen
  \bibfield  {author} {\bibinfo {author} {\bibfnamefont {Neil}\ \bibnamefont
  {Cornish}}, \bibinfo {author} {\bibfnamefont {Laura}\ \bibnamefont
  {Sampson}}, \bibinfo {author} {\bibfnamefont {Nicolas}\ \bibnamefont
  {Yunes}}, \ and\ \bibinfo {author} {\bibfnamefont {Frans}\ \bibnamefont
  {Pretorius}},\ }\bibfield  {title} {\enquote {\bibinfo {title}
  {{Gravitational Wave Tests of General Relativity with the Parameterized
  Post-Einsteinian Framework}},}\ }\href {\doibase 10.1103/PhysRevD.84.062003}
  {\bibfield  {journal} {\bibinfo  {journal} {Phys. Rev. D}\ }\textbf {\bibinfo
  {volume} {84}},\ \bibinfo {pages} {062003} (\bibinfo {year} {2011})},\
  \Eprint {http://arxiv.org/abs/1105.2088} {arXiv:1105.2088 [gr-qc]}
  \BibitemShut {NoStop}%
\bibitem [{\citenamefont {Vallisneri}(2012)}]{Vallisneri:2012qq}%
  \BibitemOpen
  \bibfield  {author} {\bibinfo {author} {\bibfnamefont {Michele}\ \bibnamefont
  {Vallisneri}},\ }\bibfield  {title} {\enquote {\bibinfo {title} {{Testing
  general relativity with gravitational waves: a reality check}},}\ }\href
  {\doibase 10.1103/PhysRevD.86.082001} {\bibfield  {journal} {\bibinfo
  {journal} {Phys. Rev. D}\ }\textbf {\bibinfo {volume} {86}},\ \bibinfo
  {pages} {082001} (\bibinfo {year} {2012})},\ \Eprint
  {http://arxiv.org/abs/1207.4759} {arXiv:1207.4759 [gr-qc]} \BibitemShut
  {NoStop}%
\bibitem [{\citenamefont {Sathyaprakash}\ and\ \citenamefont
  {Dhurandhar}(1991)}]{Sathyaprakash:1991mt}%
  \BibitemOpen
  \bibfield  {author} {\bibinfo {author} {\bibfnamefont {B.~S.}\ \bibnamefont
  {Sathyaprakash}}\ and\ \bibinfo {author} {\bibfnamefont {S.~V.}\ \bibnamefont
  {Dhurandhar}},\ }\bibfield  {title} {\enquote {\bibinfo {title} {{Choice of
  filters for the detection of gravitational waves from coalescing
  binaries}},}\ }\href {\doibase 10.1103/PhysRevD.44.3819} {\bibfield
  {journal} {\bibinfo  {journal} {Phys. Rev. D}\ }\textbf {\bibinfo {volume}
  {44}},\ \bibinfo {pages} {3819--3834} (\bibinfo {year} {1991})}\BibitemShut
  {NoStop}%
\bibitem [{\citenamefont {Ajith}\ \emph {et~al.}(2014)\citenamefont {Ajith},
  \citenamefont {Fotopoulos}, \citenamefont {Privitera}, \citenamefont
  {Neunzert},\ and\ \citenamefont {Weinstein}}]{Ajith:2012mn}%
  \BibitemOpen
  \bibfield  {author} {\bibinfo {author} {\bibfnamefont {P.}~\bibnamefont
  {Ajith}}, \bibinfo {author} {\bibfnamefont {N.}~\bibnamefont {Fotopoulos}},
  \bibinfo {author} {\bibfnamefont {S.}~\bibnamefont {Privitera}}, \bibinfo
  {author} {\bibfnamefont {A.}~\bibnamefont {Neunzert}}, \ and\ \bibinfo
  {author} {\bibfnamefont {A.~J.}\ \bibnamefont {Weinstein}},\ }\bibfield
  {title} {\enquote {\bibinfo {title} {{Effectual template bank for the
  detection of gravitational waves from inspiralling compact binaries with
  generic spins}},}\ }\href {\doibase 10.1103/PhysRevD.89.084041} {\bibfield
  {journal} {\bibinfo  {journal} {Phys. Rev. D}\ }\textbf {\bibinfo {volume}
  {89}},\ \bibinfo {pages} {084041} (\bibinfo {year} {2014})},\ \Eprint
  {http://arxiv.org/abs/1210.6666} {arXiv:1210.6666 [gr-qc]} \BibitemShut
  {NoStop}%
\bibitem [{\citenamefont {Virtanen}\ \emph {et~al.}(2020)\citenamefont
  {Virtanen}, \citenamefont {Gommers}, \citenamefont {Oliphant}, \citenamefont
  {Haberland}, \citenamefont {Reddy}, \citenamefont {Cournapeau}, \citenamefont
  {Burovski}, \citenamefont {Peterson}, \citenamefont {Weckesser},
  \citenamefont {Bright}, \citenamefont {{van der Walt}}, \citenamefont
  {Brett}, \citenamefont {Wilson}, \citenamefont {Millman}, \citenamefont
  {Mayorov}, \citenamefont {Nelson}, \citenamefont {Jones}, \citenamefont
  {Kern}, \citenamefont {Larson}, \citenamefont {Carey}, \citenamefont {Polat},
  \citenamefont {Feng}, \citenamefont {Moore}, \citenamefont {{VanderPlas}},
  \citenamefont {Laxalde}, \citenamefont {Perktold}, \citenamefont {Cimrman},
  \citenamefont {Henriksen}, \citenamefont {Quintero}, \citenamefont {Harris},
  \citenamefont {Archibald}, \citenamefont {Ribeiro}, \citenamefont
  {Pedregosa}, \citenamefont {{van Mulbregt}},\ and\ \citenamefont {{SciPy 1.0
  Contributors}}}]{2020SciPy-NMeth}%
  \BibitemOpen
  \bibfield  {author} {\bibinfo {author} {\bibfnamefont {Pauli}\ \bibnamefont
  {Virtanen}}, \bibinfo {author} {\bibfnamefont {Ralf}\ \bibnamefont
  {Gommers}}, \bibinfo {author} {\bibfnamefont {Travis~E.}\ \bibnamefont
  {Oliphant}}, \bibinfo {author} {\bibfnamefont {Matt}\ \bibnamefont
  {Haberland}}, \bibinfo {author} {\bibfnamefont {Tyler}\ \bibnamefont
  {Reddy}}, \bibinfo {author} {\bibfnamefont {David}\ \bibnamefont
  {Cournapeau}}, \bibinfo {author} {\bibfnamefont {Evgeni}\ \bibnamefont
  {Burovski}}, \bibinfo {author} {\bibfnamefont {Pearu}\ \bibnamefont
  {Peterson}}, \bibinfo {author} {\bibfnamefont {Warren}\ \bibnamefont
  {Weckesser}}, \bibinfo {author} {\bibfnamefont {Jonathan}\ \bibnamefont
  {Bright}}, \bibinfo {author} {\bibfnamefont {St{\'e}fan~J.}\ \bibnamefont
  {{van der Walt}}}, \bibinfo {author} {\bibfnamefont {Matthew}\ \bibnamefont
  {Brett}}, \bibinfo {author} {\bibfnamefont {Joshua}\ \bibnamefont {Wilson}},
  \bibinfo {author} {\bibfnamefont {K.~Jarrod}\ \bibnamefont {Millman}},
  \bibinfo {author} {\bibfnamefont {Nikolay}\ \bibnamefont {Mayorov}}, \bibinfo
  {author} {\bibfnamefont {Andrew R.~J.}\ \bibnamefont {Nelson}}, \bibinfo
  {author} {\bibfnamefont {Eric}\ \bibnamefont {Jones}}, \bibinfo {author}
  {\bibfnamefont {Robert}\ \bibnamefont {Kern}}, \bibinfo {author}
  {\bibfnamefont {Eric}\ \bibnamefont {Larson}}, \bibinfo {author}
  {\bibfnamefont {C~J}\ \bibnamefont {Carey}}, \bibinfo {author} {\bibfnamefont
  {{\.I}lhan}\ \bibnamefont {Polat}}, \bibinfo {author} {\bibfnamefont
  {Yu}~\bibnamefont {Feng}}, \bibinfo {author} {\bibfnamefont {Eric~W.}\
  \bibnamefont {Moore}}, \bibinfo {author} {\bibfnamefont {Jake}\ \bibnamefont
  {{VanderPlas}}}, \bibinfo {author} {\bibfnamefont {Denis}\ \bibnamefont
  {Laxalde}}, \bibinfo {author} {\bibfnamefont {Josef}\ \bibnamefont
  {Perktold}}, \bibinfo {author} {\bibfnamefont {Robert}\ \bibnamefont
  {Cimrman}}, \bibinfo {author} {\bibfnamefont {Ian}\ \bibnamefont
  {Henriksen}}, \bibinfo {author} {\bibfnamefont {E.~A.}\ \bibnamefont
  {Quintero}}, \bibinfo {author} {\bibfnamefont {Charles~R.}\ \bibnamefont
  {Harris}}, \bibinfo {author} {\bibfnamefont {Anne~M.}\ \bibnamefont
  {Archibald}}, \bibinfo {author} {\bibfnamefont {Ant{\^o}nio~H.}\ \bibnamefont
  {Ribeiro}}, \bibinfo {author} {\bibfnamefont {Fabian}\ \bibnamefont
  {Pedregosa}}, \bibinfo {author} {\bibfnamefont {Paul}\ \bibnamefont {{van
  Mulbregt}}}, \ and\ \bibinfo {author} {\bibnamefont {{SciPy 1.0
  Contributors}}},\ }\bibfield  {title} {\enquote {\bibinfo {title} {{{SciPy}
  1.0: Fundamental Algorithms for Scientific Computing in Python}},}\ }\href
  {\doibase 10.1038/s41592-019-0686-2} {\bibfield  {journal} {\bibinfo
  {journal} {Nature Methods}\ }\textbf {\bibinfo {volume} {17}},\ \bibinfo
  {pages} {261--272} (\bibinfo {year} {2020})}\BibitemShut {NoStop}%
\bibitem [{\citenamefont {Vitale}\ and\ \citenamefont
  {Del~Pozzo}(2014)}]{Vitale:2013bma}%
  \BibitemOpen
  \bibfield  {author} {\bibinfo {author} {\bibfnamefont {Salvatore}\
  \bibnamefont {Vitale}}\ and\ \bibinfo {author} {\bibfnamefont {Walter}\
  \bibnamefont {Del~Pozzo}},\ }\bibfield  {title} {\enquote {\bibinfo {title}
  {{How serious can the stealth bias be in gravitational wave parameter
  estimation?}}}\ }\href {\doibase 10.1103/PhysRevD.89.022002} {\bibfield
  {journal} {\bibinfo  {journal} {Phys. Rev. D}\ }\textbf {\bibinfo {volume}
  {89}},\ \bibinfo {pages} {022002} (\bibinfo {year} {2014})},\ \Eprint
  {http://arxiv.org/abs/1311.2057} {arXiv:1311.2057 [gr-qc]} \BibitemShut
  {NoStop}%
\bibitem [{\citenamefont {Behnel}\ \emph {et~al.}(2011)\citenamefont {Behnel},
  \citenamefont {Bradshaw}, \citenamefont {Citro}, \citenamefont {Dalcin},
  \citenamefont {Seljebotn},\ and\ \citenamefont {Smith}}]{behnel2011cython}%
  \BibitemOpen
  \bibfield  {author} {\bibinfo {author} {\bibfnamefont {Stefan}\ \bibnamefont
  {Behnel}}, \bibinfo {author} {\bibfnamefont {Robert}\ \bibnamefont
  {Bradshaw}}, \bibinfo {author} {\bibfnamefont {Craig}\ \bibnamefont {Citro}},
  \bibinfo {author} {\bibfnamefont {Lisandro}\ \bibnamefont {Dalcin}}, \bibinfo
  {author} {\bibfnamefont {Dag~Sverre}\ \bibnamefont {Seljebotn}}, \ and\
  \bibinfo {author} {\bibfnamefont {Kurt}\ \bibnamefont {Smith}},\ }\bibfield
  {title} {\enquote {\bibinfo {title} {Cython: The best of both worlds},}\
  }\href@noop {} {\bibfield  {journal} {\bibinfo  {journal} {Computing in
  Science \& Engineering}\ }\textbf {\bibinfo {volume} {13}},\ \bibinfo {pages}
  {31--39} (\bibinfo {year} {2011})}\BibitemShut {NoStop}%
\bibitem [{\citenamefont {Harris}\ \emph {et~al.}(2020)\citenamefont {Harris},
  \citenamefont {Millman}, \citenamefont {van~der Walt}, \citenamefont
  {Gommers}, \citenamefont {Virtanen}, \citenamefont {Cournapeau},
  \citenamefont {Wieser}, \citenamefont {Taylor}, \citenamefont {Berg},
  \citenamefont {Smith}, \citenamefont {Kern}, \citenamefont {Picus},
  \citenamefont {Hoyer}, \citenamefont {van Kerkwijk}, \citenamefont {Brett},
  \citenamefont {Haldane}, \citenamefont {del R{\'{i}}o}, \citenamefont
  {Wiebe}, \citenamefont {Peterson}, \citenamefont {G{\'{e}}rard-Marchant},
  \citenamefont {Sheppard}, \citenamefont {Reddy}, \citenamefont {Weckesser},
  \citenamefont {Abbasi}, \citenamefont {Gohlke},\ and\ \citenamefont
  {Oliphant}}]{harris2020array}%
  \BibitemOpen
  \bibfield  {author} {\bibinfo {author} {\bibfnamefont {Charles~R.}\
  \bibnamefont {Harris}}, \bibinfo {author} {\bibfnamefont {K.~Jarrod}\
  \bibnamefont {Millman}}, \bibinfo {author} {\bibfnamefont {St{\'{e}}fan~J.}\
  \bibnamefont {van~der Walt}}, \bibinfo {author} {\bibfnamefont {Ralf}\
  \bibnamefont {Gommers}}, \bibinfo {author} {\bibfnamefont {Pauli}\
  \bibnamefont {Virtanen}}, \bibinfo {author} {\bibfnamefont {David}\
  \bibnamefont {Cournapeau}}, \bibinfo {author} {\bibfnamefont {Eric}\
  \bibnamefont {Wieser}}, \bibinfo {author} {\bibfnamefont {Julian}\
  \bibnamefont {Taylor}}, \bibinfo {author} {\bibfnamefont {Sebastian}\
  \bibnamefont {Berg}}, \bibinfo {author} {\bibfnamefont {Nathaniel~J.}\
  \bibnamefont {Smith}}, \bibinfo {author} {\bibfnamefont {Robert}\
  \bibnamefont {Kern}}, \bibinfo {author} {\bibfnamefont {Matti}\ \bibnamefont
  {Picus}}, \bibinfo {author} {\bibfnamefont {Stephan}\ \bibnamefont {Hoyer}},
  \bibinfo {author} {\bibfnamefont {Marten~H.}\ \bibnamefont {van Kerkwijk}},
  \bibinfo {author} {\bibfnamefont {Matthew}\ \bibnamefont {Brett}}, \bibinfo
  {author} {\bibfnamefont {Allan}\ \bibnamefont {Haldane}}, \bibinfo {author}
  {\bibfnamefont {Jaime~Fern{\'{a}}ndez}\ \bibnamefont {del R{\'{i}}o}},
  \bibinfo {author} {\bibfnamefont {Mark}\ \bibnamefont {Wiebe}}, \bibinfo
  {author} {\bibfnamefont {Pearu}\ \bibnamefont {Peterson}}, \bibinfo {author}
  {\bibfnamefont {Pierre}\ \bibnamefont {G{\'{e}}rard-Marchant}}, \bibinfo
  {author} {\bibfnamefont {Kevin}\ \bibnamefont {Sheppard}}, \bibinfo {author}
  {\bibfnamefont {Tyler}\ \bibnamefont {Reddy}}, \bibinfo {author}
  {\bibfnamefont {Warren}\ \bibnamefont {Weckesser}}, \bibinfo {author}
  {\bibfnamefont {Hameer}\ \bibnamefont {Abbasi}}, \bibinfo {author}
  {\bibfnamefont {Christoph}\ \bibnamefont {Gohlke}}, \ and\ \bibinfo {author}
  {\bibfnamefont {Travis~E.}\ \bibnamefont {Oliphant}},\ }\bibfield  {title}
  {\enquote {\bibinfo {title} {Array programming with {NumPy}},}\ }\href
  {\doibase 10.1038/s41586-020-2649-2} {\bibfield  {journal} {\bibinfo
  {journal} {Nature}\ }\textbf {\bibinfo {volume} {585}},\ \bibinfo {pages}
  {357--362} (\bibinfo {year} {2020})}\BibitemShut {NoStop}%
\bibitem [{\citenamefont {{LIGO Scientific
  Collaboration}}(2020)}]{2020ascl.soft12021L}%
  \BibitemOpen
  \bibfield  {author} {\bibinfo {author} {\bibnamefont {{LIGO Scientific
  Collaboration}}},\ }\href@noop {} {\enquote {\bibinfo {title} {{LALSuite:
  LIGO Scientific Collaboration Algorithm Library Suite}},}\ }\bibinfo
  {howpublished} {Astrophysics Source Code Library, record ascl:2012.021}
  (\bibinfo {year} {2020}),\ \Eprint {http://arxiv.org/abs/2012.021}
  {ascl:2012.021} \BibitemShut {NoStop}%
\bibitem [{\citenamefont {Romero-Shaw}\ \emph {et~al.}(2020)\citenamefont
  {Romero-Shaw} \emph {et~al.}}]{Romero-Shaw:2020owr}%
  \BibitemOpen
  \bibfield  {author} {\bibinfo {author} {\bibfnamefont {I.~M.}\ \bibnamefont
  {Romero-Shaw}} \emph {et~al.},\ }\bibfield  {title} {\enquote {\bibinfo
  {title} {{Bayesian inference for compact binary coalescences with bilby:
  validation and application to the first LIGO\textendash{}Virgo
  gravitational-wave transient catalogue}},}\ }\href {\doibase
  10.1093/mnras/staa2850} {\bibfield  {journal} {\bibinfo  {journal} {Mon. Not.
  Roy. Astron. Soc.}\ }\textbf {\bibinfo {volume} {499}},\ \bibinfo {pages}
  {3295--3319} (\bibinfo {year} {2020})},\ \Eprint
  {http://arxiv.org/abs/2006.00714} {arXiv:2006.00714 [astro-ph.IM]}
  \BibitemShut {NoStop}%
\bibitem [{\citenamefont {{Talbot}}\ \emph {et~al.}(2019)\citenamefont
  {{Talbot}}, \citenamefont {{Smith}}, \citenamefont {{Thrane}},\ and\
  \citenamefont {{Poole}}}]{2019PhRvD.100d3030T}%
  \BibitemOpen
  \bibfield  {author} {\bibinfo {author} {\bibfnamefont {Colm}\ \bibnamefont
  {{Talbot}}}, \bibinfo {author} {\bibfnamefont {Rory}\ \bibnamefont
  {{Smith}}}, \bibinfo {author} {\bibfnamefont {Eric}\ \bibnamefont
  {{Thrane}}}, \ and\ \bibinfo {author} {\bibfnamefont {Gregory~B.}\
  \bibnamefont {{Poole}}},\ }\bibfield  {title} {\enquote {\bibinfo {title}
  {{Parallelized inference for gravitational-wave astronomy}},}\ }\href
  {\doibase 10.1103/PhysRevD.100.043030} {\bibfield  {journal} {\bibinfo
  {journal} {\prd}\ }\textbf {\bibinfo {volume} {100}},\ \bibinfo {eid}
  {043030} (\bibinfo {year} {2019})},\ \Eprint
  {http://arxiv.org/abs/1904.02863} {arXiv:1904.02863 [astro-ph.IM]}
  \BibitemShut {NoStop}%
\bibitem [{\citenamefont {Hunter}(2007)}]{Hunter:2007}%
  \BibitemOpen
  \bibfield  {author} {\bibinfo {author} {\bibfnamefont {J.~D.}\ \bibnamefont
  {Hunter}},\ }\bibfield  {title} {\enquote {\bibinfo {title} {Matplotlib: A 2d
  graphics environment},}\ }\href {\doibase 10.1109/MCSE.2007.55} {\bibfield
  {journal} {\bibinfo  {journal} {Computing in Science \& Engineering}\
  }\textbf {\bibinfo {volume} {9}},\ \bibinfo {pages} {90--95} (\bibinfo {year}
  {2007})}\BibitemShut {NoStop}%
\bibitem [{\citenamefont {Kluyver}\ \emph {et~al.}(2016)\citenamefont
  {Kluyver}, \citenamefont {Ragan-Kelley}, \citenamefont {P{\'e}rez},
  \citenamefont {Granger}, \citenamefont {Bussonnier}, \citenamefont
  {Frederic}, \citenamefont {Kelley}, \citenamefont {Hamrick}, \citenamefont
  {Grout}, \citenamefont {Corlay}, \citenamefont {Ivanov}, \citenamefont
  {Avila}, \citenamefont {Abdalla},\ and\ \citenamefont
  {Willing}}]{Kluyver2016jupyter}%
  \BibitemOpen
  \bibfield  {author} {\bibinfo {author} {\bibfnamefont {Thomas}\ \bibnamefont
  {Kluyver}}, \bibinfo {author} {\bibfnamefont {Benjamin}\ \bibnamefont
  {Ragan-Kelley}}, \bibinfo {author} {\bibfnamefont {Fernando}\ \bibnamefont
  {P{\'e}rez}}, \bibinfo {author} {\bibfnamefont {Brian}\ \bibnamefont
  {Granger}}, \bibinfo {author} {\bibfnamefont {Matthias}\ \bibnamefont
  {Bussonnier}}, \bibinfo {author} {\bibfnamefont {Jonathan}\ \bibnamefont
  {Frederic}}, \bibinfo {author} {\bibfnamefont {Kyle}\ \bibnamefont {Kelley}},
  \bibinfo {author} {\bibfnamefont {Jessica}\ \bibnamefont {Hamrick}}, \bibinfo
  {author} {\bibfnamefont {Jason}\ \bibnamefont {Grout}}, \bibinfo {author}
  {\bibfnamefont {Sylvain}\ \bibnamefont {Corlay}}, \bibinfo {author}
  {\bibfnamefont {Paul}\ \bibnamefont {Ivanov}}, \bibinfo {author}
  {\bibfnamefont {Dami{\'a}n}\ \bibnamefont {Avila}}, \bibinfo {author}
  {\bibfnamefont {Safia}\ \bibnamefont {Abdalla}}, \ and\ \bibinfo {author}
  {\bibfnamefont {Carol}\ \bibnamefont {Willing}},\ }\bibfield  {title}
  {\enquote {\bibinfo {title} {Jupyter notebooks -- a publishing format for
  reproducible computational workflows},}\ }in\ \href@noop {} {\emph {\bibinfo
  {booktitle} {Positioning and Power in Academic Publishing: Players, Agents
  and Agendas}}},\ \bibinfo {editor} {edited by\ \bibinfo {editor}
  {\bibfnamefont {F.}~\bibnamefont {Loizides}}\ and\ \bibinfo {editor}
  {\bibfnamefont {B.}~\bibnamefont {Schmidt}}}\ (\bibinfo {organization} {IOS
  Press},\ \bibinfo {year} {2016})\ pp.\ \bibinfo {pages} {87 --
  90}\BibitemShut {NoStop}%
\bibitem [{\citenamefont {Shapiro}\ and\ \citenamefont
  {Wilk}(1965)}]{shapiro1965analysis}%
  \BibitemOpen
  \bibfield  {author} {\bibinfo {author} {\bibfnamefont {Samuel~Sanford}\
  \bibnamefont {Shapiro}}\ and\ \bibinfo {author} {\bibfnamefont {Martin~B}\
  \bibnamefont {Wilk}},\ }\bibfield  {title} {\enquote {\bibinfo {title} {An
  analysis of variance test for normality (complete samples)},}\ }\href@noop {}
  {\bibfield  {journal} {\bibinfo  {journal} {Biometrika}\ }\textbf {\bibinfo
  {volume} {52}},\ \bibinfo {pages} {591--611} (\bibinfo {year}
  {1965})}\BibitemShut {NoStop}%
\bibitem [{\citenamefont {Mendes}\ and\ \citenamefont
  {Pala}(2003)}]{mendes2003type}%
  \BibitemOpen
  \bibfield  {author} {\bibinfo {author} {\bibfnamefont {Mehmet}\ \bibnamefont
  {Mendes}}\ and\ \bibinfo {author} {\bibfnamefont {Akin}\ \bibnamefont
  {Pala}},\ }\bibfield  {title} {\enquote {\bibinfo {title} {Type i error rate
  and power of three normality tests},}\ }\href@noop {} {\bibfield  {journal}
  {\bibinfo  {journal} {Pakistan Journal of Information and Technology}\
  }\textbf {\bibinfo {volume} {2}},\ \bibinfo {pages} {135--139} (\bibinfo
  {year} {2003})}\BibitemShut {NoStop}%
\bibitem [{\citenamefont {Keskin}(2006)}]{keskin2006comparison}%
  \BibitemOpen
  \bibfield  {author} {\bibinfo {author} {\bibfnamefont {Siddik}\ \bibnamefont
  {Keskin}},\ }\bibfield  {title} {\enquote {\bibinfo {title} {Comparison of
  several univariate normality tests regarding type i error rate and power of
  the test in simulation based small samples},}\ }\href@noop {} {\bibfield
  {journal} {\bibinfo  {journal} {Journal of Applied Science Research}\
  }\textbf {\bibinfo {volume} {2}},\ \bibinfo {pages} {296--300} (\bibinfo
  {year} {2006})}\BibitemShut {NoStop}%
\bibitem [{\citenamefont {Razali}\ \emph {et~al.}(2011)\citenamefont {Razali},
  \citenamefont {Wah} \emph {et~al.}}]{razali2011power}%
  \BibitemOpen
  \bibfield  {author} {\bibinfo {author} {\bibfnamefont {Nornadiah~Mohd}\
  \bibnamefont {Razali}}, \bibinfo {author} {\bibfnamefont {Yap~Bee}\
  \bibnamefont {Wah}},  \emph {et~al.},\ }\bibfield  {title} {\enquote
  {\bibinfo {title} {Power comparisons of shapiro-wilk, kolmogorov-smirnov,
  lilliefors and anderson-darling tests},}\ }\href@noop {} {\bibfield
  {journal} {\bibinfo  {journal} {Journal of statistical modeling and
  analytics}\ }\textbf {\bibinfo {volume} {2}},\ \bibinfo {pages} {21--33}
  (\bibinfo {year} {2011})}\BibitemShut {NoStop}%
\bibitem [{\citenamefont {Abbott}\ \emph
  {et~al.}(2016{\natexlab{b}})\citenamefont {Abbott} \emph
  {et~al.}}]{Abbott:2016xvh}%
  \BibitemOpen
  \bibfield  {author} {\bibinfo {author} {\bibfnamefont {Benjamin~P.}\
  \bibnamefont {Abbott}} \emph {et~al.},\ }\bibfield  {title} {\enquote
  {\bibinfo {title} {{Sensitivity of the Advanced LIGO detectors at the
  beginning of gravitational wave astronomy}},}\ }\href {\doibase
  10.1103/PhysRevD.93.112004} {\bibfield  {journal} {\bibinfo  {journal} {Phys.
  Rev. D}\ }\textbf {\bibinfo {volume} {93}},\ \bibinfo {pages} {112004}
  (\bibinfo {year} {2016}{\natexlab{b}})},\ \bibinfo {note} {[Addendum:
  Phys.Rev.D 97, 059901 (2018)]},\ \Eprint {http://arxiv.org/abs/1604.00439}
  {arXiv:1604.00439 [astro-ph.IM]} \BibitemShut {NoStop}%
\bibitem [{\citenamefont {Valdes}\ \emph {et~al.}(2022)\citenamefont {Valdes},
  \citenamefont {Hines}, \citenamefont {Nelson}, \citenamefont {Zhang},\ and\
  \citenamefont {Guzman}}]{Valdes:2022pnm}%
  \BibitemOpen
  \bibfield  {author} {\bibinfo {author} {\bibfnamefont {Guillermo}\
  \bibnamefont {Valdes}}, \bibinfo {author} {\bibfnamefont {Adam}\ \bibnamefont
  {Hines}}, \bibinfo {author} {\bibfnamefont {Andrea}\ \bibnamefont {Nelson}},
  \bibinfo {author} {\bibfnamefont {Yanqi}\ \bibnamefont {Zhang}}, \ and\
  \bibinfo {author} {\bibfnamefont {Felipe}\ \bibnamefont {Guzman}},\
  }\bibfield  {title} {\enquote {\bibinfo {title} {{A characterization method
  for low-frequency seismic noise in LIGO}},}\ }\href {\doibase
  10.1063/5.0122495} {\bibfield  {journal} {\bibinfo  {journal} {Appl. Phys.
  Lett.}\ }\textbf {\bibinfo {volume} {121}},\ \bibinfo {pages} {234102}
  (\bibinfo {year} {2022})},\ \Eprint {http://arxiv.org/abs/2209.04452}
  {arXiv:2209.04452 [astro-ph.IM]} \BibitemShut {NoStop}%
\bibitem [{\citenamefont {Hanna}\ \emph {et~al.}(2009)\citenamefont {Hanna},
  \citenamefont {Megevand}, \citenamefont {Ochsner},\ and\ \citenamefont
  {Palenzuela}}]{Hanna:2008um}%
  \BibitemOpen
  \bibfield  {author} {\bibinfo {author} {\bibfnamefont {Chad}\ \bibnamefont
  {Hanna}}, \bibinfo {author} {\bibfnamefont {Miguel}\ \bibnamefont
  {Megevand}}, \bibinfo {author} {\bibfnamefont {Evan}\ \bibnamefont
  {Ochsner}}, \ and\ \bibinfo {author} {\bibfnamefont {Carlos}\ \bibnamefont
  {Palenzuela}},\ }\bibfield  {title} {\enquote {\bibinfo {title} {{A method
  for estimating time-frequency characteristics of compact binary mergers to
  improve searches for inspiral, merger and ring-down phases separately}},}\
  }\href {\doibase 10.1088/0264-9381/26/1/015009} {\bibfield  {journal}
  {\bibinfo  {journal} {Class. Quant. Grav.}\ }\textbf {\bibinfo {volume}
  {26}},\ \bibinfo {pages} {015009} (\bibinfo {year} {2009})},\ \Eprint
  {http://arxiv.org/abs/0801.4297} {arXiv:0801.4297 [gr-qc]} \BibitemShut
  {NoStop}%
\end{thebibliography}%

\end{document}